\documentclass[12pt]{article}
\usepackage{latexsym,amsmath,amsfonts,epsf}
\def\beq{\begin{eqnarray}}
\def\eeq{\end{eqnarray}}
\def\nn{\nonumber\\}
\def\eff{\mbox{$\Gamma\lbrack\varphi\rbrack$}}
\def\x{\mbox{${\mathbf x}$}}
\def\reals{\mbox{$\mathbb R$}}
\def\vol{\mbox{$d\sigma_{\mathbf x}$}}
\def\Int{\mbox{$\displaystyle{\int_{\Sigma}\!\!\!d\sigma_{\mathbf x}\,}$}}
\def\mom{\mbox{$\displaystyle{\int\frac{d^Dk}{(2\pi)^D}}$}}
\def\bp{\mbox{$\Bar{\varphi}$}}
\def\V1{\mbox{$\displaystyle{V^{(1)}_{\rm eff}}$}}
\def\Z{\mbox{${\mathfrak Z}$}}
\def\cint{\mbox{$\displaystyle{\frac{1}{2\pi i}
\int_{c-i\infty}^{c+i\infty}\!\!\!\!\!\!\!\!\!d\alpha\,}$}}
\def\bpsi{\mbox{$\Bar{\Psi}$}}
\def\bphi{\mbox{$\Bar{\Phi}$}}
\begin{document}
\numberwithin{equation}{section}
\begin{center}
{\bf The Effective Action at Finite Temperature and Density With Application to Bose-Einstein Condensation}
\end{center}
\vspace{1cm}
\begin{center}
{\sc DAVID J. TOMS}\\
{\em Physics Department, The University of Newcastle upon Tyne,\\
Newcastle upon Tyne, U. K. NE1 7RU}
\end{center}
\vspace{1.5cm}
\begin{center}
{\bf Abstract}
\end{center}

\begin{center}
\parbox{29pc}{\footnotesize A simple pedagogical introduction to the effective action method of quantum field theory is given at a level suitable for beginning postgraduate students. It is shown how to obtain the effective potential at zero temperature from a regularized zero-point energy. The results are applicable to curved as well as to flat space. The generalization to finite temperature is also given. It is shown how to obtain high temperature expansions of the thermodynamic potential for the neutral free Bose gas and the charged Bose gas in both the relativistic and non-relativistic limits. The results are obtained for an arbitrary spatial dimension and in curved space. Results are also obtained for the self-interacting relativistic gas in three spatial dimensions. A detailed discussion of how the formalism may be applied to study Bose-Einstein condensation is given. The interpretation of Bose-Einstein condensation as symmetry breaking is discussed. Application is also given to the study of charged bosons, both relativistic and non-relativistic, in a constant magnetic field. The Meissner effect is obtained for the non-relativistic Bose gas in three spatial dimensions. The final application is to the study of non-relativistic bosons in a harmonic oscillator potential trap. A number of analytical approaches to this are discussed.}
\end{center}

\setcounter{section}{-1}
\section{Foreword}
These notes represent the lectures given at The 15th Symposium on Theoretical Physics August 22--27, 1996 at Seoul National University. The intent of these lectures was to present an introduction to the use of the effective action method in quantum field theory at a level appropriate to a beginning postgraduate student. I assume that the prospective reader has at least a nodding acquaintance with canonical quantization in quantum field theory but no detailed knowledge of the more advanced aspects. I purposely avoid the use of path integral methods and the formal properties of the effective action since they are not necessary for the examples I will discuss. It is hoped that these notes will make it easier for a student to make the bridge between the elementary concepts of quantum field theory and the more advanced ideas used in contemporary research. Particular emphasis is placed in these lectures on the application to Bose-Einstein condensation (BEC), a topic of much current investigation. A much more comprehensive introduction to the effective action method will be contained in a forthcoming book \cite{DJTbook}.

\section{Action Principle in Classical Physics}
\label{sec1}
It is well known how the action principle can be used in classical mechanics to derive the equations of motion and to discuss symmetries. (See Ref.~\cite{Lanczos} for example.) The Euler-Lagrange equations can be found by demanding that the classical action functional
\beq
S\lbrack x(t)\rbrack=\int_{t_1}^{t_2}dt\,L(t)\;,\label{1.1}
\eeq
be stationary under all possible variations of the path $x(t)$. In Eq.~\ref{1.1} $L(t)$ is the Lagrangian of the system, which is defined to be the difference between the kinetic and potential energies. The square brackets in Eq.~\ref{1.1} are used to denote a functional, which may be regarded as a real or complex valued function whose domain is a space of functions. The Euler-Lagrange equations follow from the demand that $S\lbrack x(t)+\delta x(t)\rbrack=S\lbrack x(t)\rbrack$ to first order in the infinitesimal variation of the path $\delta x(t)$. If instead of the Lagrange form of the equations we are interested in the Hamiltonian form, then we may take
\beq
S\lbrack x(t),p(t)\rbrack=\int_{t_1}^{t_2}dt\lbrace p(t)\Dot{x}(t)-H(t)\rbrace\;,\label{1.2}
\eeq
and demand that it be stationary under independent variations of the path $x(t)$ and the momentum $p(t)$. $H(t)$ is the Hamiltonian for the system. (See \cite{Lanczos} for more details.)

The action principle extends immediately from classical particle mechanics to classical field theory. The prototype is the vibrating string in one spatial dimension. Let $\psi(t,x)$ represent the displacement of a point on the string at position $x$ at time $t$. If we consider a small section of string of length $d\ell$ and assume that the displacement of the string from its equilibrium position along the $x$-axis is small then $d\ell\simeq dx$. This little section of string has kinetic energy $\displaystyle{\frac{1}{2}(\rho dx)\left(\frac{\partial\psi(t,x)}{\partial t}\right)^2}$ and potential energy
$\displaystyle{\frac{1}{2}(\tau dx)\left(\frac{\partial\psi(t,x)}{\partial x}\right)^2}$ where $\rho$ is the mass per unit length and $\tau$ is the tension. The Lagrangian for this system is therefore
\beq
L(t)=\int dx\left\lbrace
\frac{1}{2}\rho\left(\frac{\partial\psi(t,x)}{\partial t}\right)^2-
\frac{1}{2}\tau\left(\frac{\partial\psi(t,x)}{\partial x}\right)^2\right\rbrace\;,\label{1.3}
\eeq
where the integration extends over the total length of the string. The integrand of Eq.~\ref{1.3} is referred to as the Lagrangian density. Variation of the action defined as in Eq.~\ref{1.1} with $L(t)$ given in Eq.~\ref{1.3} with respect to an arbitrary variation $\delta\psi(t,x)$ leads to the familiar one-dimensional wave equation. The extension to any number of spatial dimensions is obvious.

As already mentioned, one advantage of using an action principle approach is that it is possible to discuss symmetries of the theory in a simple way. This is done by the requirement that the action functional be a scalar under the symmetry transformation. For example, for a particle moving on a curved background (eg. a particle constrained to move on a surface) the action should be a  scalar under a general coordinate transformation on the curved background. This reflects the physical requirement that the motion of the particle should not depend on the arbitrary choice of coordinates used to describe the curved background. If we consider relativistic field theory, then the action should be a scalar under Lorentz transformations for a theory in Minkowski spacetime. For fields in curved spacetime, the action should be a scalar under local Lorentz transformations as well as under general coordinate transformations. For gauge theories we would demand that the action be a scalar under the action of the gauge group. The requirement of invariance leads via Noether's theorem to a conservation law. (See \cite{AbersandLee} for example.)

\section{Action Principle in Quantum Physics}
\label{sec2}

Because classical physics is assumed to be an approximation to the more fundamental quantum physics, it is natural to ask if there is a quantum analogue to the classical action principle. In particular we can ask the question {\it Is there a quantum version of the classical action functional ?}. One approach to this was initiated by Schwinger in the early 1950's in a series of papers \cite{Schwinger}. In this work it was shown how a quantum version of the classical action principle could be formulated, and which led to the Heisenberg equations of motion for the quantum operators. Instead of pursuing this formalism directly, we will proceed in a more heuristic manner. Put simply, we wish to find an action functional \eff\ for the basic field variables $\varphi$ which includes all possible quantum corrections to the classical action $S\lbrack\varphi\rbrack$. Thus we must have
\beq
\eff=S\lbrack\varphi\rbrack+{\mathcal O}(\hbar)\;,\label{2.1}
\eeq
where the terms of order $\hbar$ denote the quantum corrections to the classical theory It is customary to call \eff\ the effective action.

The effective action would be expected to share as many symmetries of the classical theory as possible, although one should be aware that some classical symmetries (such as chiral or conformal invariance) may be broken by quantum effects. (The physical reason for the breaking of some classical symmetries in the quantum theory is that in quantum field theory the process of regularization necessarily introduces a mass scale into the theory; therefore any classical symmetries which depend on the property of masslessness in the classical theory may be broken.) Particular symmetries which we would expect \eff\ to enjoy are Lorentz invariance if the theory is in Minkowski spacetime, or general coordinate and local Lorentz invariance if the theory is in curved spacetime. In general we would also expect \eff\ to be gauge invariant if the classical theory is gauge invariant.

Although it is possible to be more general, we will limit our attention to spacetimes which are ultrastatic \cite{Fulling}. In this case the line element may be written as
\beq
ds^2=dt^2-g_{ij}(\x)dx^idx^j\;,\label{2.2}
\eeq
with $t$ the time and $x^i\;(i=1,\ldots,D)$ the spatial coordinates. We will assume that the spacetime ${\mathcal M}\simeq\reals\times\Sigma$ with $\Sigma$ a $D$-dimensional Riemannian manifold which is compact with or without boundary. In the case where we are interested in Minkowski spacetime, rather than dealing with a space of infinite volume, it often proves convenient to restrict attention to a finite box with some boundary conditions, usually periodic, imposed on the fields on the box. Our assumption that ${\mathcal M}\simeq\reals\times\Sigma$ is just a generalization of this procedure. In general if we are interested in the case where $\Sigma$ is really non-compact, we will assume that we have added some spatial boundary to obtain a compact spacetime with boundary, and then look at what happens when the boundary is removed to infinity at the end of any calculation. The restriction to static spacetimes is to ensure that there is an unambiguous definition of a vacuum state, and what is meant by the notion of a particle.

Given the line element in Eq.~\ref{2.2} the invariant volume element for the spacetime is
\beq
dv_x=\sqrt{g(\x)}\,dtd^Dx\;,\label{2.3}
\eeq
where $g(\x)=\mbox{det}\,g_{ij}(\x)$. We will often use
\beq
\vol=\sqrt{g(\x)}\,d^Dx\;,\label{2.4}
\eeq
to denote the invariant volume element on $\Sigma$.

The classical action is defined in terms of the Lagrangian as in Eq.~\ref{1.1}. In some cases we may wish to define the Lagrangian density ${\mathcal L}(t,\x)$ in terms of the Lagrangian by
\beq
L(t)=\Int{\mathcal L}(t,\x)\;,\label{2.5}
\eeq
so that
\beq
S\lbrack\varphi\rbrack=\int_{t_1}^{t_2}dt\Int{\mathcal L}(t,\x)\;.\label{2.6}
\eeq
For the effective action \eff\ we can define an effective Lagrangian density in a completely analogous way~:
\beq
\eff=\int_{t_1}^{t_2}dt\Int{\mathcal L}_{\rm eff}(t,\x)\;.\label{2.7}
\eeq
Given an expression for \eff\ we can obtain the effective field equations from the requirement that $\Gamma\lbrack\varphi+\delta\varphi\rbrack=\eff$ to first order in the arbitrary variation $\delta\varphi$.

Because the Lagrangian is the difference between the kinetic and potential energies, we can define a kinetic energy density and a potential energy density in an obvious way. In many situations of interest the potential energy contains the interesting physics. This is the case if the field is assumed to be constant, or more generally independent of time. In such a situation we can write
\beq
\eff=-(t_2-t_1)V_{\Sigma}V_{\rm eff}\;,\label{2.9}
\eeq
where $V_{\Sigma}$ is the volume of $\Sigma$, and $V_{\rm eff}$ is the effective potential. ($V_{\rm eff}$ is really the effective potential density, but it is customary to just call it the effective potential.)

\section{Massive Real Scalar Field}
\label{sec3}

The classical action functional describing a massive real scalar field is
\beq
S\lbrack\varphi\rbrack=\int_{t_1}^{t_2}dt\Int\left\lbrace
\frac{1}{2}\partial^\mu\varphi\partial_\mu\varphi-\frac{1}{2}m^2\varphi^2
-\frac{1}{2}\xi R\varphi^2-U(\varphi)\right\rbrace\;.\label{3.1}
\eeq
Here $R$ is the scalar curvature of $\Sigma$, $m^2$ is the mass associated with the field, $U(\varphi)$ represents the potential for any self-interaction, and $\xi$ is a dimensionless coupling constant which couples the scalar field to the curvature. Setting the variation of $S\lbrack\varphi\rbrack$ to zero results in
\beq
\Box\varphi+m^2\varphi+\xi R\varphi+U'(\varphi)=0\;,\label{3.1b}
\eeq
where $\Box=\displaystyle{\frac{\partial^2}{\partial t^2}-\nabla^2}$ with $\nabla^2=\displaystyle{\frac{1}{\sqrt{g}}\partial_i(\sqrt{g}g^{ij}\partial_j)}$ the covariant Laplacian on $\Sigma$.

We will initially assume that $U'(\varphi)=0$, so that $U(\varphi)$ is an arbitrary constant which does not affect the field equations. It is impossible to solve the classical field equation \ref{3.1b} exactly. However we may proceed in a way which will prove useful later. Assume that we have the solutions to
\beq
\left(-\nabla^2+\xi R\right)f_n(\x)=\sigma_nf_n(\x)\;,\label{3.2}
\eeq
and that the set of solutions $\lbrace f_n(\x)\rbrace$ is a complete set. The index $n$ on $f_n(\x)$ and the eigenvalue $\sigma_n$ is a multi-index which labels the different solutions. We can normalize the solutions by
\beq
\Int f_n(\x)f_{n'}^\ast(\x)=\delta_{nn'}\;.\label{3.3}
\eeq
It is easy to see that a complete set of solutions to Eq.~\ref{3.1b} is then obtained from
\beq
\varphi_n(t,\x)=e^{-i\omega_nt}f_n(\x)\;,\label{3.4}
\eeq
where
\beq
\omega_n=\sqrt{\sigma_n+m^2}\;.\label{3.5}
\eeq

The classical contribution to the effective potential is obtained very simply from Eq.~\ref{3.1} with $U(\varphi)=c$, where $c$ is a constant, to be 
\beq
V^{(0)}_{\rm eff}=\frac{1}{2}(m^2+\xi R)\varphi^2+c\;.\label{3.6}
\eeq
If we wish to calculate the quantum corrections to this classical result we merely need to compute the energy density associated with the energy levels $\hbar\omega_n$ with $\omega_n$ given in Eq.~\ref{3.5}. Because the scalar field may be considered as a collection of simple harmonic oscillators labelled by the multi-index $n$, we have
\beq
V^{(1)}_{\rm eff}=\frac{1}{V_\Sigma}\;\frac{1}{2}\hbar\sum_n\omega_n\;,
\label{3.7}
\eeq
since each oscillator has a zero-point energy of $\frac{1}{2}\hbar\omega_n$. Because $\omega_n$ will not be bounded as $n\rightarrow\infty$, the sum over zero-point energies in Eq.~\ref{3.7} will diverge, and therefore must be regulated. One beautiful way to do this is by $\zeta$-function regularization \cite{DowkerCritchley,Hawking}. 

The basic idea behind $\zeta$-function regularization is to define the divergent sum in Eq.~\ref{3.7} by analytic continuation of a convergent sum. We may define an energy $\zeta$-function $\omega(s)$ by
\beq
\omega(s)=\sum_n\omega_n(\ell\omega_n)^{-s}\;,\label{3.8}
\eeq
where $s$ is a complex variable, and $\ell$ is a constant with units of length introduced so that the energy $\zeta$-function $\omega(s)$ has the same dimensions for arbitrary $s$ as the original sum in Eq.~\ref{3.7}. The idea now is to use the unboundedness of $\omega_n$ as $n\rightarrow\infty$ to our advantage by evaluating $\omega(s)$ in a region of the complex $s$-plane where the sum converges, and then evaluating $\omega(0)$ by analytic continuation of the result. We therefore define the regulated expression for $V^{(1)}_{\rm eff}$ by
\beq
V^{(1)}_{\rm eff}=\frac{\hbar}{2V_\Sigma}\omega(0)\;.\label{3.9}
\eeq
It may be that $\omega(s)$ has a pole at $s=0$. In this case $\omega(0)$ denotes the analytic continuation of $\omega(s)$ back to a small neighbourhood of $s=0$. We can give a very loose justification for assuming that there is a region of the complex $s$-plane where the energy $\zeta$-function converges as follows. (We will see this in a more general way later.) If we consider the behaviour of $\omega_n$ for large $n$, then we can approximate the discrete set of eigenvalues $\sigma_n$ with the continuous value of ${\mathbf k}^2$. Then the convergence of $\omega(s)$ is determined by the convergence of $\int d^Dk({\mathbf k}^2+m^2)^{(1-s)/2}$. This integral is convergent for $\Re(s)>D+1$ and we therefore expect that the energy $\zeta$-function Eq.~\ref{3.8} converges in this region of the complex plane.

\subsection{\it Free Scalar Field in Flat Space}
\label{sec3.1}
It is helpful to illustrate how this regularization works in some simple examples. We will first consider the scalar field in a box in flat spacetime with periodic boundary conditions imposed on the box walls. If $L_1,\ldots,L_D$ represent the sides of the box, then the eigenvalues $\sigma_n$ defined in Eq.~\ref{3.2}
are just
\beq
\sigma_{\mathbf n}=\sum_{i=1}^{D}
\left(\frac{2\pi n_i}{L_i}\right)^2\;,\label{3.11}
\eeq
with $n_i=0,\pm1,\pm2,\ldots$ and ${\mathbf n}$ standing for the $D$-tuple $(n_1,\ldots,n_D)$. The energy $\zeta$-function in this case is
\beq
\omega(s)=\sum_{n_1=-\infty}^{\infty}\cdots\sum_{n_d=-\infty}^{\infty}
\ell^{-s}\left\lbrack\sum_{i=1}^{D}\left(\frac{2\pi n_i}{L_i}\right)^2+m^2\right\rbrack^{(1-s)/2}\;.\label{3.12}
\eeq
If we are interested in the case where the sides of the box become very large, ultimately we will take the infinite volume limit, then we can approximate $\omega(s)$ by replacing the sums over $n_i$ with integrals. (We will return to this at the end of the lectures.) Changing variables of integration leads in a straightforward way to
\beq
\omega(s)=V_\Sigma\ell^{-s}\mom\;
({\mathbf k}^2+m^2)^{(1-s)/2}\;.\label{3.13}
\eeq
The integral may be evaluated using the identity
\beq
a^{-z}=\frac{1}{\Gamma(z)}\int_{0}^{\infty}dt\,t^{z-1}e^{-at}\;,\label{3.14}
\eeq
which is valid for $\Re(z)>0$ and $\Re(a)>0$. This gives
\beq
\omega(s)&=&V_\Sigma\ell^{-s}\frac{1}{\Gamma((s-1)/2)}\mom\int_{0}^{\infty} dt\,t^{(s-3)/2}e^{-({\mathbf k}^2+m^2)t}\nn
&=&V_\Sigma\ell^{-s}\frac{1}{\Gamma((s-1)/2)}(4\pi)^{-D/2}\int_{0}^{\infty} dt\,t^{(s-3-D)/2}e^{-m^2t}\nn
&=&V_\Sigma\ell^{-s}\frac{1}{\Gamma((s-1)/2)}(4\pi)^{-D/2} \Gamma((s-1-D)/2) (m^2)^{(D+1-s)/2}\;.\label{3.15}
\eeq
The middle line follows from performing the integration over ${\mathbf k}$, which just involves a product of gaussian integrals, and the last line follows from using the identity Eq.~\ref{3.14} again. We must assume that $\Re(s)>D+1$ to justify the steps leading up to Eq.~\ref{3.15}.

The result for our energy $\zeta$-function in Eq.~\ref{3.15} can be seen to have simple poles whenever $(s-1-D)/2=-j$ for $j=0,1,2,\ldots$ coming from the poles of $\Gamma((s-1-D)/2)$. We can analytically continue the result in Eq.~\ref{3.15} throughout the complex plane. $\omega(s)$ is analytic at $s=0$ if $D$ is even and has a simple pole at $s=0$ if $D$ is odd. We can examine the cases of even and odd dimensional spaces separately.

For $D$ even we have
\begin{displaymath}
\omega(0)=V_\Sigma(4\pi)^{-D/2}\frac{\Gamma(-(1+D)/2)}{\Gamma(-1/2)}
(m^2)^{(D+1)/2}\;,
\end{displaymath}
which leads to
\beq
V^{(1)}_{\rm eff}=\frac{\hbar}{2}\left(-\frac{1}{\pi}\right)^{D/2}\frac{(D/2)!}{(D+1)!} 
(m^2)^{(D+1)/2}\;,\label{3.16}
\eeq
as the quantum correction to the classical potential. (The result has been simplified using some properties of the $\Gamma$-function.)

For odd $D$ we must expand about the pole of the $\Gamma$-function. It may be shown that
\beq
V^{(1)}_{\rm eff}&=&-\frac{\hbar}{2}\frac{1}{\left(\frac{D+1}{2}\right)!}
\left(-\frac{m^2}{4\pi}\right)^{(D+1)/2}\Big\lbrace\frac{2}{s}-1+\frac{1}{2}+\frac{1}{3}+\cdots\nn
&&\quad\quad+\frac{2}{D+1}-\ln(\frac{\ell^2m^2}{4})\Big\rbrace
\label{3.17}
\eeq
is the regularized expression for the quantum correction to the classical potential.

There is one obvious difference between the results found for even and odd dimensions. When $D$ is even, the regularized expression for $V^{(1)}_{\rm eff}$ is finite at $s=0$, whereas for odd $D$ the result diverges as $s\rightarrow0$. In order to deal with the divergent term we must renormalize the effective potential. This can be done very simply in the present case by imposing the renormalization condition
\beq
V_{\rm eff}(\varphi=0)=0\;.\label{3.18}
\eeq
This leads to
\beq
0&=&V_{\rm eff}(\varphi=0)=V^{(0)}_{\rm eff}(\varphi=0)+V^{(1)}_{\rm eff}( \varphi=0)\nn
&=&c+V^{(1)}_{\rm eff}\label{3.19}
\eeq
since $V^{(1)}_{\rm eff}$ does not depend on $\varphi$ in the present case. The renormalization condition results in the constant term in the effective potential cancelling the quantum part of the effective potential. For $D$ odd, this also removes the divergent part of Eq.~\ref{3.17}. We are left with the final expression
\beq
V_{\rm eff}(\varphi)=\frac{1}{2}m^2\varphi^2\;,\label{3.20}
\eeq
as the renormalized effective potential. In this case the result is not particularly interesting.

\subsection{\it Casimir Effect}
\label{sec3.2}

A slightly less trivial example occurs if we consider a flat space with some boundary conditions imposed on the field. A prototype for such an example is the Casimir effect \cite{Casimir} which is concerned with vacuum fluctuations in the electromagnetic field between a pair of neutral parallel conducting plates. As a scalar field analogue of this we will consider the massive scalar field which satisfies Dirichlet boundary conditions on two parallel planes at $x^1=0$ and $x^1=L_1$, but which is unconfined in the remaining spatial directions. We will choose the spatial dimension to be $D=3$ here for simplicity of presentation. The mode functions which obey Eq.~\ref{3.2} (with $R=0$ again) may be chosen to be
\beq
f_n(\x)=A_n\sin\big(\frac{\pi n_1x^1}{L_1}\big)\prod_{j=2}^{3}
e^{\frac{2\pi i}{L_j}n_jx^j}\;,\label{3.2.1}
\eeq
where $A_n$ is a normalization constant, $n_1=1,2,\ldots$, and $n_j=0,\pm1,\ldots$ for $j=2,3$. We will look at the limit $L_2,L_3\rightarrow\infty$, but will keep $L_1$, which represents the separation of the plates, finite. Periodic boundary conditions have been imposed in the directions transverse to the plates. Note that we cannot allow $n_1=0$ because this value does not lead to a normalizable eigenfunction. Negative values of $n_1$ can be neglected because the sign of the normalization constant is free.

We now have
\beq
\sigma_n=\left(\frac{\pi n_1}{L_1}\right)^2+\sum_{j=2}^{3}\left(\frac{2\pi n_j}{L_j}\right)^2\label{3.2.2}
\eeq
in place of Eq.~\ref{3.11}. From Eq.~\ref{3.8} we have
\begin{displaymath}
\omega(s)=\ell^{-s}\sum_{n_1=1}^{\infty}\sum_{n_2=-\infty}^{\infty}\sum_{n_3=-\infty}^{\infty}\left\lbrack\left(\frac{\pi n_1}{L_1}\right)^2+\sum_{j=2}^{3}\left(\frac{2\pi n_j}{L_j}\right)^2+m^2\right\rbrack^{(1-s)/2}\;.
\end{displaymath}
If we take the limit $L_2,L_3\rightarrow\infty$, and replace the sums over $n_2$ and $n_3$ with integrals as before, it is easy to show that 
\beq
\omega(s)=\frac{L_2L_3}{4\pi}\ell^{-s}\frac{\Gamma((s-3)/2)}{\Gamma((s-1)/2)}
\sum_{n_1=1}^{\infty}\left\lbrack m^2+\left(\frac{\pi n_1}{L_1}\right)^2\right\rbrack^{(3-s)/2}\;.\label{3.2.4}
\eeq
It is not possible to replace the remaining sum over $n_1$ with an integral as we did for the sums over $n_2$ and $n_3$ because we are not interested in the limit $L_1\rightarrow\infty$. There are a number of ways to evaluate the sum appearing in Eq.~\ref{3.2.4} in terms of a contour integral \cite{Fordetal}. Instead of trying to obtain an exact result, we will look at the expression in the limit $mL_1<<1$. This limit is of interest if we wish to study the massless limit.

When $mL_1<<1$ we can expand the summand of Eq.~\ref{3.2.4} in powers of $mL_1$ in an obvious way. If we keep only the first three terms, we have
\beq
\omega(s)&=&\ell^{-s}\frac{L_2L_3}{4\pi}\left(\frac{2}{s-3}\right)\left(
\frac{\pi}{L_1}\right)^{3-s}\sum_{n_1=1}^{\infty}n_1^{3-s}\left\lbrace
1+\left(\frac{3-s}{2}\right)\left(\frac{mL_1}{\pi}\right)^2\frac{1}{n_1^2}
\right.\nn
&&\left.\quad+\frac{1}{2}\left(\frac{3-s}{2}\right)\left(\frac{1-s}{2}\right)\left(\frac{mL_1}{\pi}\right)^4\frac{1}{n_1^4}\right\rbrace\nn
&=&\frac{L_2L_3}{4\pi}\left(\frac{\pi}{L_1}\right)^{3-s}\ell^{-s}
\left\lbrace\frac{2}{(s-3)}\zeta_R(s-3)-\left(\frac{mL_1}{\pi}\right)^2
\zeta_R(s-1)\right.\nn
&&\left.\quad-\frac{1}{4}(1-s)\left(\frac{mL_1}{\pi}\right)^4\zeta_R(s+1)\right\rbrace\;.\label{3.2.5}
\eeq
The definition of the Riemann $\zeta$-function $\zeta_R(s)=\displaystyle{\sum_{n=1}^{\infty}n^{-s}}$ for $\Re(s)>1$ has been used here. The only pole as $s\rightarrow0$ arises from the term involving $\zeta_R(s+1)$ which has a simple pole. Using properties of the Riemann $\zeta$-function \cite{WW} we have (noting that $V_\Sigma=L_1L_2L_3$)
\beq
V^{(1)}_{\rm eff}=\frac{\hbar\pi^2}{8L_1^4}\left\lbrace-\frac{1}{180}+
\frac{1}{12}\left(\frac{mL_1}{\pi}\right)^2-\frac{1}{4}\left(\frac{mL_1}{\pi}\right)^4\left\lbrack\frac{1}{s}+\gamma-\ln(\pi\ell/L_1)\right\rbrack\right\rbrace\;.\label{3.2.7}
\eeq
If this result is compared with that in Eq.~\ref{3.17} it can be observed that the pole parts of both expressions are the same. Thus the divergent part of the effective potential is unaffected by the boundary conditions imposed on the field. This is a special case of a more general result \cite{TomsPRDold} which we will establish in the next section. A natural thing to do here is to choose the constant $c$ to be the same as that found for the same theory in Minkowski spacetime. From Eqs.~\ref{3.17} and \ref{3.19} we find that the renormalized effective potential is
\beq
V_{\rm eff}(\varphi)&=&
\frac{1}{2}m^2\varphi^2-\frac{\pi^2\hbar}{1440L_1^4}\left\lbrack 
1-15\left(\frac{mL_1}{\pi}\right)^2\right\rbrack\nn
&&\quad-\frac{\hbar m^4}{32\pi^2}\left\lbrack\frac{1}{4}+\gamma+\frac{1}{2}\ln\left(\frac{m^2L_1^2}{4\pi^2}\right)\right\rbrack\;.\label{3.2.8}
\eeq
In the massless limit we are left simply with
\beq
V_{\rm eff}=-\frac{\pi^2\hbar}{1440L_1^4}\;,\label{3.2.9}
\eeq
which is half of the result found for the electromagnetic field.

\subsection{$\lambda\varphi^4$ \it Theory}
\label{sec3.3}

Suppose that we take
\beq
U(\varphi)=\frac{\lambda}{4!}\varphi^4+c\;.\label{3.3.1}
\eeq
The classical field equation \ref{3.1b} becomes
\beq
\Box\varphi+m^2\varphi+\xi R\varphi+\frac{\lambda}{6}\varphi^3=0\;.\label{3.3.2}
\eeq
In order to use the energy $\zeta$-function method we must know the energy levels associated with quantum fluctuations about the classical solution to the equation of motion. To this end, write
\beq
\varphi=\Bar{\varphi}+\psi\;,\label{3.3.3}
\eeq
where $\psi$ is the quantum fluctuation in the field. If we are only interested in working out the lowest order correction to the classical effective potential, then we may substitute Eq.~\ref{3.3.3} into Eq.~\ref{3.3.2} and linearize in $\psi$. This results in
\beq
\Box\psi+m^2\psi+\xi R\psi+\frac{\lambda}{2}\Bar{\varphi}^2\psi=0\;.\label{3.3.4}
\eeq
We have not assumed that $\Bar{\varphi}$ is constant here, but in conformity with our assumption on the static nature of the spacetime we will assume that $\Bar{\varphi}=\Bar{\varphi}(\x)$ only.

The classical potential is
\beq
V^{(0)}_{\rm eff}(\Bar{\varphi})=c+\frac{1}{2}|\nabla\Bar{\varphi}|^2+
\frac{1}{2}m^2\Bar{\varphi}^2+\frac{1}{2}\xi R\Bar{\varphi}^2+
\frac{\lambda}{4!}\Bar{\varphi}^4\;.\label{3.3.5}
\eeq
In place of Eq.~\ref{3.2}, let
\beq
\left(-\nabla^2+\xi R+\frac{\lambda}{2}\bp^2\right)f_n(\x)=\sigma_nf_n(\x)\;.\label{3.3.6}
\eeq
Take $\psi_n(t,\x)=e^{-i\omega_nt}f_n(\x)$ as in Eq.~\ref{3.4}, so that $\omega_n$ is again given by Eq.~\ref{3.5}. We can still define $V^{(1)}_{\rm eff}$ by Eq.~\ref{3.9} with $\omega(s)$ given by Eq.~\ref{3.8}. For general $\Sigma$ and arbitrary $\bp$ it is impossible to solve Eq.~\ref{3.3.6} explicitly for the eigenvalues $\sigma_n$; nevertheless, we can apply a very general method to obtain approximate results.

First of all we will examine the divergent part of the effective potential in curved space to study the renormalizability of the theory. From Eq.~\ref{3.8} we can write
\begin{displaymath}
\omega(s)=\ell^{-s}\sum_n(\sigma_n+m^2)^{(1-s)/2}\;.
\end{displaymath}
Using Eq.~\ref{3.14} we can obtain
\beq
\omega(s)=\frac{\ell^{-s}}{\Gamma((s-1)/2)}\int_{0}^{\infty}dt\,t^{(s-3)/2}
\Theta(t)\;,\label{3.3.9}
\eeq
where we have defined
\beq
\Theta(t)=\sum_ne^{-t(\sigma_n+m^2)}\;.\label{3.3.8}
\eeq
The advantage of this procedure is that a great deal of information is known about $\Theta(t)$. Our first observation is that $\Theta(t)\sim e^{-t(\sigma_0+m^2)}$ as $t\rightarrow\infty$ where $\sigma_0$ is the smallest eigenvalue. Assuming that $\sigma_0\ge0$ this shows that the integrand of Eq.~\ref{3.3.9} is well-behaved as $t\rightarrow\infty$ for all $s$. This means that if $\omega(s)$ has any poles, they can come only from the $t\rightarrow0$ limit of the integrand. We may split up the integration range in Eq.~\ref{3.3.9} into two parts as follows~:
\beq
\omega(s)=\omega_p(s)+\omega_r(s)\;,\label{3.3.10}
\eeq
with
\beq
\omega_p(s)&=&\frac{\ell^{-s}}{\Gamma((s-1)/2)}\int_{0}^{t_0}dt\,t^{(s-3)/2}
\Theta(t)\;,\label{3.3.11}\\
\omega_r(s)&=&\frac{\ell^{-s}}{\Gamma((s-1)/2)}\int_{t_0}^{\infty}dt\,t^{(s-3)/2}
\Theta(t)\;.\label{3.3.12}
\eeq
From our above discussion because of the way we have split up the integral, $\omega_r(s)$ should be an analytic function of $s$, and any poles in $\omega(s)$ should come only from $\omega_p(s)$. Here $t_0$ is an arbitrary positive constant which we can specify freely. It may be taken to be arbitrarily small, so that to see if $\omega(s)$ has a pole at $s=0$ we only need to know the behaviour of $\Theta(t)$ for small $t$. Fortunately this is a well studied problem \cite{Minaketal}. It can be shown that $\Theta(t)$ has the asymptotic expansion
\beq
\Theta(t)\simeq(4\pi t)^{-D/2}\sum_{k=0,\frac{1}{2},1,\ldots}t^k\theta_k\;,\label{3.3.14}
\eeq
for some expansion coefficients $\theta_k$ which are independent of $t$. Using this in Eq.~\ref{3.3.11} results in
\beq
\omega_p(s)=\frac{\ell^{-s}}{\Gamma((s-1)/2)}(4\pi)^{-D/2}
\sum_{k=0,\frac{1}{2},1,\ldots}\theta_k\frac{t_0^{k+(s-D-1)/2}}{k+(s-D-1)/2}\;,\label{3.3.15}
\eeq
where we initially assume $\Re(s)>D+1$. (This establishes our earlier claim that $\omega(s)$ is analytic in $s$ if $\Re(s)>D+1$ as stated earlier.) Analytic continuation back to $s=0$ shows that $\omega(s)$ has a pole only if $k=(D+1)/2$. We have
\beq
\omega(s)=-(4\pi)^{-(D+1)/2}\theta_{(D+1)/2}\,\frac{1}{s}+\cdots\;,\label{3.3.16}
\eeq
in a neighbourhood of $s=0$. Keeping only the pole term, we find
\beq
V^{(1)}_{\rm eff}=-\frac{\hbar}{2V_\Sigma}(4\pi)^{-(D+1)/2}\theta_{(D+1)/2}\,\frac{1}{s}+\cdots\;.\label{3.3.17}
\eeq
All that we require is a knowledge of the coefficients $\theta_k$ appearing in Eq.~\ref{3.3.14}.

For any operator of the form $(-\nabla^2+Q)$, where $Q$ is any arbitrary function of $\x$, the first few coefficients are known for $\Sigma$ any smooth compact Riemannian manifold with smooth boundary. If we concentrate on manifolds with no boundary, then
\beq
\theta_0&=&V_\Sigma\;,\label{3.3.18}\\
\theta_1&=&\Int\,\left(\frac{1}{6}R-Q\right)\;,\label{3.3.19}\\
\theta_2&=&\Int\,\left(\frac{1}{30}\nabla^2R+\frac{1}{72}R^2-\frac{1}{180}R_{ij}R^{ij}+\frac{1}{180}R_{ijkl}R^{ijkl}\right.\nn
&&\left.\quad-\frac{1}{6}RQ+\frac{1}{2}Q^2-\frac{1}{6}\nabla^2Q\right)\;.\label{3.3.20}
\eeq
The expressions for $\theta_3$ and $\theta_4$ are known but are too lengthy to give here. If $\Sigma$ has no boundary, then $\theta_{1/2},\theta_{3/2},\ldots$ all vanish. From Eq.~\ref{3.3.17} this means that there is no pole term when $D$ is even. For the physically interesting case of $D=3$ we find
\beq
V^{(1)}_{\rm eff}=-\frac{\hbar\theta_2}{32\pi^2V_\Sigma}\,\frac{1}{s}
+\cdots\;\label{3.3.21}
\eeq
It is straightforward to show that when $U(\varphi)=\lambda\varphi^4/4!$ the theory is renormalizable in curved spacetime with counterterms in agreement with those of Ref.~\cite{TomsPRDold}.

Although we have examined the problem of computing the infinite part of the effective potential, for general $\bp(\x)$ it is difficult to evaluate the finite part. As an example we can consider the case where $\Sigma=\reals^D$ with $\bp$ constant. In this case it is evident from Eq.~\ref{3.3.6} that the presence of $\bp$ acts just like an additional term in the mass. We may simply obtain $V^{(1)}_{\rm eff}$ from our previous result for the non-interacting theory in Sec.~\ref{sec3.1} by replacing $m^2$ with $m^2+\frac{\lambda}{2}\bp^2$. If we take $D=3$ in Eq.~\ref{3.17} we find
\beq
\V1=-\frac{\hbar}{64\pi^2}\left(m^2+\frac{\lambda}{2}\bp^2\right)^2
\left\lbrace\frac{2}{s}-\frac{1}{2}-\ln\left\lbrack\frac{\ell^2}{4}
\left(m^2+\frac{\lambda}{2}\bp^2\right)\right\rbrack\right\rbrace\;,
\label{3.3.30}
\eeq
when we replace $m^2$ with $m^2+\frac{\lambda}{2}\bp^2$. The total effective potential is
\beq
V_{\rm eff}=
c+\frac{1}{2}m^2\bp^2+\frac{\lambda}{4!}\bp^4+\V1\;.\label{3.3.31}
\eeq

In order to renormalize the theory we must impose some renormalization conditions on $V_{\rm eff}$ as we did for the non-interacting theory. We will write
\beq
c&=&c_R+\hbar\delta c\;,\label{3.3.32}\\
m^2&=&m_R^2+\hbar\delta m^2\;,\label{3.3.33}\\
\lambda&=&\lambda_R+\hbar\delta\lambda\;,\label{3.3.34}
\eeq
where the subscript "R" denotes a renormalized quantity. Because $\V1$ is already of order $\hbar$ we may take all quantities which occur in it as renormalized. We will choose the renormalization conditions
\beq
V_{\rm eff}(\bp=0)&=&c_R\;,\label{3.3.35}\\
\left.\frac{\partial^2}{\partial\bp^2}
V_{\rm eff}\right|_{\bp=0}&=&m_R^2\;,\label{3.3.36}\\
\left.\frac{\partial^4}{\partial\bp^4}
V_{\rm eff}\right|_{\bp=0}&=&\lambda_R\;.\label{3.3.37}
\eeq
The values on the RHS of these conditions are chosen to agree with what is found for the classical potential. It is now straightforward to obtain the counterterms and hence the renormalized effective potential. The result is
\beq
V_{\rm eff}(\bp)&=&c_R+\frac{1}{2}m_R^2\bp^2+\frac{\lambda_R}{4!}\bp^4
-\frac{\hbar\lambda_R}{128\pi^2}m_R^2\bp^2\nn
&&\quad-\frac{\hbar\lambda_R^2}{512\pi^2}\bp^4+\frac{\hbar}{64\pi^2}
\left(m_R^2+\frac{\lambda_R}{2}\bp^2\right)^2\ln\left(1+
\frac{\lambda_R\bp^2}{2m_R^2}\right)\;.\label{3.3.41}
\eeq

If we try to let $m_R^2\rightarrow0$ in Eq.~\ref{3.3.41} to obtain the result for the massless theory there is seen to be a problem. This is due to our choice of the renormalization condition Eq.~\ref{3.3.37}. Although the presence of $\bp$ alters the mass from its tree level value of zero to $\frac{\lambda}{2}\bp^2$ when we impose the renormalization condition at $\bp=0$ there can be problems due to infrared divergences. To deal with this we can replace the condition Eq.~\ref{3.3.37} with
\beq
\left.\frac{\partial^4}{\partial\bp^4}V_{\rm eff}\right|_{\bp=M}=\lambda_R\;,\label{3.3.42}
\eeq
where $M$ is an arbitrary renormalization point \cite{CW}. This leads to the result
\beq
V_{\rm eff}(\bp)=c_R+\frac{\lambda}{4!}\bp^4-
\frac{\hbar\lambda_R^2}{256\pi^2}\bp^4\left\lbrace\ln\left(\frac{\bp^2}{M^2}\right)-\frac{25}{6}\right\rbrace\;.\label{3.3.44}
\eeq
This expression is usually referred to as the Coleman-Weinberg effective potential. Ideally we would now solve the effective field equation $V'_{\rm eff}(\bp)=0$ for $\bp$. It is obvious that $\bp=0$ is one solution to the effective field equation. Unfortunately as discussed in Ref.~\cite{CW} other solutions lie outside the range where perturbation theory is valid, so that there are no non-trivial self-consistent solutions to the effective field equations apart from $\bp=0$.

The fact that $V_{\rm eff}(\bp)$ in Eq.~\ref{3.3.44} depends on an arbitrary renormalization constant $M$ might seem to be a weakness in the result; however it is actually a great strength since it leads to a study of the renormalization group \cite{CW}. If a different choice is made for $M$ then it is necessary to redefine the renormalized parameters in the theory in a way which makes all physical predictions unchanged. We will leave this interesting subject as outside the main aim of the lectures.

\section{Some Thermodynamics and Statistical Mechanics}
\label{secII2.1}

So far we have ignored the effects of a non-zero temperature on the effective action. We now wish to see in a simple way how to include finite temperature corrections to the results obtained in previous sections. We will only consider the case of thermal equilibrium here. More details can be found in the standard references \cite{LLStatPhys,Pathria,Huang}.

We have evaluated the effective potential at $T=0$ by summing up the zero-point energies. The simplest way to include the effects of finite temperature is to appeal to standard thermodynamics. Suppose that we have a system described by a Hamiltonian operator $\hat{H}$ which has a conserved quantity described by an operator $\hat{Q}$. $\hat{Q}$ and $\hat{H}$ necessarily commute. For a non-relativistic system we may be interested in a conserved number of particles for neutral systems, or a total conserved charge. For relativistic systems the total particle number cannot be conserved due to pair creation, but the total charge should be fixed. All of the relevant thermodynamics is encoded into the grand partition function $\Z$ defined by
\beq
\Z={\rm tr}\,\left\lbrack e^{-\beta(\hat{H}-\mu\hat{Q})}\right\rbrack\;\label{2.1.1}
\eeq
Here $\beta=(kT)^{-1}$ with $T$ the temperature, and $\mu$ is the chemical potential associated with the conserved quantity described by $\hat{Q}$. The grand partition function is regarded as a function of the thermodynamic variables $T,V,\mu$. For curved space and background fields, $\Z$ is also regarded as a function of the metric tensor on $\Sigma$ and the background fields which are held fixed in the computation.

In place of $\Z$ it is often convenient to define the $q$-potential by \cite{Pathria}
\beq
q=\ln\Z\;.\label{2.1.2}
\eeq
In the limit where the volume becomes very large, 
\beq
q=\beta PV\;,\label{2.1.2b}
\eeq
where $P$ is the pressure. The connection between statistical mechanics and thermodynamics can be made from the first law which reads
\beq
dU=TdS-PdV+\mu dQ\;,\label{2.1.3}
\eeq
where $Q$ is the conserved quantity described by the operator $\hat{Q}$, and $U$ is the internal energy. (Other terms can be present in Eq.~\ref{2.1.3} if the system is, for example, rotating or in an external field.)

A key thermodynamic quantity is the Helmholtz free energy $A$ defined by
\beq
A=U-TS\;.\label{2.1.4}
\eeq
Computing $dA$ and using the first law in the form of Eq.~\ref{2.1.3} results in
\beq
dA=-SdT-PdV+\mu dQ\;.\label{2.1.5}
\eeq
This demonstrates that $A$ is regarded as a function of $T,V,Q$. The importance of $A$ lies in the fact that in thermal equilibrium for fixed $T,V,Q$ the Helmholtz free energy must be a minimum. (See for example \cite{Gug,Mun}.)

Finally, the grand potential (sometimes just called the thermodynamic potential) $\Omega$ is defined by \cite{Mun}
\beq
\Omega&=&U-TS-\mu Q\label{2.1.6}\\
&=&A-\mu Q\;.\label{2.1.7}
\eeq
Computing $d\Omega$ and using Eq.~\ref{2.1.5} gives
\beq
d\Omega=-SdT-PdV-Qd\mu\;,\label{2.1.8}
\eeq
showing that $\Omega$ depends on $T,V,\mu$. The connection between macroscopic thermodynamics and microscopic statistical mechanics is made by noting from Eq.~\ref{2.1.6} and the non-differential form of Eq.~\ref{2.1.3}, $U=TS-PV+\mu Q$, which follows from Euler's theorem on homogeneous functions and the fact that $S,V,Q$ are extensive variables \cite{Mun}
\beq
\Omega=-PV\;.\label{2.1.9a}
\eeq
Thus
\beq
q=-\beta\Omega\;.\label{2.1.9}
\eeq

In terms of the energy levels $E_n$ of the system we have \cite{Pathria}
\beq
q=-\sum_n\ln\left\lbrack1-e^{-\beta(E_n-\mu)}\right\rbrack\;.\label{2.1.10}
\eeq
From Eq.~\ref{2.1.8} we have
\beq
Q&=&-\left(\frac{\partial\Omega}{\partial\mu}\right)_{T,V}\label{2.1.11}\\
&=&\frac{1}{\beta}\left(\frac{\partial q}{\partial\mu}\right)_{T,V}\ \ ({\rm from\ Eq.}\ \ref{2.1.9})\label{2.1.12}\\
&=&\sum_n{\left\lbrack e^{\beta(E_n-\mu)}-1\right\rbrack}^{-1}\;.\label{2.1.13}
\eeq
This is the standard Bose-Einstein distribution function. Another way to show consistency of all of our definitions is to use the definition of $Q$ as a thermal average
\beq
Q=\frac{1}{\Z}{\rm tr}\,\left\lbrack\hat{Q}e^{-\beta(\hat{H}-\mu\hat{Q})}\right\rbrack\;,\label{2.1.14}
\eeq
and to note from this that
\begin{displaymath}
Q=\frac{1}{\beta}\frac{1}{\Z}\left(\frac{\partial\Z}{\partial\mu}\right)_{T,V}=\frac{1}{\beta}\left(\frac{\partial q}{\partial\mu}\right)_{T,V}\;,
\end{displaymath}
using Eq.~\ref{2.1.2}.

A simple and important example is provided by the simple harmonic oscillator. Taking $\hat{Q}$ to be the number operator we have
\beq
\Z&=&{\rm tr}\,\left\lbrack e^{-\beta(\hat{H}-\mu\hat{Q})}\right\rbrack\nn
&=&\sum_{n=0}^{\infty}\langle n|e^{-\beta(\hat{H}-\mu\hat{Q})}|n\rangle\nn
&=&\sum_{n=0}^{\infty}e^{-\beta(E_n-\mu n)}\;,\label{2.1.15}
\eeq
where we have defined $\hat{H}|n\rangle=E_n|n\rangle$ and $\hat{Q}|n\rangle=n|n\rangle$. If we use $E_n=(n+1/2)\omega$, then the sum in Eq.~\ref{2.1.15} is just a geometric series which is easily summed to give
\beq
\Z=e^{-\beta\omega/2}\left\lbrack1-e^{-\beta(\omega-\mu)}\right\rbrack^{-1}\;.\label{2.1.16}
\eeq
Using Eq.~\ref{2.1.9} and Eq.~\ref{2.1.2} we find
\beq
\Omega=\frac{1}{2}\omega+\frac{1}{\beta}\ln\left\lbrack1-e^{-\beta(\omega-\mu)}\right\rbrack\;,\label{2.1.17}
\eeq
as the grand thermodynamic potential. From Eq.~\ref{2.1.8} we have
\beq
Q&=&-\left(\frac{\partial \Omega}{\partial\mu}\right)_{T,V}\nn
&=&\left\lbrack e^{\beta(\omega-\mu)}-1\right\rbrack^{-1}\;.\label{2.1.18}
\eeq
Note that as $T\rightarrow0$ (so that $\beta\rightarrow\infty$) the second term in $\Omega$ vanishes and $\Omega\rightarrow\frac{1}{2}\omega$ which is just the zero-point energy.

In quantum field theory we can consider the theory as an infinite collection of simple harmonic oscillators exactly as we did before. If $\omega_n$ denotes the frequency of the mode $f_n(\x)$ as in Sec.~\ref{sec3}, then we have
\beq
\Omega=\frac{1}{2}\sum_n\omega_n+\frac{1}{\beta}\sum_n\ln\left\lbrack1-
e^{-\beta(\omega_n-\mu)}\right\rbrack\;.\label{2.1.19}
\eeq
The first term appearing on the RHS is observed to take the same form as we had in the zero temperature case considered earlier. It is only this part of $\Omega$ which is divergent and requires regularization. We may write
\beq
\Omega=\Omega_{T=0}+\Omega_{T\ne0}\;,\label{2.1.20}
\eeq
where
\beq
\Omega_{T\ne0}=\frac{1}{\beta}\sum_n\ln\left\lbrack1-
e^{-\beta(\omega_n-\mu)}\right\rbrack\;,\label{2.1.21}
\eeq
is the finite temperature correction to the zero temperature theory. The zero temperature part is
\beq
\Omega_{T=0}=\frac{1}{2}\sum_n\omega_n\;.\label{2.1.22}
\eeq
From Eq.~\ref{3.7} this can be recognized as $\V1 V_\Sigma$. Thus we can define a finite temperature effective potential by
\beq
\V1=\frac{\Omega}{V_\Sigma}\;.\label{2.1.23}
\eeq
There is a natural split of this into a zero temperature part and a finite temperature correction to the zero temperature part. Finally we note that we require $\mu\le\omega_0$ to ensure that $\Omega_{T\ne0}$ is real, and to avoid a negative occupation number.

\section{Free Bose Gas in Flat Space}
\label{secII2.2}

\subsection{\it Non-relativistic Gas}

We will examine the case of a single real degree of freedom for which the dispersion relation is non-relativistic. Here
\beq
\omega_{\bf n}=\frac{1}{2m}\sum_{j=1}^{D}\left(\frac{2\pi n_j}{L_j}\right)^2\;,
\label{2.2.1}
\eeq
if we resort to the familiar device of imposing box normalization with the infinite box limit taken. If we do not have any conserved quantities then
\beq
\Omega_{T\ne0}=\frac{1}{\beta}\int dn_1\cdots dn_D\,\ln\left\lbrack1-
e^{-\beta\omega_{\bf n}}\right\rbrack\;,\label{2.2.2}
\eeq
after the large box limit is taken. Making a change of integration variable $n_j\rightarrow\frac{L_j}{2\pi}n_j$ and expanding 
\beq
\ln(1-z)=-\sum_{k=1}^{\infty}\frac{z^k}{k}\;,\label{2.2.3}
\eeq
gives
\beq
\Omega_{T\ne0}&=&-\frac{1}{\beta}V_{\Sigma}\sum_{k=1}^{\infty}\frac{1}{k}
\int\frac{d^Dn}{(2\pi)^D}\,e^{-k\frac{\beta}{2m}{\bf n}^2}\nn
&=&-\frac{V_\Sigma}{\beta}\left(\frac{m}{2\pi\beta}\right)^{D/2}
\zeta_R(1+\frac{D}{2})\;,\label{2.2.4}
\eeq
after performing the sum over $k$. If the $T=0$ contribution is ignored we have
\beq
\V1=-\frac{1}{\beta}\left(\frac{m}{2\pi\beta}\right)^{D/2}
\zeta_R(1+\frac{D}{2})\;,\label{2.2.5}
\eeq
as the thermal part of the effective potential.

\subsection{\it Relativistic Gas}
\label{secIIRelgas}

For the relativistic gas we have from Eqs.~\ref{3.5} and \ref{3.11} that
\beq
\omega_{\bf n}=\left\lbrack\sum_{j=1}^{D}\left(\frac{2\pi n_j}{L_j}\right)^2+m^2\right\rbrack^{1/2}\;.
\label{2.2.6}
\eeq
This gives the same expression as in Eq.~\ref{2.2.2} in the large box limit. Changing variables of integration as before we have
\beq
\Omega_{T\ne0}=\frac{V_\Sigma}{\beta}\int\frac{d^Dn}{(2\pi)^D}\,\ln
\left\lbrack1-e^{-\beta({\bf n}^2+m^2)^{1/2}}\right\rbrack\;.\label{2.2.8}
\eeq
It is not possible to evaluate this in such a simple way as in the non-relativistic case. Instead we will concentrate on just obtaining the high temperature expansion introducing a technique which we will use later.

Begin by expanding the logarithm using Eq.~\ref{2.2.3}~:
\beq
\Omega_{T\ne0}=-\frac{V_\Sigma}{\beta}\sum_{k=1}^{\infty}\frac{1}{k}
\int\frac{d^Dn}{(2\pi)^D}\,e^{-k\beta({\bf n}^2+m^2)^{1/2}}\;.\label{2.2.9}
\eeq
It is still not possible to evaluate the integral in simple terms. For high $T$, specifically $\beta m<<1$ or $kT>>m$, we may obtain a useful expansion by using the Mellin-Barnes integral representation for the exponential function~:
\beq
e^{-x}=\cint\Gamma(\alpha)x^{-\alpha}\;.\label{MB}
\eeq
Here $c\in\reals$ with $c>0$ chosen so that the integration contour is to the right of the poles of the $\Gamma$-function. The proof of Eq.~\ref{MB} is simple~: just close the contour in the left hand of the complex plane. $\Gamma(\alpha)$ has simple poles at $\alpha=-n$ for $n=0,1,2,\ldots$ with residues $(-1)^n/n!$. The residue theorem gives
\begin{displaymath}
\cint\;\Gamma(\alpha)x^{-\alpha}=\sum_{n=0}^{\infty}\frac{(-1)^n}{n!}x^n\;,
\end{displaymath}
which may be recognized as the Maclaurin series for $e^{-x}$.

Using the representation Eq.~\ref{MB} in Eq.~\ref{2.2.9} leads to
\beq
\Omega_{T\ne0}=-\frac{V_\Sigma}{\beta}\sum_{k=1}^{\infty}\frac{1}{k}\int
\frac{d^Dn}{(2\pi)^D}\cint\;\Gamma(\alpha)k^{-\alpha}\beta^{-\alpha}(
{\bf n}^2+m^2)^{-\alpha/2}\;.\label{2.2.10}
\eeq
Noting that
\beq
\int
\frac{d^Dn}{(2\pi)^D}({\bf n}^2+m^2)^{-\alpha/2}=(4\pi)^{-D/2}
\frac{\Gamma\left(\frac{\alpha-D}{2}\right)}{\Gamma(\alpha/2)}(m^2)^{(D-\alpha)/2}
\;,\label{2.2.11}
\eeq
which can be established as in Eq.~\ref{3.15} assuming $\Re(\alpha)>D$, if we deform the integration contour in Eq.~\ref{2.2.10} to the right so that $c>D$, we can interchange the orders of integration and obtain
\beq
\Omega_{T\ne0}&=&-\frac{(4\pi)^{-D/2}V_\Sigma}{\beta}
\cint\;\Gamma(\alpha)\beta^{-\alpha}\zeta_R(1+\alpha)
\frac{\Gamma\left(\frac{\alpha-D}{2}\right)}{\Gamma(\alpha/2)}(m^2)^{(D-\alpha)/2}\nn
&=&-\frac{V_\Sigma}{\beta}\left(\frac{m^2}{4\pi}\right)^{D/2}
\cint\;
\frac{\Gamma(\alpha)\Gamma\left(\frac{\alpha-D}{2}\right)}{\Gamma(\alpha/2)}
\zeta_R(1+\alpha)
(\beta m)^{-\alpha}
\;.\label{2.2.12}
\eeq
This can be simplified by making use of the duplication formula for the $\Gamma$-function 
\beq
\Gamma(2z)=(4\pi)^{-1/2}2^{2z}\Gamma(z)\Gamma\left(z+\frac{1}{2}\right)\;.
\label{dup}
\eeq
We find the following representation for the thermodynamic potential~:
\beq
\Omega_{T\ne0}&=&-\frac{V_\Sigma}{\beta\sqrt{4\pi}}\left(\frac{m^2}{4\pi}\right)^{D/2}
\cint\;\Gamma\left(\frac{\alpha+1}{2}\right)\Gamma\left({\frac{\alpha-D}{2}}\right)\nn
&&\quad\quad\quad\times\zeta_R(\alpha+1)(\beta m/2)^{-\alpha}\;.\label{2.2.13}
\eeq

If we close the contour in the left hand of the complex plane we encounter three sources of poles. $\Gamma\left(\frac{\alpha+1}{2}\right)$ has simple poles at $\alpha=-1-2n$ for $n=0,1,2,\ldots$; $\Gamma\left(\frac{\alpha-d}{2}\right)$ has simple poles at $\alpha=D-2n$ for $n=0,1,2,\ldots$; $\zeta_R(1+\alpha)$ has a simple pole at $\alpha=0$. Because $\zeta_R(-2n)=0$ for $n=1,2,\ldots$ it follows that $\Gamma\left(\frac{\alpha+1}{2}\right)\zeta_R(\alpha+1)$ has simple poles only at $\alpha=0,-1$. It is easy to show that when $D$ is even the integrand of Eq.~\ref{2.2.13} has a double pole at $\alpha=0$, and when $D$ is odd, there is a double pole at $\alpha=-1$. In either case all remaining poles are simple ones. It is therefore expedient to examine the cases of even and odd $D$ separately again.

For even $D$ it is straightforward to show using the residue theorem that
\beq
\Omega_{T\ne0}&=&-\frac{V_\Sigma}{\pi^{(D+1)/2}\beta^{D+1}}
\sum_{\substack{n=0\\n\ne D/2}}^{\infty}\frac{1}{n!}\Gamma\left(\frac{D+1}{2}-n\right)\zeta_R(D+1-2n)
\left(-\frac{1}{4}\beta^2m^2\right)^n\nn
&&\quad-\frac{V_\Sigma}{2\beta}\left(-\frac{m^2}{4\pi}\right)^{D/2}
\frac{1}{(D/2)!}\left\lbrack\sum_{k=1}^{D/2}\frac{1}{k}-\ln(\beta^2m^2)\right\rbrack\nn
&&\quad\quad\quad+\frac{1}{2}V_\Sigma\left(\frac{m^2}{4\pi}\right)^{(D+1)/2}\Gamma\left(\frac{-1-D}{2}\right)\;.\label{2.2.19}
\eeq

For odd $D$
\beq
\frac{\Omega_{T\ne0}}{V_\Sigma}&=&-\pi^{-(D+1)/2}\beta^{-D-1}
\sum_{n=0}^{(D-1)/2}\frac{(-1)^n}{n!}\Gamma\left(\frac{D+1}{2}-n\right)
\zeta_R(D+1-2n)\left(\frac{\beta m}{2}\right)^{2n}\nn
&&-\frac{1}{2}\left(-\frac{m^2}{4\pi}\right)^{(D+1)/2}\frac{1}{\left(
\frac{D+1}{2}\right)!}\left\lbrack2\gamma-\sum_{k=1}^{(D+1)/2}+2
\ln\left(\frac{\beta m}{4\pi}\right)\right\rbrack\nn
&&-\frac{1}{2}\beta^{-1}\left(\frac{m^2}{4\pi}\right)^{D/2}\Gamma\left(
-\frac{D}{2}\right)
-\pi^{-(D+1)/2}\beta^{-D-1}\nn
&&\times\sum_{n=(D+3)/2}^{\infty}\frac{(-1)^n}{n!}
(-\pi^2)^{n+1+(D+1)/2}\frac{\zeta_R(D-2n)}{\Gamma(2-D/2)}\left(
\frac{\beta m}{2}\right)^{2n}\;.\label{2.2.21}
\eeq
It is possible to simplify the result using properties of the $\Gamma$- and $\zeta$-functions. In the physically interesting case of $D=3$ we find
\beq
\frac{\Omega_{T\ne0}}{V_\Sigma}&=&-\frac{\pi^2}{90}\beta^{-4}+\frac{1}{24}\beta^{-2}m^2-\frac{1}{12\pi}\beta^{-1}m^3\nn
&&\quad-\frac{m^4}{64\pi^2}\left\lbrack2\gamma-\frac{3}{2}+2\ln\left(\frac{\beta m}{4\pi}\right)\right\rbrack\nn
&&\quad-\frac{1}{2}\pi^{-7/2}\beta^{-4}\sum_{n=3}^{\infty}\frac{B_{n-1}}{(n-1)n!}\left(-\frac{\pi^2}{4}\beta^2m^2\right)^n\;.\label{2.2.23}
\eeq
$B_n$ are the Bernoulli numbers.

\section{Charged Bose Gas}
\label{secII2.3}

In order to illustrate how the calculations proceed for $\mu\ne0$ we will consider the case of a conserved electric charge. We will also set up the formalism to deal with a static background magnetic field described by a vector potential ${\mathbf A}(\x)$. The classical action functional for the magnetic field will be written as
\beq
S_{\rm em}=\int_{t_1}^{t_2}dt\Int\left\lbrace\frac{1}{4}F_{ij}F^{ij}-J_{\rm ext}^iA_i\right\rbrace\;.\label{2.3.1}
\eeq
Here $F_{ij}=\nabla_iA_j-\nabla_jA_i$ is the field strength tensor which describes the magnetic field. ${\mathbf J}_{\rm ext}$ is the externally applied current responsible for setting up the background magnetic field. If we want to keep the spatial dimension $D$ general then the magnetic field should not be described by a vector field, but rather by the antisymmetric tensor field $F_{ij}$. In the special case of $D=3$ we may introduce a vector field ${\mathbf B}$ with components $B^i$ through the duality relation $F_{ij}=\epsilon_{ijk}B^k$ where $\epsilon_{ijk}$ is the Levi-Civita tensor.

\subsection{\it Non-relativistic Field}
\label{chargenorel}

We will consider the non-relativistic theory first. This is described by a complex Schr\"{o}dinger field $\Psi$ and we take the action functional to be
\beq
S\lbrack\Psi,\Psi^\dagger\rbrack=\int_{t_1}^{t_2}dt\Int\left\lbrace
\frac{i}{2}\left(\Psi^\dagger\Dot{\Psi}-\Dot{\Psi}^\dagger\Psi\right)-
\frac{1}{2m}|{\mathbf D}\Psi|^2-U(\x)|\Psi|^2\right\rbrace\;.\label{2.3.2}
\eeq
It is easy to verify that under independent variations of $\Psi$ and $\Psi^\dagger$ the Schr\"{o}dinger equation and its complex conjugate are obtained. We have set $\hbar=1$ as usual. $U(\x)$ represents any static potential which might be present. ${\mathbf D}=\nabla-ie{\mathbf A}$ is the usual gauge covariant derivative. We will adopt the gauge choice
\beq
\nabla\cdot{\mathbf A}=0\;.\label{2.3.3}
\eeq

The theory is invariant under the local $U(1)$ gauge transformation
\beq
\Psi(t,\x)&\rightarrow&e^{ie\theta(\x)}\Psi(t,\x)\;,\label{2.3.4a}\\
{\mathbf A}(\x)&\rightarrow&{\mathbf A}(\x)+\nabla\theta(\x)\;,\label{2.3.4b}
\eeq
for arbitrary $\theta(\x)$. Associated with this gauge invariance is a conserved Noether current which can be found by varying the matter part of the action with respect to $A_i$ and defining the current by
\beq
\delta S=-\int_{t_1}^{t_2}dt\Int J^i\delta A_i\;.\label{2.3.4c}
\eeq
The minus sign in Eq.~\ref{2.3.4c} is for conformity with that in the current which appears in Eq.~\ref{2.3.1} and ensures that the current appears with the correct sign in the Maxwell equations. It is found that
\beq
{\mathbf J}=\frac{ie}{2m}\left(\nabla\Psi^\dagger\Psi-\Psi^\dagger\nabla\Psi\right)-
\frac{e^2}{m}{\mathbf A}|\Psi|^2\;.\label{2.3.5}
\eeq
The charge associated with this current is
\beq
Q=e\Int|\Psi|^2\;.\label{2.3.6}
\eeq
For this non-relativistic theory we can deal with a fixed particle number rather than a fixed total charge. In this case the results of Eq.~\ref{2.3.5} and \ref{2.3.6} are divided by $e$, in which case ${\mathbf J}$ may be recognized as the probability current of wave mechanics.

The Hamiltonian for the theory with action Eq.~\ref{2.3.2} is simply
\beq
H=\Int\left\lbrace\frac{1}{2m}|{\mathbf D}\Psi|^2+U(\x)|\Psi|^2\right\rbrace\;.\label{2.3.7}
\eeq
To incorporate the conserved charge we introduce the Lagrange multiplier $\mu$ as described in Sec.~\ref{secII2.1} and define
\beq
\Bar{H}=H-\mu Q=\Int\left\lbrace\frac{1}{2m}|{\mathbf D}\Psi|^2-e\mu|\Psi|^2+U(\x)|\Psi|^2\right\rbrace\;.\label{2.3.8}
\eeq
Associated with $\Bar{H}$ we can define a Lagrangian $\Bar{L}$ by a Legendre transformation and obtain the action functional
\beq
\Bar{S}\lbrack\Psi,\Psi^\dagger\rbrack=\int_{t_1}^{t_2}\!\!\!dt\Int\left\lbrace
\frac{i}{2}\left(\Psi^\dagger\Dot{\Psi}-\Dot{\Psi}^\dagger\Psi\right)-
\frac{1}{2m}|{\mathbf D}\Psi|^2+e\mu|\Psi|^2-U(\x)|\Psi|^2\right\rbrace\label{2.3.9}
\eeq
which incorporates the fact that the total charge is fixed. We can now use this action in the usual way. If $\bpsi$ represents the background field, assumed to be independent of time, it is easily seen that the effective potential density has a classical part which is given by
\beq
V^{(0)}_{\rm eff}=\frac{1}{2m}
|{\mathbf D}\bpsi|^2
+e\mu|\bpsi|^2-U(\x)|\bpsi|^2\;,\label{2.3.10}
\eeq

In order to calculate the quantum part of the effective potential we may proceed as we have done before and calculate the energy levels for the system. The field equation which follow from varying Eq.~\ref{2.3.9} with respect to $\Psi^\dagger$ is
\beq
0=i\Dot{\Psi}+\frac{1}{2m}{\mathbf D}^2\Psi+e\mu\Psi-U(\x)\Psi\;.\label{2.3.11}
\eeq
Apart from the term in $\mu$ which accounts for charge conservation this is just the Schr\"{o}dinger equation. Let $\lbrace f_n(\x)\rbrace$ be a complete set of solutions to
\beq
\left\lbrack-\frac{1}{2m}{\mathbf D}^2+U(\x)\right\rbrack f_n(\x)=\sigma_n
f_n(\x)\;.\label{2.3.12}
\eeq
Then $\sigma_n$ are seen to be the energy levels of the time independent Schr\"{o}dinger equation. If we write
\beq
\Psi(t,\x)=\sum_{n}A_ne^{-iE_nt}f_n(\x)\label{2.3.13}
\eeq
for some coefficients $A_n$, we see that
\beq
E_n=\sigma_n-e\mu\;.\label{2.3.14}
\eeq
The thermodynamic potential $\Omega_{T\ne0}$ is, from Sec.~\ref{secII2.1},
\beq
\Omega_{T\ne0}&=&\frac{1}{\beta}\sum_n\ln\left\lbrack1-e^{-\beta E_n}\right\rbrack\nn
&=&\frac{1}{\beta}\sum_n\ln\left\lbrack1-e^{-\beta(\sigma_n-e\mu)}\right\rbrack
\;.\label{2.3.16}
\eeq
This last result could have been written down immediately in a straightforward way from Sec.~\ref{secII2.1}; however the same procedure when applied to the relativistic theory is less trivial.

As an illustration, we will evaluate the thermodynamic potential when there is no magnetic field present. Then $\sigma_n$ is simply the same as $\omega_n$ in Eq.~\ref{2.2.1}. If we ignore $e$ in Eq.~\ref{2.3.16}, so that we talk about a fixed particle number rather than a fixed charge, then in the large box limit we have
\beq
\Omega_{T\ne0}=\frac{1}{\beta}\int\!\!dn_1\cdots dn_D\,\ln\left\lbrack1-
e^{-\beta(\omega_{\bf n}-\mu)}\right\rbrack\;,\label{2.3.17}
\eeq
in place of Eq.~\ref{2.2.2}. The presence of the chemical potential complicates things slightly. Making use of Eq.~\ref{2.2.3} we have
\beq
\Omega_{T\ne0}=-\frac{V_\Sigma}{\beta}\int\!\!\frac{d^Dn}{(2\pi)^D}\sum_{k=1}^{\infty}\frac{1}{k}e^{-k\beta\left(\frac{1}{2m}{\mathbf n}^2-\mu\right)}\;
\label{2.3.18}
\eeq
The integral over ${\mathbf n}$ may be performed as before, but this time the sum over $k$ is not simply a Riemann $\zeta$-function. We have
\beq
\Omega_{T\ne0}=-\frac{V_\Sigma}{\beta}\left(\frac{m}{2\pi\beta}\right)^{D/2}
\sum_{k=1}^{\infty}\frac{e^{k\beta\mu}}{k^{1+D/2}}\;.\label{2.3.19}
\eeq
Note that as discussed in Sec.~\ref{secII2.1} we need $\mu\le\omega_0=0$ here in order that there be no negative occupation numbers. The sum in Eq.~\ref{2.3.19} is therefore convergent. It can be written in terms of the polylogarithm function defined as
\beq
Li_p(z)=\sum_{n=1}^{\infty}\frac{z^n}{n^p}\;,\label{poly}
\eeq
for all $p$ if $|z|<1$, and for $\Re(p)>1$ if $|z|=1$. We can write
\beq
\Omega_{T\ne0}=-\frac{V_\Sigma}{\beta}\left(\frac{m}{2\pi\beta}\right)^{D/2}
Li_{1+D/2}\left(e^{\beta\mu}\right)\;.\label{2.3.20}
\eeq

\subsection{\it Relativistic Field}

The action for a charged relativistic scalar field $\Phi$ is
\beq
S=\int_{t_1}^{t_2}\!\!\!dt\Int\!\left\lbrace\left(D^\mu\Phi\right)^\dagger\left(D_\mu\Phi\right)-m^2\Phi^\dagger\Phi-U(\x)\Phi^\dagger\Phi\right\rbrace\;.\label{2.3.21}
\eeq
(We will keep $A_0$ temporarily non-zero here.) $D_\mu=\partial_\mu-ieA_\mu$ is the relativistic covariant derivative. The action Eq.~\ref{2.3.21} is invariant under the $U(1)$ gauge symmetry
\beq
\Phi&\rightarrow&e^{ie\theta}\Phi\;,\label{2.3.22a}\\
A_\mu&\rightarrow&A_\mu+\partial_\mu\theta\;.\label{2.3.22b}
\eeq
The Noether current which follows from Eq.~\ref{2.3.21} is
\beq
J_\mu=ie\left(\Phi^\dagger\partial_\mu\Phi-\partial_\mu\Phi^\dagger\Phi\right)-2ieA_\mu\Phi^\dagger\Phi\;.\label{2.3.23}
\eeq
The conserved charge is
\beq
Q=\Int J_0=ie\Int\left(\Phi^\dagger\Dot{\Phi}-\Dot{\Phi}^\dagger\Phi\right)
\;,\label{2.3.24}
\eeq
 after choosing the gauge $A_0=0$ and ${\mathbf A}$ satisfying Eq.~\ref{2.3.3}.

The momentum canonically conjugate to $\Phi$ is
\beq
\Pi=\frac{\partial{\mathcal L}}{\partial\Dot{\Phi}}=\Dot{\Phi}^\dagger\label{2.3.25}
\eeq
where ${\mathcal L}$ is the integrand of Eq.~\ref{2.3.21}. The Hamiltonian density is
\beq
{\mathcal H}&=&\Pi\Dot{\Phi}+\Pi^\dagger\Dot{\Phi}^\dagger-{\mathcal L}\nn
&=&\Pi^\dagger\Pi+|{\mathbf D}\Phi|^2+m^2|\Phi|^2+U(\x)|\Phi|^2\;.\label{2.3.26}
\eeq
To incorporate the conserved charge we form
\beq
\Bar{\mathcal H}&=&{\mathcal H}-\mu J_0\label{2.3.27}\\
&=&\Pi^\dagger\Pi+|{\mathbf D}\Phi|^2+m^2|\Phi|^2+U(\x)|\Phi|^2-ie\mu(\Pi^\dagger\Phi^\dagger-\Pi\Phi)\;.\label{2.3.28}
\eeq
$\Bar{H}$ is just $\Bar{H}=\Int\Bar{\mathcal H}$. 

To go back to a Lagrangian description we form $\Bar{\mathcal L}$ defined in terms of the Legendre transformation of $\Bar{\mathcal H}$ by
\beq
\Bar{\mathcal L}=\Pi\Dot{\Phi}+\Pi^\dagger\Dot{\Phi}^\dagger-\Bar{\mathcal H}\label{2.3.29}
\eeq
with
\beq
\Dot{\Phi}=\frac{\partial\Bar{\mathcal H}}{\partial\Pi}=\Pi^\dagger+ie\mu\Phi\;.\label{2.3.30a}
\eeq
Similarly,
\beq
\Dot{\Phi}^\dagger=\frac{\partial\Bar{\mathcal H}}{\partial\Pi^\dagger}=\Pi-ie\mu\Phi^\dagger\;.\label{2.3.30b}
\eeq
If we eliminate $\Pi$ and $\Pi^\dagger$ Using Eqs.~\ref{2.3.30a} and \ref{2.3.30b} we find
\beq
\Bar{\mathcal L}=\Dot{\Phi}^\dagger\Dot{\Phi}+ie\mu(\Phi^\dagger\Dot{\Phi}-\Dot{\Phi}^\dagger\Phi)+e^2\mu^2\Phi^\dagger\Phi-|{\mathbf D}\Phi|^2-m^2\Phi^\dagger\Phi-U(\x)\Phi^\dagger\Phi
\;,\label{2.3.31}
\eeq
as the Lagrangian density which describes the theory with a conserved charge.

The Euler-Lagrange equations which follow from Eq.~\ref{2.3.31} are
\beq
0&=&-\Ddot{\Phi}+2ie\mu\Dot{\Phi}+{\mathbf D}^2\Phi-(m^2-e^2\mu^2)\Phi-U(\x)\Phi\;,\label{2.3.32a}\\
0&=&-\Ddot{\Phi}^\dagger-2ie\mu\Dot{\Phi}^\dagger+{\mathbf D}^2\Phi^\dagger-(m^2-e^2\mu^2)\Phi^\dagger-U(\x)\Phi^\dagger
\;.\label{2.3.32b}
\eeq
If we let $\lbrace f_n(\x)\rbrace$ be a complete set of solutions to
\beq
\left\lbrack-{\mathbf D}^2+U(\x)\right\rbrack f_n(\x)=\sigma_n f_n(\x)\;,\label{2.3.33}
\eeq
we may expand
\beq
\Phi(t,\x)=\sum_nA_ne^{-iE_nt}f_n(\x)\;,\label{2.3.34}
\eeq
for expansion coefficients $A_n$. Substitution of this expansion for $\Phi$ into Eq.~\ref{2.3.32a} results in
\begin{displaymath}
0=E_n^2+2e\mu E_n-\sigma_n-(m^2-e^2\mu^2)\;,
\end{displaymath}
giving
\beq
E_n=\sqrt{\sigma_n+m^2}-e\mu\;.\label{2.3.35}
\eeq
A similar analysis applied to Eq.~\ref{2.3.32b} gives
\beq
E_n=\sqrt{\sigma_n+m^2}+e\mu\;.\label{2.3.36}
\eeq
The two results for $E_n$ are seen to differ only in the sign of the electric charge. Physically they correspond to the contributions from particles and antiparticles. The thermodynamic potential in Eqs.~\ref{2.1.20}---\ref{2.1.22} is therefore
\beq
\Omega_{T=0}=\sum_n\omega_n\;,\label{2.3.37}
\eeq
and
\beq
\Omega_{T\ne0}=\frac{1}{\beta}\sum_n\ln\left\lbrack1-
e^{-\beta(\omega_n-e\mu)}\right\rbrack+
\frac{1}{\beta}\sum_n\ln\left\lbrack1-
e^{-\beta(\omega_n+e\mu)}\right\rbrack\;,\label{2.3.38}
\eeq
where
\beq
\omega_n=\sqrt{\sigma_n+m^2}\;.\label{2.3.39}
\eeq
$\Omega_{T=0}$ in Eq.~\ref{2.3.37} differs from that found for the real scalar field by a factor of 2, which is easily understood due to the fact that a complex scalar field has two real degrees of freedom (the real and imaginary parts). The two terms in Eq.~\ref{2.3.38} can be thought of as the separate contributions from particles and antiparticles. We note that $|e\mu|\le\omega_0=\sqrt{\sigma_0+m^2}$ here to avoid the problem of negative occupation numbers. For flat space with $\sigma_0=0$ this requires $|e\mu|\le m$.

The high temperature expansion of Eq.~\ref{2.3.38} presents some difficulties. There are a number of methods in the literature \cite{HW,Actor,TomsBEC,Daicicetal} which are quite involved. Here we will show in a very simple way how the Mellin-Barnes integral representation in Eq.~\ref{MB} can be used to obtain the leading terms in the expansion. This method also has the advantage of being directly generalized to curved space as we will show in the next section. The first step, as for $\mu=0$, is to expand the logarithms using Eq.~\ref{2.2.3}. This gives
\beq
\Omega_{T\ne0}&=&-\frac{1}{\beta}\sum_{k=1}^{\infty}\frac{1}{k}\sum_n
e^{-k\beta\omega_n}\left(e^{k\beta e\mu}+e^{-k\beta e\mu}\right)
\label{2.3.40a}\\
&=&-\frac{2}{\beta}\sum_{k=1}^{\infty}\sum_ne^{-k\beta\omega_n}\frac{\cosh(k\beta e\mu)}{k}\;.\label{2.3.40b}
\eeq
We cannot just use the Mellin-Barnes representation for the exponential in Eq.~\ref{2.3.40b} because the sum over $k$ will diverge due to the presence of $\cosh(k\beta e\mu)$. This means that we must deviate slightly from the procedure used in the non-relativistic case. Instead we will argue that because we are interested in $\beta m<<1$ we also have $\beta|e\mu|<<1$ since $|e\mu|\le m$. We therefore expand $\cosh(k\beta e\mu)$ in powers of $\mu$ and keep only the first few terms~:
\beq
\Omega_{T\ne0}\simeq-\frac{2}{\beta}\sum_{k=1}^{\infty}\sum_ne^{-k\beta\omega_n}\left\lbrace\frac{1}{k}+\frac{1}{2}\beta^2e^2\mu^2k+
\frac{1}{24}\beta^4e^4\mu^4k^3\right\rbrace\;.\label{2.3.41}
\eeq
The first term on the RHS of Eq.~\ref{2.3.41} is just the result for the thermodynamic potential found earlier for $\mu=0$ (apart from an overall  factor of 2). An equivalent way of arriving at eq.~\ref{2.3.41} is to start from Eq.~\ref{2.3.38} and expand the thermodynamic potential directly in powers of the small quantity $\beta e\mu$. Since $\Omega_{T\ne0}$ is an even function of $\mu$, only even powers of $\mu$ will arise.

The only difference between the various terms in Eq.~\ref{2.3.41} is the power of $k$ which enters the sum. We can define
\beq
f_l=\sum_{k=1}^{\infty}\sum_ne^{-k\beta\omega_n}k^{2l-1}\;,\label{2.3.42}
\eeq
where $l$ is an integer. We evaluated $f_0$ in Sec.~\ref{secIIRelgas}. Exactly the same procedure used there can be applied to calculate $f_l$ for $l=1,2$. Use of Eq.~\ref{MB} results in the contour integral representation
\beq
f_l=\cint\Gamma(\alpha)\beta^{-\alpha}\zeta_R(\alpha+1-2l)
\Big(\sum_n\omega_n^{-\alpha}\Big)\;,\label{2.3.43}
\eeq
after performing the sum on $k$ in terms of the Riemann $\zeta$-function. We have already evaluated $\sum_n\omega_n^{-\alpha}$ in the large box limit in Eq.~\ref{2.2.11}. After using this expression in Eq.~\ref{2.3.44} along with the duplication formula Eq.~\ref{dup} for the $\Gamma$-function we find
\beq
f_l=\frac{V_\Sigma}{\sqrt{4\pi}}\left(\frac{m^2}{4\pi}\right)^{D/2}
\!\!\cint\Gamma\Big(\frac{\alpha-D}{2}\Big)
\Gamma\Big(\frac{\alpha+1}{2}\Big)
\zeta_R(\alpha+1-2l)\Big(\frac{1}{2}\beta m\Big)^{-\alpha}
\!\!.\label{2.3.44}
\eeq
This is valid for $c>{\rm max}\;(D/2,2l-1)$.

It now only remains to study the pole structure of the integrand and evaluate the residues. $\Gamma\big(\frac{\alpha+1}{2}\big)\zeta_R(\alpha+1-2l)$ has only a single simple pole at $\alpha=2l$ coming from the Riemann $\zeta$-function. The residue at this pole is $\Gamma(l+1/2)$. (Note that the potential poles of $\Gamma\big(\frac{\alpha+1}{2}\big)$ coming at $\alpha=-1,-3,-5,\ldots$ are cancelled by the zeros of $\zeta_R(\alpha+1-2l)$ at these values of $\alpha$. We are assuming that $l\ge1$ here since $l=0$ was considered in Sec.~\ref{secIIRelgas}.) As before the poles of $\Gamma\big(\frac{\alpha-D}{2}\big)$ occur at $\alpha=D-2n$ for $n=0,1,2,\ldots$. For odd $D$ none of these poles coincide with the pole at $\alpha=2l$ and therefore the integrand of Eq.~\ref{2.3.44} has only a series of simple poles. For even $D$ we can have a double pole at $\alpha=2l$ corresponding to $n=D/2-l$ when $l\le D/2$. This presents enough information to evaluate $f_l$ for all $D$. Rather than write down the general result, we will simply quote the results for $D=1,2,3$. For $D=1$ we have
\beq
\frac{\Omega_{T\ne0}}{V_\Sigma}\simeq-\frac{\pi}{3}\beta^{-2}+\beta^{-1}m^{-1}(m^2-\frac{1}{2}e^2\mu^2)+\cdots\;.\label{2.3.45}
\eeq
For $D=2$ we have
\beq
\frac{\Omega_{T\ne0}}{V_\Sigma}&\simeq&-\frac{\zeta_R(3)}{\pi}\beta^{-3}-
\frac{1}{4\pi}\beta^{-1}(m^2+2e^2m^2)\nn
&&+\frac{1}{2\pi}\beta^{-1}(m^2+e^2\mu^2)\ln(\beta m)+\cdots\;.\label{2.3.46}
\eeq
For $D=3$ the result is
\beq
\frac{\Omega_{T\ne0}}{V_\Sigma}\simeq-\frac{\pi^2}{45}\beta^{-4}
+\frac{1}{12}\beta^{-2}(m^2-2e^2m^2)
-\frac{1}{6\pi}\beta^{-1}m(m^2-\frac{3}{2}e^2\mu^2)+\cdots\;.\label{2.3.47}
\eeq
For $D\ge4$ the leading terms in the expansion are
\beq
\frac{\Omega_{T\ne0}}{V_\Sigma}&\simeq&-2\pi^{-(D+1)/2}\Gamma\Big(\frac{D+1}{2}\Big)\zeta_R(D+1)\beta^{-D-1}\label{2.3.48}\\
&+&\!\!\!\frac{1}{2}\pi^{-(D+1)/2}\Gamma\Big(\frac{D-1}{2}\Big)
\zeta_R(D-1)\beta^{1-D}\lbrack m^2-(D-1)e^2\mu^2\rbrack+\cdots.\nonumber
\eeq
We will derive a more general version of this result in the next section.

\section{Curved Space}
\label{curvedspace}

We will now show how to obtain the high temperature expansion for the thermodynamic potential of the non-relativistic and relativistic ideal gas for a general space $\Sigma$ with a possible boundary.

\subsection{\it Uncharged Non-relativistic Gas}

For the non-relativistic gas we have
\beq
\Omega_{T\ne0}=\frac{1}{\beta}\sum_n\ln\left\lbrack1-e^{-\beta\omega_n}
\right\rbrack\;,\label{2.4.1}
\eeq
with $\omega_n=\sigma_n/(2m)$. Using Eq.~\ref{2.2.3} to expand the logarithm, and making use of the Mellin-Barnes integral representation Eq.~\ref{MB} leads immediately to
\beq
\Omega_{T\ne0}=-\frac{1}{\beta}\cint\Gamma(\alpha)\beta^{-\alpha}\zeta_R(\alpha+1)\Big(\sum_n\omega_n^{-\alpha}\Big)\;.\label{2.4.2}
\eeq
The remaining sum in Eq.~\ref{2.4.2} is just the energy $\zeta$-function of Sec~3, which was
\beq
\omega(s)=\sum_n\omega_n^{1-s}\;,\label{2.4.3}
\eeq
if we ignore the renormalization length. We can therefore write
\beq
\Omega_{T\ne0}=-\frac{1}{\beta}\cint\Gamma(\alpha)\beta^{-\alpha}\zeta_R(\alpha+1)\omega(\alpha+1)\;,\label{2.4.4}
\eeq
where we take $c>D$ so that the integration contour is to the right of the poles of $\omega(\alpha+1)$. We can analyze the pole structure of $\omega(\alpha+1)$ exactly as in Sec.~3.3. Note that
\begin{eqnarray*}
\omega(\alpha+1)&=&\sum_n\omega_n^{-\alpha}=(2m)^\alpha
\sum_n\sigma_n^{-\alpha}\\
&=&(2m)^\alpha\sum_n\frac{1}{\Gamma(\alpha)}\int_{0}^{\infty}\!\!dt\,
t^{\alpha-1}e^{-t\sigma_n}\\
&=&\frac{(2m)^\alpha}{\Gamma(\alpha)}\int_{0}^{\infty}\!\!dt\,
t^{\alpha-1}\Theta(t)\;,
\end{eqnarray*}
where $\Theta(t)=\sum_ne^{-t\sigma_n}$. We therefore have
\beq
\Gamma(\alpha)\omega(\alpha+1)&=&(2m)^\alpha\int_{0}^{\infty}\!\!dt\,
t^{\alpha-1}\Theta(t)\label{2.4.5}\\
&=&(2m)^\alpha I_1(\alpha)+(2m)^\alpha I_2(\alpha)\label{2.4.6}
\eeq
where
\beq
I_1(\alpha)&=&\int_{0}^{t_0}\!\!dt\,
t^{\alpha-1}\Theta(t)\;,\label{2.4.7a}\\
I_2(\alpha)&=&\int_{t_0}^{\infty}\!\!dt\,
t^{\alpha-1}\Theta(t)\;.\label{2.4.7b}
\eeq
As before, $t_0$ is an arbitrary positive constant. Repeating the argument given in Sec.~\ref{sec3.3} shows that $I_2(\alpha)$ should be an analytic function of $\alpha$ for any $t_0>0$. Therefore any poles of $\Gamma(\alpha)\omega(\alpha+1)$ can come only from $I_1(\alpha)$. As we did earlier, we can take $t_0$ to be arbitrarily small and use the asymptotic expansion Eq.~\ref{3.3.14} for $\Theta(t)$ in the integrand of $I_1(\alpha)$. This gives
\beq
I_1(\alpha)=(4\pi)^{-D/2}\sum_{k=0,\frac{1}{2},1,\ldots}^{\infty}\theta_k
\frac{t_0^{\alpha+k-D/2}}{\alpha+k-D/2}\;,\label{2.4.8}
\eeq
showing poles at $\alpha=D/2-k$ for $k=0,\frac{1}{2},1,\ldots$.

The only other source of poles in the integrand of Eq.~\ref{2.4.4} is from the simple pole at $\alpha=0$ of $\zeta_R(\alpha+1)$. This may coincide with the pole of $\Gamma(\alpha)\omega(\alpha+1)$ at $k=D/2$ if $D$ is even, giving a double pole for the integrand of Eq.~\ref{2.4.4}. In any case, even if $D$ is odd, we require the finite part of  $\Gamma(\alpha)\omega(\alpha+1)$ at $\alpha=0$ which is impossible to write down in the general case. In order to avoid this calculational problem we will keep only the terms up to $k=(D-1)/2$ in the high temperature expansion of $\Omega_{T\ne0}$. We therefore find
\beq
\Omega_{T\ne0}\simeq-\frac{1}{\beta}\left(\frac{m}{2\pi\beta}\right)^{D/2}\sum_{k=0,\frac{1}{2},1,\ldots}^{D/2-1}\left(\frac{\beta}{2m}\right)^k\zeta_R(1+D/2-k)\theta_k\;.\label{2.4.9}
\eeq
Given that $\theta_0=V_\Sigma$, the first term in this result is seen to agree with that of Eq.~\ref{2.2.4} found for flat $\Sigma$. The remaining terms in the sum over $k$ in Eq.~\ref{2.4.9} represent the corrections present if $\Sigma$ has a boundary or is curved. In particular cases they can be written down from the known coefficients $\theta_k$.

\subsection{\it Uncharged Relativistic Gas}

For the relativistic gas we still have Eq.~\ref{2.4.4} holding, but with $\omega_n=\sqrt{\sigma_n+m^2}$ this time. The results of Sec.~3.3 are directly relevant here. If we split up $\omega(\alpha+1)$ as in Eq.~\ref{3.3.10} and use Eq.~\ref{3.3.15} we find
\beq
\omega(\alpha)=\frac{(4\pi)^{-D/2}}{\Gamma(\alpha/2)}
\sum_{k=0,\frac{1}{2},1,\ldots}^{\infty}\theta_k
\frac{t_0^{k+(\alpha-D)/2}}{k+(\alpha-D)/2}+\;{\rm analytic\ terms}
\;.\label{2.4.10}
\eeq
As in the non-relativistic case, if we are to avoid terms which require a knowledge of the finite part of $\omega(\alpha+1)$, we want to only keep those poles with $\alpha>0$; namely, $k=0,\frac{1}{2},\ldots,\frac{D-1}{2}$. This gives
\beq
\Omega_{T\ne0}\simeq-\pi^{-(D+1)/2}\beta^{-D-1}
\sum_{k=0,\frac{1}{2},1,\ldots}^{(D-1)/2}
\Gamma\Big(\frac{D+1}{2}-k\Big)\left(\frac{\beta}{2}\right)^{2k}
\zeta_R(1+D-2k)\theta_k\;.\label{2.4.11}
\eeq
Again if we use $\theta_0=V_\Sigma$ the first term gives a result in agreement with our flat space calculation.

\subsection{\it Charged Non-relativistic Gas}

We can begin with eq.~\ref{2.3.16} and expand the logarithm as usual to obtain
\beq
\Omega_{T\ne0}=-\frac{1}{\beta}\sum_{k=1}^{\infty}\frac{e^{k\beta e\mu}}{k}\sum_ne^{-\beta k\sigma_n}\;.\label{2.4.12}
\eeq
The only difference with the expression in Sec.~7.1 is the presence of $e^{k\beta e\mu}$. It is easy to see that in place of Eq.~\ref{2.4.4} we will get
\beq
\Omega_{T\ne0}=-\frac{1}{\beta}\cint\Gamma(\alpha)\beta^{-\alpha}Li_{1+\alpha}\big(e^{\beta e\mu}\big)\omega(\alpha+1)\;,\label{2.4.13}
\eeq
where $Li$ is the polylogarithm function defined in Eq.~\ref{poly}.

The advantage of this result over that in Eq.~\ref{2.4.4} is that $Li_{1+\alpha}$ is an analytic function of $\alpha$ provided that $\mu<0$. Therefore so long as $\mu<0$ we have
\beq
\Omega_{T\ne0}\simeq-\frac{1}{\beta}\left(\frac{m}{2\pi\beta}\right)^{D/2}
\sum_{k=0,\frac{1}{2},1,\ldots}^{\infty}\left(\frac{\beta}{2m}\right)^k\theta_k Li_{1-k+D/2}\big(e^{\beta e\mu}\big)\;.\label{2.4.14}
\eeq
The first term of this result is in agreement with that in Eq.~\ref{2.3.20}.

\subsection{\it Charged Relativistic Gas}

For the relativistic gas we had from Eq.~\ref{2.3.41} and \ref{2.3.42}
\beq
\Omega_{T\ne0}\simeq-\frac{2}{\beta}\left\lbrace f_0+\frac{1}{2}\beta^2e^2
\mu^2f_1+\frac{1}{24}\beta^4e^4\mu^4f_2\right\rbrace\;.\label{2.4.15}
\eeq
The result in Eq.~\ref{2.3.43} still holds, so we have
\beq
f_l=\cint\Gamma(\alpha)\beta^{-\alpha}\zeta_R(\alpha+1-2l)\omega(\alpha+1)\;.\label{2.4.16}
\eeq
The first term on the RHS of Eq.~\ref{2.4.15} is just the $\mu=0$ result, and therefore should be twice the result found for the real field in Eq.~\ref{2.4.11}. The leading terms in $f_1$ and $f_2$ may be evaluated by using Eq.~\ref{2.4.10} in a straightforward manner. In order to avoid a knowledge of the finite part of $\omega(\alpha+1)$ we must only keep those poles of $\omega(\alpha+1)$ which lie to the right of those of $\zeta_R(\alpha+1-2l)$; namely, those with $\alpha>2l$. For $f_1$ this means that we can only keep pole terms with $\alpha=D-2k$ for $k=0,\frac{1}{2},1,\ldots,\frac{D-3}{2}$. It is therefore not possible to obtain general results if $D<3$. For $f_2$ we must have $D\ge5$. For specific spaces these restrictions can be removed; however in the general case we are unable to write down the general term.

For $f_1$ assuming $D\ge3$ we have
\beq
f_1&\simeq&\pi^{-(D+1)/2}\beta^{-D}\sum_{k=0,\frac{1}{2},1,\ldots}^{(D-3)/2}\left(\frac{\beta}{2}\right)^{2k}\Gamma\Big(\frac{D+1}{2}-k\Big)
\zeta_R(D-1-2k)\theta_k\label{2.4.17}\\
&\simeq&\pi^{-(D+1)/2}\beta^{-D}\Gamma\Big(\frac{D+1}{2}\Big)
\zeta_R(D-1)V_\Sigma+\cdots\label{2.4.18}
\eeq
if we keep only the leading order term and use $\theta_0=V_\Sigma$. This last result is general and holds for any space $\Sigma$ irrespective of whether there is a boundary or not, and independently of any boundary conditions imposed on the field. We have therefore established that the leading terms in the high temperature expansion of the thermodynamic potential for $D\ge3$ are
\beq
\Omega_{T\ne0}&\simeq&-2\pi^{-(D+1)/2}\Gamma\Big(\frac{D+1}{2}\Big)\zeta_R(D+1)\beta^{-D-1}V_\Sigma\nn
&&-\pi^{-(D+1)/2}\Gamma(D/2)\zeta_R(D)\beta^{-D}\theta_{1/2}\label{2.4.19}\\
&&-\frac{1}{2}\pi^{-(D+1)/2}\Gamma\Big(\frac{D-1}{2}\Big)\zeta_R(D-1)\beta^{1-D}\lbrack\theta_1+(D-1)e^2\mu^2V_\Sigma\rbrack
+\cdots\;.\nonumber
\eeq
For flat space with no boundary $\theta_1=-m^2V_\Sigma$ and we recover the result of Eq.~\ref{2.3.48}. The result in Eq.~\ref{2.4.19} was obtained in Ref.~\cite{TomsBEC} using a different method to that used here.

\section{Charged Interacting Gas}
\label{interacting}

We will now consider the inclusion of a self-interaction term to the ideal relativistic gas. The classical action will be written as
\beq
S=\int_{t_1}^{t_2}\!\!dt\Int\left\lbrace\Big(D^\mu\Phi\Big)^\dagger\Big(D_\mu\Phi\Big)-m^2\Phi^\dagger\Phi-U(\x)\Phi^\dagger\Phi-V_{\rm tree}(|\Phi|^2)\right\rbrace\;,\label{2.5.1}
\eeq
where $V_{\rm tree}(|\Phi|^2)$ is some potential term which represents the interaction. We will assume that the potential depends only on $|\Phi|^2$ to ensure invariance under the $U(1)$ gauge transformation Eq.~\ref{2.3.22a}. Following through the same steps as those which led to Eq.~\ref{2.3.31} we find
\beq
\Bar{\mathcal L}&=&\Dot{\Phi}^\dagger\Dot{\Phi}+ie\mu(\Phi^\dagger\Dot{\Phi}-\Dot{\Phi}^\dagger\Phi)+e^2\mu^2\Phi^\dagger\Phi-|{\mathbf D}\Phi|^2\nn
&&\quad-m^2\Phi^\dagger\Phi-U(\x)\Phi^\dagger\Phi-V_{\rm tree}(|\Phi|^2)
\;,\label{2.5.2}
\eeq
as the Lagrangian density describing the theory with a conserved charge. Variation with respect to $\Phi$ and $\Phi^\dagger$ results in
\beq
0&=&-\Ddot{\Phi}+2ie\mu\Dot{\Phi}+{\mathbf D}^2\Phi-(m^2-e^2\mu^2)\Phi-U(\x)\Phi-V'_{\rm tree}\Phi\;,\label{2.5.3}\\
0&=&-\Ddot{\Phi}^\dagger-2ie\mu\Dot{\Phi}^\dagger+{\mathbf D}^2\Phi^\dagger-(m^2-e^2\mu^2)\Phi^\dagger-U(\x)\Phi^\dagger
-V'_{\rm tree}\Phi^\dagger\;,\label{2.5.4}
\eeq
where $V'_{\rm tree}=\frac{\partial}{\partial|\Phi|^2}V_{\rm tree}(|\Phi|^2)$.

In order to evaluate the thermodynamic potential we will calculate the excitation energies \cite{Bern} by looking at the split of $\Phi$ into the background field $\Bar{\Phi}$ and the quantum fluctuation~:
\beq
\Phi=\Bar{\Phi}+\psi\;.\label{2.5.5}
\eeq
This split may be substituted into the field equations \ref{2.5.3} and \ref{2.5.4} and the result linearized in $\psi$. (Compare with the zero temperature result in Sec.~3.3.) The fluctuations $\psi$ obey
\beq
0&=&-\Ddot{\psi}+2ie\mu\Dot{\psi}+{\mathbf D}^2\psi-(m^2-e^2\mu^2+U(\x))\psi\nn
&&-V'_{\rm tree}(|\Bar{\Phi}|^2)\psi-V''_{\rm tree}(|\Bar{\Phi}|^2)\left\lbrack|\Bar{\Phi}|^2\psi+\Bar{\Phi}^2\psi^\dagger\right\rbrack\;,\label{2.5.6}
\eeq
and the complex conjugate equation. If we assume that $\Bar{\Phi}$ is constant as we did for the zero temperature theory (or less strongly that the quantum corrections to the derivative part of the potential are negligible in comparison with the tree level part) then we may proceed in much the same way as we did in Sec.~3.3. 

Expand
\beq
\psi&=&\sum_nA_ne^{-iE_nt}f_n(\x)\;,\label{2.5.7}\\
\psi^\dagger&=&\sum_nB_ne^{-iE_nt}f_n(\x)\;,\label{2.5.8}
\eeq
where $f_n(\x)$ obeys Eq.~\ref{2.3.33}. Here $A_n$ and $B_n$ are arbitrary expansion coefficients. Substitution into Eq.~\ref{2.5.6} and its complex conjugate leads to two coupled equations for the expansion coefficients if we use the linear independence of the $f_n(\x)$~:
\beq
0&=&\big\lbrack(E_n+e\mu)^2-\sigma_n-m^2-V'_{\rm tree}-|\Bar{\Phi}|^2V''_{\rm tree}\big\rbrack A_n-\Bar{\Phi}^2V''_{\rm tree}B_n\;,\label{2.5.9}\\
0&=&\big\lbrack(E_n-e\mu)^2-\sigma_n-m^2-V'_{\rm tree}-|\Bar{\Phi}|^2V''_{\rm tree}\big\rbrack B_n-(\Bar{\Phi}^\dagger)^2V''_{\rm tree}A_n\;.\label{2.5.10}
\eeq
In order that these equations have a non-trivial solution for $A_n$ and $B_n$ the $2\times2$ matrix of coefficients of $A_n$ and $B_n$ found from Eqs.~\ref{2.5.9} and \ref{2.5.10} must have a vanishing determinant. This leads to $E_n=E^{\pm}_{n}$ where
\beq
E^{\pm}_{n}=\left\lbrace M_n^2+e^2\mu^2\pm\left\lbrack4e^2\mu^2M_n^2+
|\Bar{\Phi}|^4(V''_{\rm tree})^2\right\rbrack^{1/2}\right\rbrace^{1/2}\;,
\label{2.5.11}
\eeq
with
\beq
M_n^2=\sigma_n+m^2+V'_{\rm tree}+|\Bar{\Phi}|^2
V''_{\rm tree}\;.\label{2.5.12}
\eeq
This agrees with the results of Refs.~\cite{Bern,Benson} if we specialize to flat space and write the results in terms of real fields. As a check on the results if we take $V_{\rm tree}\rightarrow0$, the result in Eq.~\ref{2.5.11} reduces to those found in Eqs.~\ref{2.3.35} and \ref{2.3.36} for the non-interacting theory.

The thermodynamic potential is therefore
\beq
\Omega_{T\ne0}=\frac{1}{\beta}\sum_n\left\lbrace\ln\left(1-e^{-\beta E_n^+}\right)+\ln\left(1-e^{-\beta E_n^-}\right)\right\rbrace\;,\label{2.5.13}
\eeq
with $E_n^\pm$ given by Eq.~\ref{2.5.11}. An exact evaluation of this expression is impossible for general $\Sigma$. Considerable progress can be made in flat space, but we will keep $\Sigma$ general and show how to obtain the leading order behaviour at high temperature.

Define
\beq
M_n^2=\omega_n^2+\delta M^2\;,\label{2.5.14}
\eeq
where
\beq
\delta M^2=V'_{\rm tree}(|\Bar{\Phi}|^2)+|\Bar{\Phi}|^2V''_{\rm tree}(|\Bar{\Phi}|^2)\;,\label{2.5.15}
\eeq
and $\omega_n=\sqrt{\sigma_n+m^2}$ as usual. By expanding $E_n^\pm$ to first order in the potential $V_{\rm tree}$ it is straightforward to show from Eq.~\ref{2.5.11} that
\beq
E_n^\pm=\omega_n\pm e\mu+\frac{\delta M^2}{2\omega_n}+\cdots\;.\label{2.5.16}
\eeq
This result may be used in Eq.~\ref{2.5.13} with the result expanded to first order in $\delta M^2$. We find
\beq
\Omega_{T\ne0}&=&\frac{1}{\beta}\sum_n\left\lbrace\ln\left\lbrack1-e^{-\beta(\omega_n-e\mu)}\right\rbrack+\ln\left\lbrack1-e^{-\beta(\omega_n+e\mu)}\right\rbrack\right\rbrace\nn
&&+\sum_n\frac{\delta M^2}{2\omega_n}\left\lbrace\left\lbrack e^{\beta(\omega_n-e\mu)}-1\right\rbrack^{-1}+\left\lbrack e^{\beta(\omega_n+e\mu)}-1\right\rbrack^{-1}\right\rbrace+\cdots\;.\label{2.5.17}
\eeq
The first term on the RHS of Eq.~\ref{2.5.17} is just the contribution from the free theory which was calculated in the previous section. The term in $\delta M^2$ can be evaluated in quite a simple manner if it is noted that
\beq
\frac{1}{2\omega_n}\left\lbrack e^{\beta(\omega_n\pm e\mu)}-1\right\rbrack^{-1}=\frac{1}{\beta}\frac{\partial}{\partial m^2}\ln\left\lbrack1-e^{-\beta(\omega_n\pm e\mu)}\right\rbrack\;.\label{2.5.18}
\eeq
If we call
\beq
\Omega_{T\ne0}^{\rm free}=
\frac{1}{\beta}\sum_n\left\lbrace\ln\left\lbrack1-
e^{-\beta(\omega_n-e\mu)}\right\rbrack+
\ln\left\lbrack1-e^{-\beta(\omega_n+e\mu)}\right\rbrack
\right\rbrace\;,\label{2.5.19}
\eeq
then
\beq
\Omega_{T\ne0}=\Omega_{T\ne0}^{\rm free}+\delta M^2\frac{\partial}{\partial m^2}\Omega_{T\ne0}^{\rm free}+\cdots\;.\label{2.5.20}
\eeq
$\Omega_{T\ne0}^{\rm free}$ was given in Eq.~\ref{2.4.19} for $D\ge3$. If we use the fact that $m^2$ first enters $\theta_1$, and that $\frac{\partial}{\partial m^2}\theta_1=-V_\Sigma$, we have
\beq
\frac{\partial}{\partial m^2}\Omega_{T\ne0}^{\rm free}=\frac{1}{2}
\pi^{-(D+1)/2}\Gamma\Big(\frac{D-1}{2}\Big)\zeta_R(D-1)\beta^{1-D}V_\Sigma+\cdots\;,\label{2.5.21}
\eeq
as the leading term at high temperature. When the leading correction due to the interaction is included in the thermodynamic potential, the high temperature expansion becomes
\beq
\Omega_{T\ne0}&\simeq&-2\pi^{-(D+1)/2}\Gamma\Big(\frac{D+1}{2}\Big)\zeta_R(D+1)\beta^{-D-1}V_\Sigma\nn
&&-\pi^{-(D+1)/2}\Gamma(D/2)\zeta_R(D)\beta^{-D}\theta_{1/2}\nn
&&-\frac{1}{2}\pi^{-(D+1)/2}\Gamma\Big(\frac{D-1}{2}\Big)\zeta_R(D-1)\beta^{1-D}\nn
&&\times\left\lbrace\theta_1+V_\Sigma\left\lbrack-\delta M^2+(D-1)e^2\mu^2\right\rbrack\right\rbrace+\cdots\;.\label{2.5.22}
\eeq

In the case of greatest physical interest, $D=3$, if we take $U(\x)=\xi R$ with $R$ treated as constant and ignore boundary terms we have $\theta_1=-\lbrack m^2+(\xi-\frac{1}{6})R\rbrack$. Dropping $\theta_{1/2}$ (since it is a boundary term) we find
\beq
\frac{\Omega_{T\ne0}}{V_\Sigma}&\simeq&-\frac{\pi^2}{45}\beta^{-4}+
\frac{1}{12}\beta^{-2}\Big\lbrack m^2+(\xi-\frac{1}{6})R+
V'_{\rm tree}(|\Bar{\Phi}|^2)\nn
&&\quad+|\Bar{\Phi}|^2V''_{\rm tree}(|\Bar{\Phi}|^2)-2e^2\mu^2\Big\rbrack+\cdots\;,\label{2.5.23}
\eeq
as the leading terms at high temperature.

\section{Bose-Einstein Condensation for the Non-rel\-at\-iv\-is\-tic Ideal Gas}
\label{secBECnon}

The fact that below a certain critical temperature an ideal gas of bosons can condense into the ground state was first predicted by Einstein \cite{Einstein}. The basic physics of the situation can be understood using a simple argument based on the Heisenberg uncertainty principle. Suppose that we have $N$ bosons confined in a volume $V$. Each boson has a volume $N/V=L^3$ available to it. Because the bosons are confined there is a zero-point energy of $E_0=p^2/(2m)$ present. The uncertainty principle relates the momentum associated with the zero-point motion $p$ to the linear size of the confining region $L$ by $pL\approx\hbar$. We therefore find $E_0\approx\hbar^2/(2mL^2)=\frac{\hbar^2}{2m}(N/V)^{2/3}$. The spacing between the ground and the first excited state would be expected to be of the order $E_0$. If $kT>E_0$ then there is sufficient thermal energy for there to be many transitions between the ground and the excited states; however if $kT<E_0$ then we would not expect this to be the case. For $T<T_c$ where $kT_c\approx E_0$ there is insufficient thermal energy to excite particles from the ground state and the bosons will condense. We estimate that the critical temperature $T_c$ for this to happen is $T_c\approx\frac{\hbar^2}{2m}(N/V)^{2/3}$.

The argument just presented is of course no substitute for a proper mathematical derivation; however we will show that the result obtained by this heuristic argument is of the correct order of magnitude. A derivation of Bose-Einstein condensation (BEC) along the lines followed in these lectures will be given. This has the feature that the connection between BEC and symmetry breaking in quantum field theory becomes very obvious. (The connection was described in relativistic field theory in Refs.~\cite{HW,Kapusta}.)

The non-relativistic gas has an effective potential determined in Eq.~\ref{2.3.1}
to be
\beq
\Omega_{\rm eff}=\Int\left\lbrace\frac{1}{2m}|{\mathbf D}\bpsi|^2+U(\x)|\bpsi|^2-e\mu|\bpsi|^2\right\rbrace+\Omega_{T\ne0}
\;,\label{3.1.1}
\eeq
with $\Omega_{T\ne0}$ given by eq.~\ref{2.3.16}. An important point is that for the ideal gas (defined as having no self-interactions) $\Omega_{T\ne0}$ does not have any dependence on $\bpsi$. When we form the background field equation there is no contribution from $\Omega_{T\ne0}$ and we find simply
\beq
\left\lbrack-\frac{1}{2m}{\mathbf D}^2+U(\x)-e\mu\right\rbrack\bpsi(\x)=0\;.\label{3.1.2}
\eeq
(We have left ${\mathbf A}$ and $U(\x)$ non-zero since they do not affect the main analysis given here.) Although we cannot solve Eq.~\ref{3.1.2} exactly, we can expand $\bpsi(\x)$ in terms of the modes $f_n(\x)$ defined in Eq.~\ref{2.3.12} as 
\beq
\bpsi(\x)=\sum_n C_nf_n(\x)\;.\label{3.1.3}
\eeq
Here $C_n$ are some expansion coefficients to be determined. Using Eq.~\ref{2.3.12} it is easy to see from substituting Eq.~\ref{3.1.3} into \ref{3.1.2} that
\beq
\sum_n(\sigma_n-e\mu)C_nf_n(\x)=0\;.\label{3.1.4}
\eeq
Because the $\lbrace f_n(\x)\rbrace$ form a complete set of functions, we must have
\beq
0=(\sigma_n-e\mu)C_n\;,\label{3.1.5}
\eeq
holding for all $n$.

To avoid negative occupation numbers we had to restrict $e\mu<\sigma_0$ where $\sigma_0$ was the smallest eigenvalue for Eq.~\ref{2.3.12}. We will define a critical value of $\mu$ by $\mu_c$ where
\beq
e\mu_c=\sigma_0\;.\label{3.1.6}
\eeq
If $\mu<\mu_c$ then the only solution to Eq.~\ref{3.1.5} is if $C_n=0$ for all values of $n$. This implies that $\bpsi=0$ in Eq.~\ref{3.1.3} vanishes and there is no symmetry breaking.

Suppose that it is possible for $\mu$ to reach the critical value of $\mu_c$ defined in Eq.~\ref{3.1.6}. In this case $C_0$ in Eq.~\ref{3.1.5} is not determined; however, $C_n=0$ for all $n\ne0$ since $\sigma_n>\sigma_0$ in this case. The background field is given by
\beq
\bpsi(\x)=C_0f_0(\x)\;,\label{3.1.7}
\eeq
and is therefore determined by the eigenfunction of lowest eigenvalue; that is, by the ground state. This corresponds to symmetry breaking.

We can make a connection with BEC by looking at the total charge, or particle number. From Eq.~\ref{2.1.11} we have
\beq
Q=Q_0+Q_1\;,\label{3.1.8}
\eeq
where
\beq
Q_0=-\frac{\partial\Omega^{(0)}}{\partial\mu}=e\Int|\bpsi|^2=e|C_0|^2\;
\label{3.1.9}
\eeq
and
\beq
Q_1=-\frac{\partial\Omega_{T\ne0}}{\partial\mu}\;.\label{3.1.10}
\eeq
This has separated off the "classical" from the "quantum" contribution. Because $C_0$ is associated with a non-zero value for $\bpsi$, it is natural to try to link $Q_0$ with the ground state charge.

We will now specialize to $\Sigma=\reals^D$ and set $e=1$ so that we talk about the particle number $N$ rather than the charge. Again we will impose periodic boundary conditions inside a box with the infinite box limit taken. $\sigma_n$ was given in Eq.~\ref{2.2.1} and we see that $\sigma_0=0$ resulting in the critical value of $\mu$ defined in Eq.~\ref{3.1.6} being $\mu_c=0$. The eigenfunction corresponding to the  lowest eigenvalue, normalized to the box volume, is just
\beq
f_0(\x)=V_\Sigma^{-1/2}\;,\label{3.1.11}
\eeq
which is seen to be constant in this case. Thus if symmetry breaking occurs it will result in a constant value for the background field.

We have already determined $\Omega_{T\ne0}$ in Eq.~\ref{2.3.19} or \ref{2.3.20}. Forming $N_1=-\frac{\partial}{\partial\mu}\Omega_{T\ne0}$ we find
\beq
N_1&=&V_\Sigma\left(\frac{m}{2\pi\beta}\right)^{D/2}\sum_{k=1}^{\infty}\frac{e^{k\beta\mu}}{k^{D/2}}\label{3.1.12}\\
&=&V_\Sigma\left(\frac{m}{2\pi\beta}\right)^{D/2}Li_{D/2}\big(e^{\beta\mu}\big)\;.\label{3.1.13}
\eeq
Suppose that we initially set $C_0=0$. Then we must have the total particle number given by $N_1$, since $N_0$ will vanish~:
\beq
N=V_\Sigma\left(\frac{m}{2\pi\beta}\right)^{D/2}Li_{D/2}\big(e^{\beta\mu}\big)\;.\label{3.1.14}
\eeq
This equation determines $\mu$ in terms of the total particle number. If $\mu<0$ then the sum in Eq.~\ref{3.1.12}, which defines the polylogarithm, converges for all values of $D$. However as $\mu\rightarrow0$ the sum only converges for $D>2$. We will assume that $D>2$ initially and examine what happens if $D\le2$ later. For $D>2$ the sum in Eq.~\ref{3.1.12} is bounded above by the value obtained when  $\mu=0$, so that $N_1\sim T^{D/2}$ for large $T$. From Eq.~\ref{3.1.14} this means that for high enough temperatures, no matter how large $N$ is, we can always solve Eq.~\ref{3.1.14} for $\mu$ with $\mu<0$. Thus at high enough temperature there is no symmetry breaking.

Now consider what happens when the temperature is reduced from very large values. Because $N$ in Eq.~\ref{3.1.14} is constant, as $T$ is reduced, $Li_{D/2}(e^{\beta\mu})$ must increase. This means that $\mu$ must increase towards the critical value of $\mu=0$. Eventually $T$ will reach a critical temperature $T_c$ at which $\mu=0$. From Eq.~\ref{3.1.4} we find
\beq
N=V_\Sigma\left(\frac{mkT_c}{2\pi}\right)^{D/2}\zeta_{R}(D/2)
\;,\label{3.1.15}
\eeq
or solving for $T_c$,
\beq
kT_c=\frac{2\pi}{m}\left\lbrack\frac{N}{V_\Sigma\zeta_R(D/2)}\right\rbrack^{2/D}\;.\label{3.1.16}
\eeq
If we take $D=3$ this gives the dependence on the density found from our simple argument at the beginning of this section.

As $T$ is reduced below $T_c$, $\mu$ cannot continue to increase beyond the critical value of $\mu=0$. Thus when $T<T_c$, $\mu$ remains frozen at the value $\mu=0$. It is therefore no longer possible to have $C_0=0$ because we can no longer solve $N=N_1$. From Eqs.~\ref{3.1.15} and \ref{3.1.13} with $\mu=0$ we have
\beq
N_1=N\left(\frac{T}{T_c}\right)^{D/2}\;.\label{3.1.17}
\eeq
Because $N=N_0+N_1$ we find
\beq
N_0=N\left\lbrack1-\left(\frac{T}{T_c}\right)^{D/2}\right\rbrack\;.\label{3.1.18}
\eeq
Using Eq.~\ref{3.1.9}, $N_0=|C_0|^2$, we find
\beq
C_0=N^{1/2}\left\lbrack1-\left(\frac{T}{T_c}\right)^{D/2}\right\rbrack^{1/2}
\;,\label{3.1.19}
\eeq
resulting in
\beq
\bpsi=\left(\frac{N}{V_\Sigma}\right)^{1/2}\left\lbrack1-\left(\frac{T}{T_c}\right)^{D/2}\right\rbrack^{1/2}
\;.\label{3.1.20}
\eeq
We can interpret $N_0$ as the number of particles in the ground state, and $N_1$ as the number of particles in excited states.

For $D\le2$ the situation is different from the one we have just described. If we try to let $\mu\rightarrow0$ in Eq.~\ref{3.1.12} we find that the sum is not bounded. This means that regardless of how large $N$ is, or how large $T$ is, we can always solve Eq.~\ref{3.1.14} for $\mu$ with a value $\mu<0$. Physically, this means that we can accommodate any number of particles in excited states, and that BEC in the sense of symmetry breaking will not occur. It is easy to see this explicitly for $D=2$ because we can perform the sum in Eq.~\ref{3.1.12} to obtain
\beq
N_1=-V_\Sigma\left(\frac{m}{2\pi\beta}\right)\ln(1-e^{\beta\mu})
\;.\label{3.1.21}
\eeq
Setting $N=N_1$ and solving for $\mu$ gives
\beq
\mu=kT\ln\left\lbrack1-e^{-\frac{2\pi N}{V_\Sigma mkT}}\right\rbrack\;.\label{3.1.22}
\eeq
Thus for any finite particle density and any non-zero temperature we can never have $\mu=0$. We always have $\mu<0$. For very small $T$ we may expand the logarithm in Eq.~\ref{3.1.22} to obtain
\beq
\mu\simeq-kT e^{-\frac{2\pi N}{V_\Sigma mkT}}\;.\label{3.1.23}
\eeq
As $T$ becomes smaller, $\mu$ can become arbitrarily close to $\mu=0$ but never actually reaches this critical value. For $D=1$, it is not possible to evaluate $Li_{1/2}$ in closed form, but divergence of the sum shows that BEC is not possible in this case.

BEC for the $D$-dimensional gas has been studied in Refs.~\cite{May,Kac}.

\section{BEC for the Relativistic Ideal Gas}
\label{secBECrel}

We can apply much the same formalism as we did in Sec.~\ref{secBECnon}. The effective thermodynamic potential is
\beq
\Omega_{\rm eff}=\Omega^{(0)}+\Omega_{T=0}+
\Omega_{T\ne0}\;,\label{33.2.1}
\eeq
where
\beq
\Omega^{(0)}=\Int\left\lbrace|{\mathbf D}\bphi|^2+m^2|\bphi|^2-e^2\mu^2|\bphi|^2+U(\x)|\bphi|^2\right\rbrace
\;,\label{33.2.2}
\eeq
is the classical contribution which follows from Eq.~\ref{2.3.31}. $\Omega_{T=0}$ is given by Eq.~\ref{2.3.37} and $\Omega_{T\ne0}$ is given by Eq.~\ref{2.3.38}. Again there is no $\bphi$ dependence in $\Omega_{T=0}$ or in $\Omega_{T\ne0}$ so that the background field equation follows from Eq.~\ref{33.2.2} as
\beq
0=-{\mathbf D}^2\bphi+(m^2-e^2\mu^2)\bphi+U(\x)\bphi\;.\label{33.2.3}
\eeq
If $f_n(\x)$ obeys Eq.~\ref{2.3.33} we may expand the background field $\bphi$ as
\beq
\bphi(\x)=\sum_nC_nf_n(\x)\;.\label{33.2.4}
\eeq
The result may be substituted into Eq.~\ref{33.2.3} and we find that
\beq
(\sigma_n+m^2-e^2\mu^2)C_n=0\;.\label{33.2.5}
\eeq
We will define the critical value of $\mu$ by
\beq
e\mu_c=\sqrt{\sigma_0+m^2}\;.\label{33.2.6}
\eeq
For $\mu<\mu_c$ it is easy to see that $C_n=0$ for all $n$. This results in $\bphi=0$ and there is no symmetry breaking. If $\mu$ can reach the critical value $\mu_c$, then $C_0$ is undetermined and we have
\beq
\bphi(\x)=C_0f_0(\x)\;.\label{33.2.7}
\eeq
The total charge can be decomposed as before in Eq.~\ref{3.1.8} where now
\beq
Q_0&=&2e^2\mu\Int|\bphi|^2=2e^2\mu|C_0|^2\;,\label{33.2.8}\\
Q_1&=&-\frac{\partial}{\partial\mu}\Omega_{T\ne0}\;.\label{33.2.9}
\eeq

We can now specialize to flat space and take the high temperature limit. $\Omega_{T\ne0}$ was given in Sec.~6.2. For $D\ge3$ we may use Eq.~\ref{2.3.48} which shows that
\beq
Q_1\simeq\pi^{-(D+1)/2}\Gamma\Big(\frac{D+1}{2}\Big)\zeta_R(D-1)\beta^{1-D}
V_\Sigma2e^2\mu\;,\label{33.2.10}
\eeq
is the leading term in the high temperature expansion. (In fact Eq.~\ref{2.4.19} shows that this is true even for a curved space, provided that we are justified in dropping the zero temperature term.) From Eq.~\ref{33.2.10} we see that for large $T$, $Q_1\approx T^{D-1}$. This means that for high enough $T$ we can always solve $Q=Q_1$ for $\mu$ with $\mu<\mu_c$. In this case $\bphi=0$ and there is no symmetry breaking. As $T$ decreases $\mu$ must increase if $Q$ is to remain fixed. However $\mu$ can never increase beyond $\mu=\mu_c$. Thus we may define a critical temperature $T_c$ by
\beq
Q\simeq\pi^{-(D+1)/2}\Gamma\Big(\frac{D+1}{2}\Big)\zeta_R(D-1)
(kT_c)^{D-1}V_\Sigma2e^2\mu_c\;,\label{33.2.11}
\eeq
or explicitly,
\beq
kT_c=\left(\frac{Q}{e^2\mu_cV_\Sigma\alpha_D}\right)^{1/(D-1)}
\;,\label{33.2.12}
\eeq
where
\beq
\alpha_D=2\pi^{-(D+1)/2}\Gamma\Big(\frac{D+1}{2}\Big)\zeta_R(D-1)
\;.\label{33.2.13}
\eeq
For $T<T_c$ we have
\beq
Q_1\simeq Q\left(\frac{T}{T_c}\right)^{D-1}\;,\label{33.2.14}
\eeq
from Eq.~\ref{33.2.10} since $\mu=\mu_c$ for this temperature range. Because $Q=Q_0+Q_1$ we have
\beq
Q_0=Q\left\lbrack1-\left(\frac{T}{T_c}\right)^{D-1}\right\rbrack
\;.\label{33.2.15}
\eeq
Using Eq.~\ref{33.2.15} in Eq.~\ref{33.2.8} with $\mu=\mu_c$ shows that
\beq
C_0=\left(\frac{1}{2}\alpha_DV_\Sigma\right)^{1/2}\left\lbrack
(kT_c)^{D-1}-(kT)^{D-1}\right\rbrack\;.\label{33.2.16}
\eeq
For flat space $\sigma_0=0$ and $e\mu_c=m$ from Eq.~\ref{33.2.6}. Also $f_0(\x)=V_{\Sigma}^{-1/2}$, so that
\beq
\bphi=\left(\frac{1}{2}\alpha_DV_\Sigma\right)^{1/2}\left\lbrack
(kT_c)^{D-1}-(kT)^{D-1}\right\rbrack^{1/2}\;,\label{33.2.17}
\eeq
gives the background field and symmetry breaking can occur.

We can analyze the flat space situation in another way which does not involve the high temperature limit, and which provides a suggestion for the general approach we will discuss later. Beginning with
\beq
\Omega_{T\ne0}=\frac{1}{\beta}\sum_n\left\lbrace\ln\left\lbrack
1-e^{-\beta(\omega_n-e\mu)}\right\rbrack+\ln\left\lbrack
1-e^{-\beta(\omega_n+e\mu)}\right\rbrack\right\rbrace\;,\label{33.2.18a}
\eeq
and expanding the logarithms gives
\beq
\Omega_{T\ne0}=-\frac{1}{\beta}\sum_{k=1}^{\infty}\frac{1}{k}\left(
e^{k\beta e\mu}+e^{-k\beta e\mu}\right)\sum_ne^{-k\beta\omega_n}
\;.\label{33.2.18}
\eeq
In the large box limit we established in Eq.~\ref{2.2.11} that
\beq
\sum_n e^{-k\beta\omega_n}=V_\Sigma\int\frac{d^Dn}{(2\pi)^D}\;
e^{-k\beta\sqrt{{\mathbf n}^2+m^2}}\;.\label{33.2.19}
\eeq
Although the integral cannot be done in terms of elementary functions, it can be done in terms of Bessel functions with the result
\beq
\sum_n e^{-k\beta\omega_n}=2\beta kV_\Sigma\left(
\frac{m}{2\pi\beta k}\right)^{(D+1)/2}K_{(D+1)/2}(k\beta m)\;.\label{33.2.20}
\eeq
We therefore obtain
\beq
\Omega_{T\ne0}=-2V_\Sigma\left(\frac{m}{2\pi\beta}\right)^{(D+1)/2}
\sum_{k=1}^{\infty}k^{-(D+1)/2}\left(e^{k\beta e\mu}+
e^{-k\beta e\mu}\right)K_{(D+1)/2}(k\beta m)\;.\label{33.2.21}
\eeq
No approximations have been made to obtain this expression other than the large box limit. Although this expression is not very well suited to examine the high temperature limit (although it can be done \cite{KirstenJPA}), it can be used to see the restriction on the allowed dimensions for BEC.

Using $Q_1=-\frac{\partial}{\partial\mu}\Omega_{T\ne0}$ we find
\beq
Q_1=2e\beta V_\Sigma\left(\frac{m}{2\pi\beta}\right)^{(D+1)/2}
\sum_{k=1}^{\infty}k^{-(D-1)/2}\left(e^{k\beta e\mu}-
e^{-k\beta e\mu}\right)K_{(D+1)/2}(k\beta m)\;.\label{33.2.22}
\eeq
Provided that $Q_1$ remains finite as $\mu\rightarrow\mu_c$ then our analysis of BEC presented above is alright. Convergence of $Q_1$ is reduced to a study of the sum over $k$, which is determined by the large $k$ behaviour of the summand. For large enough $k$ we can use the asymptotic expansion for the Bessel function,
\beq
K_\nu(z)\simeq\sqrt{\frac{\pi}{2z}}\;e^{-z}\left\lbrace1+{\mathcal O}(z^{-1})
\right\rbrace\;.\label{33.2.23}
\eeq
The convergence of the sum is determined by the behaviour
\beq
Q_1(\mu\rightarrow\mu_c)\simeq2e\beta V_\Sigma\left(\frac{m}{2\pi\beta}\right)^{(D+1)/2}\sum_{k=1}^{\infty}k^{-(D-1)/2}e^{k\beta e\mu}\sqrt{\frac{\pi}{2k\beta m}}\;e^{-k\beta m}\;,
\label{33.2.24}
\eeq
where terms which are finite as $\mu\rightarrow\mu_c$ or which are less divergent than the one indicated have been dropped. Thus,
\beq
Q_1(\mu\rightarrow\mu_c)\simeq eV_\Sigma\left(\frac{m}{2\pi\beta}\right)^{D/2}\sum_{k=1}^{\infty}k^{-D/2}
e^{k\beta(e\mu-m)}\;,\label{33.2.25}
\eeq
determines convergence. It is now easy to see that this sum only converges as $e\mu\rightarrow e\mu_c=m$ for $D\ge3$. For $D=1,2$ it does not converge. This means that for $D=1,2$ it is possible to put any amount of the total charge in the excited states and BEC in the sense of symmetry breaking will not occur. The discussion, and in particular the restriction on the spatial dimension, is the same as in the non-relativistic case.

\section{BEC for Nonrelativistic Bosons in a Constant Magnetic Field}
\label{secBECmag}

We will now study the behaviour of relativistic and non-relativistic charged spin-0 fields in a constant magnetic field in a flat space, $\Sigma=\reals^D$, of general dimension. We are interested in how the presence of an externally applied field affects the results found for free fields.

The first point worth making concerns what is meant by a constant magnetic field in a space of general dimension. The magnetic field in $D$ spatial dimensions is described by a constant antisymmetric tensor $F_{ij}$, not in general by a vector. (The electric field components $F_{0i}$ are described by a vector.) In the special case $D=3$, given $F_{ij}$ we can define a vector with components $B^i$ by $F_{ij}=\epsilon_{ijk}B^k$ where $\epsilon_{ijk}$ is the Levi-Civita tensor. ($\epsilon_{ijk}=+1(-1)$ for $ijk$ an even (odd) cyclic permutation of 123; $\epsilon_{ijk}=0$ otherwise.) If $D=3$ we are always free to choose our coordinate axes so that the vector ${\mathbf B}$ lies along one of the axes. There is therefore no loss of generality by assuming only $F_{12}=-F_{21}$ to be non-zero. For general $D$ the situation is not so simple.

One easy way to see that for general $D>3$  it is never possible to find coordinates such that only $F_{12}=-F_{21}$ is nonzero is to consider the invariants which can be formed from $F_{ij}$. For example when $D=4$ there are two independent invariants~:$F_{ij}F^{ij}$ and $\epsilon_{ijkl}F^{ij}F^{kl}$. (See Ref.~\cite{LLfields} for example.) If we keep only one component of the magnetic field non-zero then the second invariant vanishes. It should be clear that for $D=4$ we must keep two independent components of $F_{ij}$ non-zero, say $F_{12}=-F_{21}$ and $F_{34}=-F_{43}$.

For general $D$, rather than consider the invariants directly, it is easier to consider reducing $F_{ij}$ to canonical block diagonal form. Consider first the case of $D$ even. Define $D=2\delta$ for some integer $\delta=1,2,\ldots$. We can always put $F_{ij}$ into the form
\beq
F_{ij}={\rm diag}(\lambda_1J,\ldots,\lambda_\delta J)\;,\label{33.3.1}
\eeq
where
\beq
J=\left(\begin{array}{cc}0&1\\-1&0\end{array}\right)\;,\label{33.3.2}
\eeq
and $\pm i\lambda_i,\ldots,\pm i\lambda_\delta$ are the eigenvalues of $F_{ij}$. There are therefore $\delta$ independent invariants we can construct from $F_{ij}$. We lose no generality in setting all but $\delta$ independent components $F_{12},F_{34},\ldots,F_{2\delta-1\,2\delta}$ to zero. Define
\beq
B_j=F_{2j-1\,2j}\;,\label{33.3.3}
\eeq
for $j=1,2,\ldots,\delta$. Then $B_1,\ldots,B_\delta$ represent the independent components of the magnetic field. We can choose a gauge such that the non-zero components of the vector potential are
\beq
A_{2j-1}=-B_jx^{2j}\;,\label{33.3.4}
\eeq
for $j=1,\ldots,\delta$.

For $D$ odd we may define $D=2\delta+1$ for $\delta=1,2,\ldots$. ($D=1$ does not have any magnetic components for $F_{ij}$.) In this case ${\rm det}\,F_{ij}=0$ since $F_{ij}$ is antisymmetric. Thus there is always at least one zero eigenvalue. The subspace spanned by eigenvectors orthogonal to the one with zero eigenvalue is $2\delta$-dimensional. Therefore as in the even dimensional case there are $\delta$ independent invariants. We may still choose the gauge where Eq.~\ref{33.3.4} holds.

In order to study how the magnetic field affects BEC, we will keep only the first $p$ components in Eq.~\ref{33.3.4} nonzero. We require $p\le\delta$. 
With the gauge choice in Eq.~\ref{33.3.4} we require the eigenfunctions and eigenvalues of
\beq
-\frac{1}{2m}{\mathbf D}^2=-\frac{1}{2m}\sum_{j=1}^{p}
\left\lbrack\left(\partial_{2j-1}+ieB_jx^{2j}\right)^2+
\partial_{2j}^2\right\rbrack-\frac{1}{2m}\sum_{j=2p+1}^{D}\partial_j^2
\;.\label{33.3.5}
\eeq
$D=2\delta$ or $2\delta+1$ depending  upon whether $D$ is even or odd. The advantage of our gauge choice is that the first $p$ terms on the RHS of Eq.~\ref{33.3.5} are equivalent to a sum of $p$ non-interacting simple harmonic oscillator Hamiltonians. If we consider a single term, say $j$, in the sum, then the eigenvalue is $(2n_j+1)eB_j$ where $n_j=0,1,\ldots$ with degeneracy $eB_jL_{2j-1}L_{2j}/(2\pi)$. (See Ref.~\cite{LLQM} for example.) Here we are again imposing box normalization with $L_j$ the length of the box in the $j^{\rm th}$ direction. If we impose periodic boundary conditions in the remaining directions, then the eigenvalues of $-\frac{1}{2m}{\mathbf D}^2$ are
\beq
\sigma_{{\mathbf n},{\mathbf k}}=\sum_{j=1}^{p}(2n_j+1)
\frac{eB_j}{2m}+\frac{1}{2m}\sum_{j=2p+1}^{D}
\left(\frac{2\pi k_j}{L_j}\right)^2\;,\label{33.3.6}
\eeq
where $k_j=0,\pm1,\ldots$, with degeneracy
\beq
\prod_{j=1}^{p}\frac{eB_j}{2\pi}L_{2j-1}L_{2j}\;.\label{33.37}
\eeq

The thermodynamic potential involves
\beq
\Omega_{T\ne0}=\frac{V_\Sigma}{\beta}\left(\prod_{j=1}^{p}\frac{eB_j}{2\pi}
\right)\sum_{n_1=0}^{\infty}\cdots\!\!\!\sum_{n_p=0}^{\infty}\int
\frac{d^{D-2p}k}{(2\pi)^{D-2p}}\,
\ln\left\lbrack1-e^{-\beta(\omega_{{\mathbf n}{\mathbf k}}-e\mu)}
\right\rbrack\;,\label{33.3.8}
\eeq
if we take the large box limit. In Eq.~\ref{33.3.8} we have 
\beq
\omega_{{\mathbf n}{\mathbf k}}=\sum_{j=1}^{p}(2n_j+1)\frac{eB_j}{2m}
+\frac{{\mathbf k}^2}{2m}\;.\label{33.3.9}
\eeq
Expanding the logarithm in Eq.~\ref{33.3.8} as usual, and performing the integration over ${\mathbf k}$ results in
\beq
\Omega_{T\ne0}&=&-\frac{V_\Sigma}{\beta}
\left(\prod_{j=1}^{p}\frac{eB_j}{2\pi}
\right)\sum_{n_1=0}^{\infty}\cdots\!\!\!\sum_{n_p=0}^{\infty}\sum_{l=1}^{\infty}
\int\frac{d^{D-2p}k}{(2\pi)^{D-2p}}\,
e^{-l\beta(\omega_{{\mathbf n}{\mathbf k}}-e\mu)}
\nn
&=&-\frac{V_\Sigma}{\beta}
\left(\prod_{j=1}^{p}\frac{eB_j}{2\pi}
\right)\left(\frac{m}{2\pi\beta}\right)^{(D-2p)/2}\sum_{l=1}^{\infty}
e^{l\beta e\mu}\nn
&&\quad\times\sum_{n_1=0}^{\infty}\cdots\!\!\!\sum_{n_p=0}^{\infty}
e^{-\frac{l\beta}{2m}\sum_{j=1}^{p}(2n_j+1)eB_j}\;.\label{33.3.10a}
\eeq
It is easy to perform the sums over $n_1,\ldots,n_p$ since they just involve geometric series and obtain
\beq
\Omega_{T\ne0}=-\frac{V_\Sigma}{\beta}
\left(\frac{m}{2\pi\beta}\right)^{(D-2p)/2}\sum_{l=1}^{\infty}l^{p-1-D/2}
e^{l\beta e(\mu-\mu_c)}
\prod_{j=1}^{p}\frac{eB_j}{2\pi}\left\lbrack1-e^{-\frac{l\beta eB_j}{m}}\right\rbrack^{-1}\;,\label{33.3.10}
\eeq
where we have used
\beq
e\mu_c=\sigma_{{\mathbf 0}{\mathbf 0}}=
\sum_{j=1}^{p}\frac{eB_j}{2m}\;.\label{33.3.11}
\eeq
Defining $Q_1=-\frac{\partial}{\partial\mu}\Omega_{T\ne0}$ as before, we have
\beq
Q_1=eV_\Sigma
\left(\frac{m}{2\pi\beta}\right)^{(D-2p)/2}\sum_{l=1}^{\infty}l^{p-D/2}
e^{l\beta e(\mu-\mu_c)}
\prod_{j=1}^{p}\frac{eB_j}{2\pi}\left\lbrack1-e^{-\frac{l\beta eB_j}{m}}\right\rbrack^{-1}\;.\label{33.3.12}
\eeq
We can now study whether or not BEC can occur.

From our earlier discussion, in order to have BEC we must have $Q_1$ bounded as $\mu\rightarrow\mu_c$. Since $\beta eB_j/m>0$ is assumed, we may use the inequality
\beq
1<(1-e^{-lx})^{-1}\le(1-e^{-x})^{-1}\;,\label{33.3.13}
\eeq
for $l=1,2,\ldots$ to see that
\beq
Q_1>
eV_\Sigma
\left(\frac{m}{2\pi\beta}\right)^{(D-2p)/2}
\left(\prod_{j=1}^{p}\frac{eB_j}{2 m}\right)
\sum_{l=1}^{\infty}l^{p-D/2}
e^{l\beta e(\mu-\mu_c)}
\;.\label{33.3.14}
\eeq
This shows that $Q_1$ is not bounded as $\mu\rightarrow\mu_c$ for $\frac{D}{2}-p\le1$. When this inequality holds BEC does not occur because any amount of the total charge can be accommodated by setting $Q=Q_1$ and solving for $\mu$ with $\mu<\mu_c$. For $D=2\delta$ this will be the case if $p=\delta$ or $\delta-1$. For $D=2\delta+1$ this will only be the case if $p=\delta$. This shows that for a general magnetic field in a space of any dimension BEC in the sense of symmetry breaking will not occur. For special values of $F_{ij}$ BEC is still possible provided that the inequality $\frac{D}{2}-p\le1$ is violated.

In order to see that BEC is still possible for $\frac{D}{2}-p>1$ we need to prove that $Q_1$ in Eq.~\ref{33.3.12} remains bounded as $\mu\rightarrow\mu_c$. This follows easily by using the inequality in Eq.~\ref{33.3.13} again to show
\beq
Q_1<eV_\Sigma
\left(\frac{m}{2\pi\beta}\right)^{(D-2p)/2}
\left(\prod_{j=1}^{p}\frac{eB_j}{2\pi}\left\lbrack1-e^{-\frac{\beta eB_j}{m}}\right\rbrack^{-1}\right)
\sum_{l=1}^{\infty}l^{p-D/2}
e^{l\beta e(\mu-\mu_c)}
\;.\label{33.3.15}
\eeq
The sum on the RHS is bounded by $\zeta_R(D/2-p)$ if $\frac{D}{2}-p>1$. If $D=2\delta$, BEC can occur for $p=1,2,\ldots,\delta-2$ provided that $\delta\ge3$ ({\em ie} $D\ge6$). If $D=2\delta+1$ BEC can occur for $p=1,2,\ldots,\delta-1$ provided that $\delta\ge2$ ({\em ie} $D\ge5$). If we specialize to the case of a single component field, $p=1$, our results show that in order for BEC to occur we require $D\ge5$. In particular BEC in an externally applied field does not occur for $D=3$. It is easy to show that letting $B_j\rightarrow0$ in Eq.~\ref{33.3.12} reduces the expression to the result found earlier for the free Bose gas. There is therefore a substantial change in the system caused by turning on a constant magnetic field, no matter how small.

Suppose now that the conditions for BEC are met ({\em ie} $\frac{D}{2}-p>1$). The critical temperature $T_c$ is defined by
\beq
Q=eV_\Sigma
\left(\frac{mkT_c}{2\pi}\right)^{(D-2p)/2}\sum_{l=1}^{\infty}l^{p-D/2}
\prod_{j=1}^{p}\frac{eB_j}{2\pi}\left\lbrack1-e^{-\frac{l\beta_c eB_j}{m}}\right\rbrack^{-1}\;.\label{33.3.16}
\eeq
Here $\beta_c=(kT_c)^{-1}$. It is not possible to solve this expression explicitly for $T_c$ as we did for the free Bose gas. However, it is possible to obtain approximate results in the limits of strong and weak fields.

Assume first that $\beta_ceB_j>>m$. As a first rough approximation we can neglect the exponential term in Eq.~\ref{33.3.16} and obtain
\beq
Q\approx eV_\Sigma
\left(\frac{mkT_c}{2\pi}\right)^{(D-2p)/2}\zeta_R(\frac{D}{2}-p)
\prod_{j=1}^{p}\frac{eB_j}{2\pi}\;,\label{33.3.17}
\eeq
which is easily solved for $T_c$. If $T<T_c$ then using the approximation in Eq.~\ref{33.3.17} we have
\beq
Q_1&\approx& Q\left(\frac{T}{T_c}\right)^{D/2-p}\label{33.3.18}\\
Q_0&\approx& Q\left\lbrack 1-\left(\frac{T}{T_c}\right)^{D/2-p}\right\rbrack\;.\label{33.3.19}
\eeq
Comparison of these results with those in Eqs.~\ref{3.1.17} and \ref{3.1.18} obtained for $F_{ij}=0$ shows that the presence of the magnetic field has the same effect as lowering the spatial dimension from $D$ to $D-2p$. If we define $D_{\rm eff}=D-2p$, then the condition for BEC is $D_{\rm eff}\ge3$ for $p=1$, exactly as in the free Bose gas. Physically the fact that a non-zero magnetic field reduces the spatial dimension by two is clear if you keep in mind what happens classically. For $D=3$ the classical motion of a charged particle is a spiral around the direction specified by the magnetic field. The only free motion of the particle is in the direction of the magnetic field; the motion perpendicular to the magnetic field is confined. We will return to this in Sec.~\ref{secgeneral}.

Next assume that $\beta_ceB_j<<m$. A naive expansion of the exponential in Eq.~\ref{33.3.16} will not work because no matter how small $\beta_ceB_j/m$ is $l$ eventually becomes large enough so that the sum over $l$ diverges. In Ref.~\cite{TomsPRD51} we approximated the result by replacing the factor of $\lbrack 1-\exp(-l\beta_ceB_j/m)\rbrack^{-1}$ with its upper limit obtained when $l=1$. This allowed an analytic result for $T_c$ to be obtained. Here we wish to show how a more accurate analysis can be used to obtain a better approximation. (This was suggested to me several years ago by Klaus Kirsten.)

We will begin with the expression for $Q_1$ found from $Q_1=-\frac{\partial}{\partial\mu}\Omega_{T\ne0}$ with Eq.~\ref{33.3.10a} used for $\Omega_{T\ne0}$. We will also define
\beq
x_j=\beta\frac{eB_j}{m}\;.\label{33.3.20}
\eeq
In addition, because we want to let $\mu\rightarrow\mu_c$ where $\mu_c$ was defined in Eq.~\ref{33.3.11}, we will let
\beq
\mu=\mu_c(1-\epsilon)\;.\label{33.3.21}
\eeq
In terms of $x_j$ we have
\beq
\beta e\mu_c=\frac{1}{2}\sum_{j=1}^{p}x_j\;.\label{33.3.22}
\eeq
It is easy to show that
\beq
Q_1&=&eV_\Sigma\left(\frac{m}{2\pi\beta}\right)^{D/2}
\left(\prod_{j=1}^{p}x_j\right)\sum_{n_1=0}^{\infty}\cdots
\sum_{n_p=0}^{\infty}\sum_{l=1}^{\infty}l^{p-D/2}\nn
&&\quad\quad\quad\times\exp{-l\displaystyle{\sum_{j=1}^{p}}(n_j+\frac{\epsilon}{2})x_j}\;,
\label{33.3.23}
\eeq
when written in terms of the dimensionless variables $x_j$ and $\epsilon$. We can now use the Mellin-Barnes representation in Eq.~\ref{MB} for the exponential and perform the sum on $l$ in terms of the Riemann $\zeta$-function. The sums over $n_1,\ldots,n_p$ can be done in terms of a multidimensional generalization of the Hurwitz $\zeta$-function called the Barnes $\zeta$-function \cite{Barnes}. We will first outline very briefly the properties needed for our analysis here.

The Barnes $\zeta$-function is defined by
\beq
\zeta_B(s,a|\x)=\sum_{n_1=0}^{\infty}\cdots
\sum_{n_p=0}^{\infty}(a+{\mathbf n}\cdot\x)^{-s}\;,\label{Barnes1}
\eeq
where $\x$ and ${\mathbf n}$ are $p$-dimensional vectors and $a>0$. When $p=1$ the Barnes $\zeta$-function reduces to the Hurwitz $\zeta$-function. Using the identity in Eq.~\ref{3.14} we can easily perform the sums over ${\mathbf n}$ and obtain the integral representation
\beq
\zeta_B(s,a|\x)=\frac{1}{\Gamma(s)}\int_{0}^{\infty}dt\,t^{s-1}e^{-at}
\prod_{j=1}^{p}\left(1-e^{-tx_j}\right)^{-1}\;.\label{Barnes2}
\eeq
All that we require are the poles of $\zeta_B(s,a|\x)$. These can be obtained in a simple way by following the same procedure as we used in analyzing Eq.~\ref{3.3.9}. The integration range can be split up into two parts; the first extending from $0$ to $t_0$ and the second from $t_0$ to $\infty$. It can be argued that all of the poles come from the integral from $0$ to $t_0$, and as before we only require the behaviour of the integrand for small $t$. It is an easy exercise to show that the first few terms are given by
\beq
\prod_{j=1}^{p}\left(1-e^{-tx_j}\right)^{-1}&=&\frac{1}{t^p
\left(\displaystyle{\prod_{j=1}^{p}x_j}\right)}
\bigg\lbrace1+\frac{t}{2}\sum_{j=1}^{p}x_j\nn
&&\quad+\frac{t^2}{8}\Big\lbrack\Big(
\sum_{j=1}^{p}x_j\Big)^2-\frac{1}{3}\sum_{j=1}^{p}x_j^2\Big\rbrack+
\cdots\bigg\rbrace\;.\label{Barnes3}
\eeq
It is now straightforward to show that $\zeta_B(s,a|\x)$ has simple poles at $s=p,p-1,p-2,\ldots$. The residues at the first few poles are
\beq
{\rm Res}\;\zeta_B(s,a|\x)\Big|_{s=p}&=&\frac{1}{\Gamma(p)
\left(\displaystyle{\prod_{j=1}^{p}x_j}\right)}\;,\label{Barnes4}\\
{\rm Res}\;\zeta_B(s,a|\x)\Big|_{s=p-1}&=&
\frac{\displaystyle{\sum_{j=1}^{p}x_j}-2a}{2\Gamma(p-1)
\left(\displaystyle{\prod_{j=1}^{p}x_j}\right)}
\;,\label{Barnes5}\\
{\rm Res}\;\zeta_B(s,a|\x)\Big|_{s=p-2}&=&
\frac{\displaystyle{\frac{1}{4}}\Big(
\displaystyle{\sum_{j=1}^{p}x_j}\Big)^2-\frac{1}{12}
\displaystyle{\sum_{j=1}^{p}x_j^2}
-a\displaystyle{\sum_{j=1}^{p}x_j}+a^2}{2\Gamma(p-2)
\left(\displaystyle{\prod_{j=1}^{p}x_j}\right)}
\;.\label{Barnes6}
\eeq
Clearly we must have $p>1$ for the result in Eq.~\ref{Barnes5} and $p>2$ for the result in Eq.~\ref{Barnes6}.

From Eq.~\ref{33.3.23} and the definition of the Barnes $\zeta$-function, we find
\beq
Q_1&=&eV_\Sigma\left(\frac{m}{2\pi\beta}\right)^{D/2}
\left(\prod_{j=1}^{p}x_j\right)\cint\,\Gamma(\alpha)\zeta_R(\alpha-p+\frac{D}{2})\nn
&&\quad\quad\quad\times\zeta_B(\alpha,\frac{\epsilon}{2}\sum_{j=1}^{p}x_j|\x)
\;.
\label{33.3.24}
\eeq
Closing the contour in the left hand plane and evaluating the residues using the results just obtained for the Barnes $\zeta$-function we find
\beq
Q_1\simeq eV_\Sigma\left(\frac{m}{2\pi\beta}\right)^{D/2}\left\lbrace
\zeta_R\Big(\frac{D}{2}\Big)+\frac{1}{2}(1-\epsilon)\zeta_R\Big(\frac{D}{2}-1\Big)
\sum_{j=1}^{p}\left(\frac{\beta eB_j}{m}\right)\right\rbrace\;,\label{33.3.25}
\eeq
where we have kept only the first order correction in the magnetic field. (Recall that we are assuming that BEC does occur, which requires $D\ge5$, so that the Riemann $\zeta$ functions are finite.) We can now let $\epsilon\rightarrow0$ and set $Q_1=Q$ which defines the critical temperature $T_c$~:
\beq
Q\simeq eV_\Sigma\left(\frac{mkT_c}{2\pi}\right)^{D/2}\left\lbrace
\zeta_R\Big(\frac{D}{2}\Big)+\frac{1}{2}(1-\epsilon)\zeta_R\Big(\frac{D}{2}-1\Big)
\sum_{j=1}^{p}\left(\frac{eB_j}{mkT_c}\right)\right\rbrace
\;.\label{33.3.26}
\eeq
If we let $B_j\rightarrow0$, it is clear that $T_c$ reduces the result found earlier in the absence of a magnetic field. (See Eq.~\ref{3.1.15} for the particle number for the free field case.) 

In order to see more clearly how the presence of a magnetic field alters the result from that for the free field Bose gas, let $T_0$ be the critical temperature in the absence of a magnetic field. (This was the temperature called $T_c$ in Sec.~\ref{secBECnon}.) It is easily seen from Eq.~\ref{33.3.26} that
\beq
T_c\simeq T_0-\frac{1}{D}\frac{\zeta_R(\frac{D}{2}-1)}{\zeta_R(\frac{D}{2})}
\sum_{j=1}^{p}\left(\frac{eB_j}{km}\right)\;.\label{33.3.27}
\eeq
This result shows that the presence of a weak magnetic field lowers the critical temperature with respect to that found in the absence of a magnetic field. This conclusion was also reached in Ref.~\cite{TomsPRD51} using a less refined method which led to a cruder approximation for $T_c$ in terms of $T_0$.

One final comment we wish to make concerns the nature of the $B_j\rightarrow0$ limit. The steps culminating in Eq.~\ref{33.3.27} assume that $D$ is at least 5. When $D<5$ we have seen that BEC does not occur when a magnetic field is present. This means that the physically interesting case of $D=3$ is qualitatively different when a magnetic field is present (no BEC) and when a magnetic field is absent (BEC).The existence of a magnetic field, no matter how small it might be, destroys the BEC of an otherwise free gas. This was pointed out originally by Schafroth~\cite{Schafroth} who noted that it is not possible to treat the constant magnetic field by perturbing around the zero field result in any finite order of perturbation theory. It would be a very interesting problem to study the behaviour of a system of bosons when a magnetic field is turned on or off. The restriction on the spatial dimension $D\ge5$ for the single component magnetic field was first noted by May~\cite{May65}. The general field was first presented in Ref.~\cite{TomsPRD51,TomsPLB}.

\section{Meissner-Ochsenfeld Effect}
\label{secMO}

It is a well-known feature of a superconductor that when a magnetic field is applied to a sample there may be no field inside the sample if the field is larger than a certain critical value. A beautiful analysis of the Meissner-Ochsenfeld effect and its relationship with BEC was presented by Schafroth \cite{Schafroth}. Our aim here is to show how the approach we have been using in these lectures may be used to study this problem. There has been some confusion in the literature following Schafroth's (correct) analysis. One advantage of adopting the effective action approach is that there is no possible source of error concerning what Schafroth called the acting field and what he called the microscopic field. Another source of error, which seems to have originated with Ref.~\cite{May65} concerns the choice of units. We will adopt Heaviside-Lorentz rationalized units as is conventional in quantum field theory. (A discussion of the various units and their inter-relationship in this problem is contained in Ref.~\cite{KKDJTgen2}.)

The complete thermodynamic potential is (see Sec.~\ref{secII2.3})
\beq
\Omega_{\rm eff}=\Omega_{\rm em}+\Omega^{(0)}+\Omega_{T\ne0}\;,\label{3.4.1}
\eeq
where
\beq
\Omega_{\rm em}=\Int\Big\lbrace\frac{1}{4}F_{ij}F^{ij}-J_{\rm ext}^iA_i\Big\rbrace\;,\label{3.4.2}
\eeq
has been ignored until now because it is independent of $\bpsi$. We will be interested in the case where BEC does not occur, since this is the case for the physically interesting situation of $D=3$. The Maxwell equation, which follows from setting the variation of $\Omega_{\rm eff}$ with respect to $A_i$ to zero is
\beq
\nabla_jF^{ij}=J_{\rm ext}^i+J_{\rm ind}^i\;,\label{3.4.3}
\eeq
where
\beq
\delta\Omega_{T\ne0}=-\Int\,J_{\rm ind}^i\delta A_i\;,\label{3.4.4}
\eeq
defines the current density induced by quantum corrections to the classical theory. (If $\bpsi\ne0$ then there would also be a term from $\Omega^{(0)}$ on the RHS of Eq.~\ref{3.4.3}.)

We have computed $\Omega$ as a function of $F_{ij}$ rather than $A_i$, so it will be best if we can express $J_{\rm ind}^i$ in terms of a derivative with respect to $F_{ij}$ rather than $A_i$. If we vary $\Omega_{T\ne0}$ with respect to $F_{ij}$ we have
\beq
\delta\Omega_{T\ne0}&=&\Int\;\frac{\delta\Omega_{T\ne0}}{\delta F_{ij}}\;\delta
F_{ij}\label{3.4.5}\\
&=&\Int\;\frac{\delta\Omega_{T\ne0}}{\delta F_{ij}}\Big(\nabla_i\delta A_j-
\nabla_j\delta A_i\Big)\nn
&=&2\Int\;\nabla_j\left(\frac{\delta\Omega_{T\ne0}}{\delta F_{ij}}\right)
\delta A_i\;.\label{3.4.6}
\eeq
Comparison of Eq.~\ref{3.4.6} with \ref{3.4.4} shows that we can write
\beq
J_{\rm ind}^i=-2\nabla_j\left(
\frac{\delta\Omega_{T\ne0}}{\delta F_{ij}}\right)\;.\label{3.4.7}
\eeq
The Maxwell equation \ref{3.4.3} becomes
\beq
\nabla_jH^{ij}=J_{\rm ext}^i\;,\label{3.4.8}
\eeq
with
\beq
H^{ij}=F^{ij}+2\frac{\delta\Omega_{T\ne0}}{\delta F_{ij}}\;.\label{3.4.9}
\eeq
This is a generalization of the usual vector electromagnetic result of ${\mathbf H}={\mathbf B}-{\mathbf M}$ in $D=3$ where ${\mathbf M}$ is the magnetization. In fact for $D=3$ it is easy to show that Eq.~\ref{3.4.9} reduces to this standard result \cite{KKDJTgen2}.

The factor of 2 which occurs in Eq.~\ref{3.4.9} is because we have treated all of the components of $F_{ij}$ as independent when performing the variation. However when we actually computed $\Omega_{T\ne0}$ we used $F_{21}=-F_{12}$ and defined $F_{12}=B_1$, and so on for the other independent components. If we define $H_{12}=H_1$ and so forth in an analogous way, then we obtain
\beq
H_i=B_i+\frac{\delta\Omega_{T\ne0}}{\delta B_i}\;,\label{3.4.10}
\eeq
where the factor of 2 no longer appears. (The factor of 2 can also be seen to disappear between Eq.~\ref{3.4.9} and \ref{3.4.10} if we note that $\Omega_{T\ne0}$ only depends on $F_{ij}$ through $F_{ij}F^{ij}$ and use $F_{ij}F^{ij}=2(B_1^2+\cdots B_p^2)$.)

We will define the magnetization $M_i$ by
\beq
M_i=-\frac{\delta\Omega_{T\ne0}}{\delta B_i}\;.\label{3.4.11}
\eeq
From Eq.~\ref{3.4.8} it is seen that $H^{ij}$ or $H_i$ is the field set up by the external current, and therefore should be interpreted as the externally applied field. $F^{ij}$ or $B_i$ represents the field present in the sample which is felt by the particles. These two fields are of course different because the externally applied field causes the charged particles to move, which gives rise to an induced current $J_{\rm ind}^i$ in the sample. This induced current changes the overall magnetic field. For a Meissner-Ochsenfeld effect to occur we need to know if it is possible for $B_i\rightarrow0$ when $H_i$ is changed. If this can happen it corresponds to the particles in the sample feeling a vanishing magnetic field.

We will now concentrate on the case of $D=3$, and evaluate the magnetization $M_i$. From Eq.~\ref{33.3.11} we have
\beq
\mu_c=\frac{B}{2m}\label{3.4.12}
\eeq
as the critical value of the chemical potential. We know from our analysis in Sec.~\ref{secBECmag} that BEC does not occur, and therefore that $\mu$ can never reach the critical value of $\mu_c$. We will define
\beq
\mu=\mu_c(1-\epsilon)\;,\label{3.4.13}
\eeq
for dimensionless parameter $\epsilon$ with $\epsilon>0$. In order to calculate the magnetization we will first obtain the Mellin-Barnes type integral representation for $\Omega_{T\ne0}$, similarly to the representation found for the charge in Eq.~\ref{33.3.24}. From Eq.~\ref{33.3.10a} we have (with $p=1$ and $D=3$)
\beq
\Omega_{T\ne0}=-\frac{V_\Sigma}{\beta}\left(\frac{eB}{2\pi}\right)
\left(\frac{m}{2\pi\beta}\right)^{1/2}\sum_{l=1}^{\infty}l^{-3/2}
e^{l\beta e\mu}\sum_{n=0}^{\infty}e^{-\frac{l\beta}{2m}eB(2n+1)}
\;.\label{3.4.14}
\eeq
This may be written in terms of $\epsilon$, and $x=\beta eB/m$ defined in Eq.~\ref{33.3.20}. We have
\beq
\Omega_{T\ne0}=-\frac{V_\Sigma}{\beta}
\left(\frac{m}{2\pi\beta}\right)^{3/2}x\sum_{l=1}^{\infty}l^{-3/2}
\sum_{n=0}^{\infty}e^{-lx(n+\frac{\epsilon}{2})}
\;.\label{3.4.15}
\eeq

Rather than just apply the Mellin-Barnes integral representation directly to the exponential, it is better to separate off the $n=0$ term for special treatment. The result can be expressed in terms of the polylogarithm, 
\beq
-\;\frac{\Omega_{T\ne0}}{V_\Sigma}&=&\frac{1}{\beta}
\left(\frac{m}{2\pi\beta}\right)^{3/2}x\,{\rm Li}_{3/2}\Big(e^{-\frac{1}{2}\epsilon x}\Big)\nn
&&+\frac{1}{\beta}
\left(\frac{m}{2\pi\beta}\right)^{3/2}x\sum_{l=1}^{\infty}l^{-3/2}
\sum_{n=1}^{\infty}e^{-lx(n+\frac{\epsilon}{2})}
\;,\label{3.4.16}
\eeq
where the first term has arisen from the $n=0$ term in Eq.~\ref{3.4.15}. Now we will use Eq.~\ref{MB} to convert the double sum into a contour integral with the result
\beq
-\;\frac{\Omega_{T\ne0}}{V_\Sigma}&=&\frac{1}{\beta}
\left(\frac{m}{2\pi\beta}\right)^{3/2}x\,{\rm Li}_{3/2}\Big(e^{-\frac{1}{2}\epsilon x}\Big)\nn
&+&\frac{x}{\beta}
\left(\frac{m}{2\pi\beta}\right)^{3/2}\cint\Gamma(\alpha)
\zeta_R\Big(\alpha+\frac{3}{2}\Big)
\zeta_H\Big(\alpha,1+\frac{\epsilon}{2}\Big)x^{-\alpha}\;.
\label{3.4.17}
\eeq
(We take $c>1$ here.) The advantage of separating off the $n=0$ term as we have done is that we ended up with $\zeta_H(\alpha,1+\frac{\epsilon}{2})$ rather than $\zeta_H(\alpha,\frac{\epsilon}{2})$. This has the practical advantage that we can let $\epsilon\rightarrow0$ without worry in all terms arising from the contour integral since they are well-behaved in this limit. More care must be exercised if the result is written in terms of $\zeta_H(\alpha,\frac{\epsilon}{2})$. In addition we can interpret the two terms in Eq.~\ref{3.4.17} physically. The first piece involving ${\rm Li}_{3/2}$ comes from the $n=0$ term and therefore is associated with the ground state. The remaining part comes from $n\ge1$ terms and is therefore the excited states contribution.

We now close the contour in the familiar way to obtain the asymptotic expansion
\beq
-\;\frac{\Omega_{T\ne0}}{V_\Sigma}&=&\frac{eB}{m}
\left(\frac{m}{2\pi\beta}\right)^{3/2}\,{\rm Li}_{3/2}\Big(e^{-\frac{1}{2}\epsilon x}\Big)\nn
&&+\frac{1}{\beta}
\left(\frac{m}{2\pi\beta}\right)^{3/2}\bigg\lbrace\zeta_R(5/2)-
\frac{1}{2}\zeta_R(3/2)(1+\epsilon)x\nn
&&\quad\quad-2\sqrt{\pi}\;x^{3/2}\zeta_H\Big(
-\frac{1}{2},1+\frac{\epsilon}{2}\Big)+\cdots\bigg\rbrace\;.\label{3.4.18}
\eeq
(It is simple to write down more terms in the expansion if they are needed.) The charge density is
\beq
\rho=\frac{\partial}{\partial\mu}\left(-\;\frac{\Omega_{T\ne0}}{V_\Sigma}
\right)&\simeq&ex\left(\frac{m}{2\pi\beta}\right)^{3/2}\,{\rm Li}_{3/2}\Big(e^{-\frac{1}{2}\epsilon x}\Big)
+e\left(\frac{m}{2\pi\beta}\right)^{3/2}\zeta_R(3/2)\nn
&&\quad+
e\left(\frac{m}{2\pi\beta}\right)^{3/2}\sqrt{\pi x}\;
\zeta_H\Big(\frac{1}{2},1+\frac{\epsilon}{2}\Big)\;.\label{3.4.19}
\eeq
(The differentiation is performed with $B$ held fixed.) The magnetization is
\beq
M&=&\frac{\partial}{\partial B}
\left(-\;\frac{\Omega_{T\ne0}}{V_\Sigma}
\right)\nn
&=&\bigg\lbrace\frac{\beta e}{m}\frac{\partial}{\partial x}+\frac{1}{B}(1-\epsilon)\frac{\partial}{\partial\epsilon}\bigg\rbrace
\left(-\;\frac{\Omega_{T\ne0}}{V_\Sigma}
\right)\;,\label{3.4.20}
\eeq
since $\mu$ is held fixed. (It is easy to compute $\frac{\partial\epsilon}{\partial B}$ holding $\mu$ fixed if we use Eqs.~\ref{3.4.12} and \ref{3.4.13}.) It is found that
\beq
M&\simeq&-\;\frac{ex}{2m}\left(\frac{m}{2\pi\beta}\right)^{3/2}
{\rm Li}_{1/2}\left(e^{-\frac{1}{2}\epsilon x}\right)-\frac{e}{m}
\zeta_R(3/2)\left(\frac{m}{2\pi\beta}\right)^{3/2}\label{3.4.21}\\
&&-\frac{e}{m}\left(\frac{m}{2\pi\beta}\right)^{3/2}\sqrt{\pi x}
\left\lbrack3\zeta_H\Big(-\frac{1}{2},1+\frac{\epsilon}{2}\Big)
+\frac{1}{2}\zeta_H\Big(\frac{1}{2},1+\frac{\epsilon}{2}\Big)\right\rbrack
+\cdots\;.\nonumber
\eeq

Now consider what happens as $T$ is reduced from large values towards zero. If BEC could occur then this would correspond to $\epsilon\rightarrow0$ because $\mu\rightarrow\mu_c$ would be reached at some critical temperature. However $\epsilon$ can never reach zero because ${\rm Li}_{1/2}(e^{-\frac{1}{2}\epsilon x})$ which enters the charge density diverges in this limit, and we saw that this is what prohibits BEC as symmetry breaking from occurring. Suppose that we define $\rho_0$ to be the charge density in the absence of a magnetic field. Then from Eq.~\ref{3.1.15} we have
\beq
\rho_0=e\left(\frac{m}{2\pi\beta}\right)^{3/2}\zeta_R(3/2)\;.\label{3.4.22}
\eeq
Furthermore let $T_0$ be the critical temperature in the absence of a magnetic field. Then from Eq.~\ref{3.1.17} we have
\beq
\rho_0=\rho\left(\frac{T}{T_0}\right)^{3/2}\;.\label{3.4.23}
\eeq
We expect that as $T\rightarrow T_c$, $\epsilon$ should start to get very small in the presence of a weak magnetic field. We can use
\beq
{\rm Li}_{1/2}\left(e^{-\frac{1}{2}\epsilon x}\right)\simeq
\sqrt{\frac{2\pi}{\epsilon x}}+\zeta_R(1/2)+{\mathcal O}(\epsilon x)
\;.\label{3.4.24}
\eeq
This is easy to establish using the Mellin-Barnes expression in Eq.~\ref{MB} as follows. Begin with
\beq
{\rm Li}_{1/2}(e^{-\theta})&=&
\sum_{n=1}^{\infty}\frac{e^{-n\theta}}{\sqrt{n}}\nn
&=&\cint\Gamma(\alpha)\zeta_R\Big(\alpha+\frac{1}{2}\Big)
\,\theta^{-\alpha}\nn
&\simeq&\sqrt{\frac{\pi}{\theta}}+\zeta_R(1/2)-\zeta_R(-1/2)\theta+\cdots\;.\nn
\eeq
We can therefore approximate Eq~\ref{3.4.19} by
\beq
\rho\simeq\rho_0+e\left(\frac{m}{2\pi\beta}\right)^{3/2}
\sqrt{\frac{2\pi x}{\epsilon}}\;.\label{3.4.25}
\eeq
Using the same expansion for the magnetization results in
\beq
M\simeq-\;\frac{e}{2m}\left(\frac{m}{2\pi\beta}\right)^{3/2}
\sqrt{\frac{2\pi x}{\epsilon}}\;,\label{3.4.26}
\eeq
for small $\epsilon$. We can eliminate $\epsilon$ between Eq.~\ref{3.4.25} and Eq.~\ref{3.4.26} to obtain
\beq
M&\simeq&-\;\frac{1}{2m}(\rho-\rho_0)\label{3.4.27}\\
&=&-\;\frac{\rho}{2m}\left\lbrack1-\left(\frac{T}{T_c}\right)^{3/2}
\right\rbrack\label{3.4.28}
\eeq
for $T\le T_c$. Finally we may use this result in eqs.~\ref{3.4.10} and \ref{3.4.11} to see that
\beq
H=B+\frac{\rho}{2m}\left\lbrack1-\left(\frac{T}{T_c}\right)^{3/2}
\right\rbrack\;.\label{3.4.29}
\eeq
This shows that once $T$ reaches the temperature $T_c$, which is the critical temperature for BEC in the absence of a magnetic field, it is possible to have $B=0$ for $H\ge H_c$ where
\beq
H_c=\frac{\rho}{2m}\left\lbrack1-\left(\frac{T}{T_c}\right)^{3/2}
\right\rbrack\;.\label{3.4.30}
\eeq
This corresponds to an expulsion of the applied external magnetic field from the sample, which is the Meissner-Ochsenfeld effect.

\section{Relativistic Charged Bosons in the Presence of a Constant Magnetic Field}
\label{secrelmag}

The starting point is the thermodynamic potential $\Omega_{T\ne0}$ expressed by
\beq
\Omega_{T\ne0}=\frac{1}{\beta}\sum_n\ln
\left\lbrack1-e^{-\beta(\omega_n-e\mu)}\right\rbrack
\left\lbrack1-e^{-\beta(\omega_n+e\mu)}\right\rbrack\;.\label{3.X.1}
\eeq
For relativistic charged bosons in a constant magnetic field it is easy to show that, using the same notation as in Sec.~\ref{secBECmag}, 
\beq
\omega_{{\mathbf n}{\mathbf k}}=
\sqrt{\sigma_{{\mathbf n}{\mathbf k}}+m^2}\label{3.X.2}
\eeq
where
\beq
\sigma_{{\mathbf n}{\mathbf k}}=\sum_{j=1}^{p}(2n_j+1)eB_j+\sum_{j=2p+1}^{D}
\left(\frac{2\pi k_j}{L_j}\right)^2\;.\label{3.X.3}
\eeq
If we expand the logarithms in Eq.~\ref{3.X.1} and take the large box limit we have
\beq
\Omega_{T\ne0}&=&-\frac{V_\Sigma}{\beta}\sum_{l=1}^{\infty}\frac{1}{l}\left(
e^{l\beta e\mu}+e^{-l\beta e\mu}\right)\left(\prod_{j=1}^{p}
\frac{eB_j}{2\pi}\right)\sum_{n_1=0}^{\infty}\cdots\sum_{n_p=0}^{\infty}\nn
&&\times\int
\frac{d^{D-2p}k}{(2\pi)^{D-2p}}\exp\left(-l\beta\left\lbrack{\mathbf k}^2+m^2
+\sum_{j=1}^{p}(2n_j+1)eB_j\right\rbrack^{1/2}\right)\;.\label{3.X.4}
\eeq
(Compare with eq.~\ref{33.2.18} for the free Bose gas.) We can now use Eqs.~\ref{33.2.19} and \ref{33.2.20} to evaluate the integral over ${\mathbf k}$~:
\beq
\frac{\Omega_{T\ne0}}{V_\Sigma}&=&-\left(\prod_{j=1}^{p}
\frac{eB_j}{2\pi}\right)\sum_{n_1=0}^{\infty}\cdots\sum_{n_p=0}^{\infty}
\sum_{l=1}^{\infty}\left(
e^{l\beta e\mu}+e^{-l\beta e\mu}\right)(2\pi\beta l)^{-(D-2p+1)/2}\nn
&&\quad\times\left\lbrack m^2+\sum_{j=1}^{p}(2n_j+1)eB_j\right\rbrack^{(D-2p+1)/4}\nn
&&\quad\quad\quad\times
K_{(D-2p+1)/2}\left(l\beta\left\lbrack m^2+\sum_{j=1}^{p}(2n_j+1)eB_j\right\rbrack^{1/2}\right)
\;.\label{3.X.5}
\eeq
We can compute $Q_1=-\frac{\partial}{\partial\mu}\Omega_{T\ne0}$ to be
\beq
Q_1&=&V_\Sigma e(2\pi)^{(D-2p+1)/2}\beta^{-(D-2p-1)/2}
\left(\prod_{j=1}^{p}\frac{eB_j}{2\pi}\right)
\sum_{n_1=0}^{\infty}\cdots\sum_{n_p=0}^{\infty}
\sum_{l=1}^{\infty} l^{-(D-2p-1)/2}\nn
&&\times\left(
e^{l\beta e\mu}-e^{-l\beta e\mu}\right) l^{-(D-2p-1)/2}
\left\lbrack m^2+\sum_{j=1}^{p}(2n_j+1)eB_j\right\rbrack^{(D-2p+1)/4}\nn
&&\quad\quad\times K_{(D-2p+1)/2}\left(l\beta\left\lbrack m^2+\sum_{j=1}^{p}(2n_j+1)eB_j\right\rbrack^{1/2}\right)
\;.\label{3.X.6}
\eeq

To examine what happens as $\mu\rightarrow\mu_c$ we can use the same analysis as we used at the end of Sec.~\ref{secBECrel}. All we need to know is the behaviour of the sums which occur in Eq.~\ref{3.X.6}. For large $l$ we can use Eq.~\ref{33.2.23} and note that for all terms with $n_j\ne0$ the sum will converge exponentially fast. This means that the behaviour of $Q_1$ as $\mu\rightarrow\mu_c$ is determined by the $n_1=\cdots=n_p=0$ term in the sums. Thus
\beq
Q_1(\mu\rightarrow\mu_c)\simeq2eV_\Sigma(2\pi\beta)^{p-D/2}\Big\lbrack 
m^2+\sum_{j=1}^{p}eB_j\Big\rbrack^{(D-2p)/4}\sum_{l=1}^{\infty}
l^{p-D/2}e^{-l\beta e(\mu_c-\mu)}\;.\label{3.X.7}
\eeq
This expression only converges in the limit $\mu=\mu_c$ for $D/2-p>1$. This is identical to the result found for the nonrelativistic Bose gas in Sec.~\ref{secBECmag}. In particular, we can define an effective spatial dimension by $D_{\rm eff}=D-2p$. The requirement that BEC occurs requires $D_{\rm eff}\ge3$. For $p=0$ this reduces to the usual one for the free Bose gas. An analysis of the Meissner-Ochsenfeld effect can be given~\cite{Daicicetal2,Elmforsetal,Daicicetal3,Elmforsetal2,KKDJTgen2}.

\section{General Criterion for BEC Interpreted as Symmetry Breaking}
\label{secgeneral}

We have seen in Sec.~\ref{secBECmag} and Sec.~\ref{secrelmag} how the presence of a constant magnetic field with $p$ independent components alters the criterion for BEC from that found for the free Bose gas. If we define an effective dimension $D_{\rm eff}=D-2p$ then BEC occurs only for $D_{\rm eff}\ge3$. In this section we will see in a fairly simple way the general features present which alter the effective spatial dimension from $D$ to $D_{\rm eff}$. This analysis was first presented in Refs.~\cite{KKDJTgen2,KKDJTgen1}.

\subsection{\it Non-relativistic Gas}

Suppose that we consider any system for which the energy levels can be expressed as the sum of a discrete part, which we will denote by $E_{\mathbf p}^d$, and a continuous part which can be dealt with by box normalization. (A more general possibility for the continuous part can be considered by using a method we will describe later.) We will take the infinite box limit at the end. Assume that the box has dimension $q$ and write
\beq
E_n=E_{\mathbf p}^d+\frac{1}{2m}\sum_{i=1}^{q}\left(
\frac{2\pi n_i}{L_i}\right)^2\;.\label{3.5.1}
\eeq
${\mathbf p}$ just denotes the set of labels for the discrete part of the energy spectrum. The thermodynamic potential is
\beq
\Omega_{T\ne0}&=&\frac{1}{\beta}\sum_n\ln\left\lbrack1
-e^{-\beta(E_n-e\mu)}\right\rbrack\nn
&=&-\frac{V_q}{\beta}\sum_{\mathbf p}\int\frac{d^qn}{(2\pi)^q}
\sum_{k=1}^{\infty}\frac{1}{k}e^{-k\beta(E_{\mathbf p}^d+
\frac{{\mathbf n}^2}{2m}-e\mu)}\nn
&=&-\frac{V_q}{\beta}\left(\frac{m}{2\pi\beta}\right)^{q/2}\sum_{\mathbf p}
\sum_{k=1}^{\infty}k^{-1-q/2}e^{-k\beta(E_{\mathbf p}^d-e\mu)}
\;,\label{3.5.2}
\eeq
after expanding the logarithm and performing the integration over ${\mathbf n}$. The critical value of $\mu$ is given by
\beq
e\mu_c=E_{\mathbf 0}^d\;,\label{3.5.3}
\eeq
where $E_{\mathbf 0}^d$ is the smallest energy eigenvalue in the discrete part of the spectrum. Because the behaviour as $\mu\rightarrow\mu_c$ is of interest we can separate off the lowest mode contribution in Eq.~\ref{3.5.2} and write
\beq
\Omega_{T\ne0}=\Omega_{T\ne0}^{\rm gr}+\Omega_{T\ne0}^{\rm ex}
\;,\label{3.5.4}
\eeq
with $\Omega_{T\ne0}^{\rm gr}$ the contribution coming from ${\mathbf p}={\mathbf 0}$, and $\Omega_{T\ne0}^{\rm ex}$ the contribution coming from ${\mathbf p}\ne{\mathbf 0}$. Physically $\Omega_{T\ne0}^{\rm gr}$ contains the ground state contribution and $\Omega_{T\ne0}^{\rm ex}$ contains the contributions from excited states. If $d_0$ is the ground state degeneracy we have
\beq
\Omega_{T\ne0}^{\rm gr}=
-\frac{V_q}{\beta}d_0\left(\frac{m}{2\pi\beta}\right)^{q/2}
{\rm Li}_{1+q/2}\Big(e^{-\beta e(\mu_c-\mu)}\Big)\;.\label{3.5.5}
\eeq
We also have 
\beq
\Omega_{T\ne0}^{\rm ex}=
-\frac{V_q}{\beta}\left(\frac{m}{2\pi\beta}\right)^{q/2}
\sum_{k=1}^{\infty}\sum_{{\mathbf p}\ne{\mathbf 0}}k^{-1-q/2}
e^{-k\beta (E_{\mathbf p}^d-e\mu)}\;,\label{3.5.6}
\eeq
where the sum over $\mathbf p$ is restricted to excited states only. The advantage of separating off the lowest mode as we have done is that even if we let $\mu\rightarrow\mu_c$ the argument of the exponential in Eq.~\ref{3.5.6} remains negative ensuring convergence of the sum over $k$. This can be seen by noting that for large $k$ we have
\begin{displaymath}
\sum_{{\mathbf p}\ne{\mathbf 0}}e^{-k\beta (E_{\mathbf p}^d-e\mu)}\approx
e^{-k\beta(E_{\mathbf 1}^d-e\mu)}\;,
\end{displaymath}
where $E_{\mathbf 1}^d$ represents the first excited energy level. It follows that for large $k$ the summand behaves like
\begin{displaymath}
k^{-1-q/2}e^{-k\beta(E_{\mathbf 1}^d-e\mu)}\;.
\end{displaymath}
As $e\mu\rightarrow e\mu_c=E_{\mathbf 0}^d$ the sum over $k$ is exponentially damped because $E_{\mathbf 1}^d-E_{\mathbf 0}^d>0$ by assumption. This exponential damping persists even for $\mu=\mu_c$, and is also true even if we differentiate with respect to $\mu$ before taking the $\mu\rightarrow\mu_c$ limit to get the charge. Thus, whether or not a phase transition representing BEC occurs is determined by $\Omega_{T\ne0}^{\rm gr}$.

We can write
\beq
Q_1=Q_1^{\rm gr}+Q_1^{\rm ex}\;.\label{3.5.7}
\eeq
With $Q_1^{\rm gr}=-\frac{\partial}{\partial\mu}\Omega_{T\ne0}^{\rm gr}$ we have
\beq
Q_1^{\rm gr}=e
V_qd_0\left(\frac{m}{2\pi\beta}\right)^{q/2}
{\rm Li}_{q/2}\Big(e^{-\beta e(\mu_c-\mu)}\Big)\;.\label{3.5.8}
\eeq
($Q_1^{\rm ex}$ can be defined in terms of $\Omega_{T\ne0}^{\rm ex}$ in an obvious manner.) We have argued that the behaviour of $Q_1$ as $\mu\rightarrow\mu_c$ is determined by $Q_1^{\rm gr}$, so
\beq
Q_1(\mu\rightarrow\mu_c)\simeq e
V_qd_0\left(\frac{m}{2\pi\beta}\right)^{q/2}
{\rm Li}_{q/2}\Big(e^{-\beta e(\mu_c-\mu)}\Big)\;.\label{3.5.9}
\eeq
${\rm Li}_{q/2}\Big(e^{-\beta e(\mu_c-\mu)}\Big)$ is only finite as $\mu\rightarrow\mu_c$ if $q>2$. For $q=0,1,2$ $Q_1$ will diverge as $\mu\rightarrow\mu_c$ meaning that BEC cannot occur as it does for the free Bose gas. The actual divergence as $\mu\rightarrow\mu_c$ is easily obtained from the behaviour of the polylogarithm \cite{KKDJTgen2}.

For a magnetic field with $p$ independent components in $D$ spatial dimensions we had $q=D-2p$ as the dimension associated with the continuous part of the spectrum. (See Eq.~\ref{33.3.6} in the large box limit.) The restriction $q\ge3$ found here is therefore fully consistent with our earlier result in Sec.~\ref{secBECmag} for the special case of a constant magnetic field.

\subsection{\it Relativistic Gas}

In this case we have
\beq
\Omega_{T\ne0}=\frac{1}{\beta}\sum_n\left\lbrace
\ln\left\lbrack1-e^{-\beta E_n^+}\right\rbrack+
\ln\left\lbrack1-e^{-\beta E_n^-}\right\rbrack\right\rbrace\;,
\label{3.5.10}
\eeq
where
\beq
E_n^{\pm}=\sqrt{\sigma_n+m^2}\;\pm e\mu\;.\label{3.5.11}
\eeq
Assume, as in Eq.~\ref{3.5.1}, that $\sigma_n$ splits into the sum of a discrete part $\sigma_{{\mathbf p}}^d$ and a continuous part which we deal with by box normalization~:
\beq
\sigma_n=\sigma_{\mathbf p}^d+\sum_{i=1}^{q}\left(
\frac{2\pi n_i}{L_i}\right)^2\;.\label{3.5.12}
\eeq
If we expand the logarithms in Eq.~\ref{3.5.10} and take the infinite box limit we obtain
\beq
\Omega_{T\ne0}=-\frac{V_q}{\beta}\sum_{\mathbf p}\sum_{k=1}^{\infty}
\frac{1}{k}\Big(e^{k\beta e\mu}+e^{-k\beta e\mu}\Big)\int\frac{d^qn}{(2\pi)^q}e^{-k\beta\sqrt{\sigma_{\mathbf 0}^d+
{\mathbf n}^2+m^2}}\;.\label{3.5.13}
\eeq
We can do the integration over $\mathbf n$ in terms of a Bessel function as in Sec.~\ref{secBECrel} with the result
\beq
\Omega_{T\ne0}&=&-\frac{V_q}{\pi\beta}(2\pi\beta)^{(1-q)/2}
\sum_{k=1}^{\infty}
k^{-(1+q)/2}\sum_{\mathbf p}(\sigma_{{\mathbf p}}^{d}+m^2)^{(q+1)/4}\nn
&&\times K_{(q+1)/2}\Big(k\beta\sqrt{\sigma_{\mathbf p}^d+m^2}\Big)\Big(
e^{-k\beta e\mu}+e^{k\beta e\mu}\Big)\;.\label{3.5.14}
\eeq
As in the nonrelativistic case we will separate off the ${\mathbf p}={\mathbf 0}$ term and define
\beq
\Omega_{T\ne0}^{\rm gr}&=&-\frac{V_qd_0}{\pi\beta}(2\pi\beta)^{(1-q)/2}
\sum_{k=1}^{\infty}
k^{-(1+q)/2}(\sigma_{{\mathbf 0}}^{d}+m^2)^{(q+1)/4}\nn
&&\times K_{(q+1)/2}\Big(k\beta\sqrt{\sigma_{\mathbf 0}^d+m^2}\Big)\Big(
e^{-k\beta e\mu}+e^{k\beta e\mu}\Big)\;.\label{3.5.15}
\eeq
$\Omega_{T\ne0}^{\rm ex}$ is defined as in Eq.~\ref{3.5.14}, but with the sum restricted to ${\mathbf p}\ne{\mathbf 0}$. Due to the behaviour of the Bessel function for large values of its argument, $\Omega_{T\ne0}^{\rm ex}$ remains finite (as does its derivative with respect to $\mu$) even for $\mu=\mu_c$. The convergence or divergence of the charge $Q_1$ is determined by the behaviour of $\Omega_{T\ne0}^{\rm gr}$.

For large $k$ we have
\beq
K_{(q+1)/2}\Big(k\beta\sqrt{\sigma_{\mathbf 0}^d+m^2}\Big)\simeq
\left(\frac{\pi}{2k\beta}\right)^{1/2}(\sigma_{\mathbf 0}^d+m^2)^{-1/4}
e^{-k\beta\sqrt{\sigma_{\mathbf 0}^d+m^2}}\;.\label{3.5.16}
\eeq
Noting that $e\mu_c=\sqrt{\sigma_{\mathbf 0}^d+m^2}$ we have the large $k$ behaviour of the thermodynamic potential determined by
\beq
\Omega_{T\ne0}^{\rm gr}(\mu\rightarrow\mu_c)&\simeq&-\frac{V_qd_0}{\beta}
\left(\frac{e\mu_c}{2\pi\beta}\right)^{q/2}\sum_{k=1}^{\infty}k^{-1-q/2}
e^{-k\beta e(\mu_c-\mu)}\label{3.5.17}\\
&=&-\frac{V_qd_0}{\beta}
\left(\frac{e\mu_c}{2\pi\beta}\right)^{q/2}{\rm Li}_{1+q/2}
\Big(e^{-\beta e(\mu_c-\mu)}\Big)\;.\label{3.5.18}
\eeq
Just as in the nonrelativistic case, it is easily seen that $Q_1$ remains finite as $\mu\rightarrow\mu_c$ only for $q>2$. 

\section{BEC for the Self-interacting Theory}
\label{BECself}

We will assume that $D=3$ here, and that BEC does occur. (This requires a continuous energy spectrum as described in the last section.) We will also assume that $\bphi$ is constant. All of these conditions are met for $\Sigma=\reals^3$. For the self-interaction we will take
\beq
V_{\rm tree}(|\bphi|^2)=\frac{\lambda}{6}|\bphi|^4\;,\label{3.6.1}
\eeq
to be the usual $\lambda\varphi^4$ theory. Using the results of Sec.~\ref{sec3.3} with the zero-point energy term dropped as irrelevant at high temperature, we have the total effective thermodynamic potential given by
\beq
\frac{\Omega_{\rm eff}}{V_\Sigma}&\simeq&(m^2+\xi R-e^2\mu^2)|\bphi|^2+\frac{\lambda}{6}|\bphi|^4-
\frac{\pi^2}{45}\beta^{-4}\nn
&&+\frac{1}{12}\beta^{-2}\lbrack m^2+(\xi-\frac{1}{6})R-2e^2\mu^2+\frac{2\lambda}{3}|\bphi|^2\rbrack+
\cdots\;.\label{3.6.2}
\eeq
The background field $\bphi$ must satisfy $\frac{\partial}{\partial\bphi}\Omega_{\rm eff}=0$, giving
\beq
0\simeq(m^2+\xi R-e^2\mu^2)\bphi+\frac{\lambda}{3}|\bphi|^2\bphi+
\frac{\lambda}{18}\beta^{-2}\bphi\;.\label{3.6.3}
\eeq
There are two possible solutions to this~: $\bphi=0$, corresponding to the symmetric phase, and 
\beq
|\bphi|^2=\frac{3}{\lambda}(e^2\mu^2-m^2-\xi R-
\frac{\lambda}{18}\beta^{-2})\;,\label{3.6.4}
\eeq
which corresponds to the broken symmetry. The total charge density is
\beq
\rho&=&-\frac{1}{V_\Sigma}\frac{\partial}{\partial\mu}\Omega_{\rm eff}\nn
&\simeq&2e^2\mu|\bphi|^2+\frac{1}{3}e^2\mu\beta^{-2}\;\label{3.6.5}
\eeq
If we define
\beq
\rho_0&=&2e^2\mu|\bphi|^2\;,\label{3.6.6}\\
\rho_1&=&\frac{1}{3}e^2\mu\beta^{-2}\;,\label{3.6.7}
\eeq
we can interpret $\rho_0$ as the charge density of the Bose condensate. In the symmetric phase we have $\bphi=0$ so that
\beq
\rho\simeq\frac{1}{3}e^2\mu\beta^{-2}\;.\label{3.6.8}
\eeq
As $T$ decreases from large values, for fixed total charge $\mu$ will increase until we reach the critical value set by
\beq
e^2\mu_c^2=m^2+\xi R+\frac{\lambda}{18}\beta_c^{-2}\;,\label{3.6.9}
\eeq
at the critical temperature $T_c$. This is the temperature at which the symmetry is broken. For $T<T_c$ we have (see Eq.~\ref{3.6.4})
\beq
|\bphi|^2=\frac{1}{6}(\beta_c^{-2}-\beta^{-2})\;,\label{3.6.10}
\eeq
and $\mu=\mu_c$. Setting $T=T_c$ in Eq.~\ref{3.6.8} we find
\beq
e\mu_c=\frac{3\rho}{e}\beta_c^2\;.\label{3.6.11}
\eeq
Using this in the left hand side of Eq.~\ref{3.6.9} gives us an equation which completely determines $T_c$ in terms of the charge density. From Eq.~\ref{3.6.7} we have
\beq
\rho_1\simeq\rho\left(\frac{\beta_c}{\beta}\right)^2
=\rho\left(\frac{T}{T_c}\right)^2\;,\label{3.6.12}
\eeq
and hence
\beq
\rho_0\simeq\rho\left\lbrack1-\left(\frac{T}{T_c}\right)^2\right\rbrack
\;.\label{3.6.13}
\eeq
This is exactly the result found for the free Bose gas (although of course the result for $T_c$ is different).

It is possible to study the nonrelativistic limit of the theory. This has been done in a very beautiful paper \cite{Benson}. It is also possible to generalize to a non-abelian gauge symmetry \cite{KKDJTint}.

\section{BEC in an Isotropic Harmonic Oscillator Potential}
\label{harm}

We will now consider the case of non-relativistic bosons in a harmonic oscillator potential. This simple theoretical model provides a good description of the magnetic traps used in the recent experiments on BEC \cite{rub,lith,sod1,sod2}. Interactions among the atoms will be neglected. As a first approximation this would be expected to be not too bad because the atomic gases are dilute. The grand canonical ensemble will be used.

We will only look at $D=3$, although the analysis may be generalized to other spatial dimensions. We will take
\beq
U(\x)=\frac{1}{2}m(\omega_1^2x^2+\omega_2^2y^2+\omega_3^2z^2)
\;,\label{3.7.1}
\eeq
where $\omega_1,\omega_2$ and $\omega_3$ are the frequencies characterizing the anisotropic harmonic oscillator potential. The energy levels for this system are simply
\beq
E_{\mathbf n}=(n_1+\frac{1}{2})\hbar\omega_1+
(n_2+\frac{1}{2})\hbar\omega_2+(n_3+\frac{1}{2})\hbar\omega_3
\;,\label{3.7.2}
\eeq
where ${\mathbf n}=(n_1,n_2,n_3)$ with $n_i=0,1,\ldots$. (We have restored $\hbar$ here.) 

The spectrum is entirely discrete so that $q=0$ in the terminology of Sec.~\ref{secgeneral}. This means that BEC will not occur in the sense we have discussed so far; that is in the sense of symmetry breaking and as a phase transition. Nevertheless, as we will see, there is a temperature range which signals a point at which there is a sudden and dramatic increase in the number of particles in the ground state. If this criterion is adopted for BEC in place of the usual phase transition, then BEC will occur. The fact that a real physical system can exhibit such a behaviour is borne out by the recent experiments on cold alkali gases \cite{rub}--\cite{sod2}.

Because the atoms are neutral, we will use the particle number rather than charge as we have done earlier. The critical value of $\mu$ is 
\beq
\mu_c=E_{\mathbf 0}=\frac{\hbar}{2}(\omega_1+\omega_2+\omega_3)
\;.\label{3.7.3}
\eeq
There is an obvious similarity between this problem and BEC in a constant magnetic field, so that the methods used earlier should help us here. In order to present the analysis in as clear a way as possible we will first look at the case of the isotropic harmonic oscillator, returning to the anisotropic case later.

Setting $\omega_1=\omega_2=\omega_3=\omega$ in Eqs.~\ref{3.7.1}--\ref{3.7.3} gives
\beq
\mu_c=\frac{3}{2}\hbar\omega\;.\label{3.7.4}
\eeq
The $q$-potential defined in Sec.~\ref{secII2.1} is
\beq
q=-\sum_{n_1=0}^{\infty}\sum_{n_2=0}^{\infty}\sum_{n_3=0}^{\infty}
\ln\left\lbrack1-ze^{-\beta E_{\mathbf n}}\right\rbrack\;,\label{3.7.5}
\eeq
with $z=e^{\beta\mu}$ the fugacity. Because we are assuming an isotropic potential we may write the triple sum in Eq.~\ref{3.7.5} as a single one. It is advantageous to introduce dimensionless variables $x$ and $\epsilon$ defined by
\beq
x&=&\beta\hbar\omega=\frac{\hbar\omega}{kT}
\:,\label{3.7.8}\\
\mu&=&\hbar\omega\Big(\frac{3}{2}-\epsilon\Big)\;.\label{3.7.9}
\eeq
It is straightforward to show that
\beq
q=-\sum_{k=0}^{\infty}\frac{1}{2}(k+1)(k+2)\ln\left\lbrack
1-e^{-(k+\epsilon)x}\right\rbrack\;.\label{3.7.10}
\eeq
There are several possible ways to proceed here. We will deal with the most direct way first.

As we have done before, expand the logarithm in Eq.~\ref{3.7.10} in terms of its Taylor series to give
\beq
q=\sum_{k=0}^{\infty}\frac{1}{2}(k+1)(k+2)\sum_{n=1}^{\infty}
\frac{e^{-n(k+\epsilon)x}}{n}\;.\label{3.7.11}
\eeq
The sum over $k$ just involves a geometric series, and can be done with the result
\beq
q=\sum_{n=1}^{\infty}\frac{e^{-n\epsilon x}}{n}\left(1-e^{-nx}\right)^{-3}
\;.\label{3.7.13}
\eeq
The particle number is $N=\beta^{-1}\left(\frac{\partial q}{\partial\mu}\right)_{T,\omega}$. With $T$ and $\omega$ fixed, we have from Eq.~\ref{3.7.9} that $\delta\mu=-\hbar\omega\delta\epsilon$. Thus
\beq
N=-\frac{1}{x}\left(\frac{\partial q}{\partial\epsilon}\right)_x=
\sum_{n=1}^{\infty}e^{-n\epsilon x}\left(1-e^{-nx}\right)^{-3}
\;.\label{3.7.16}
\eeq

\begin{figure}[htb]
\begin{center}
\leavevmode
\epsffile{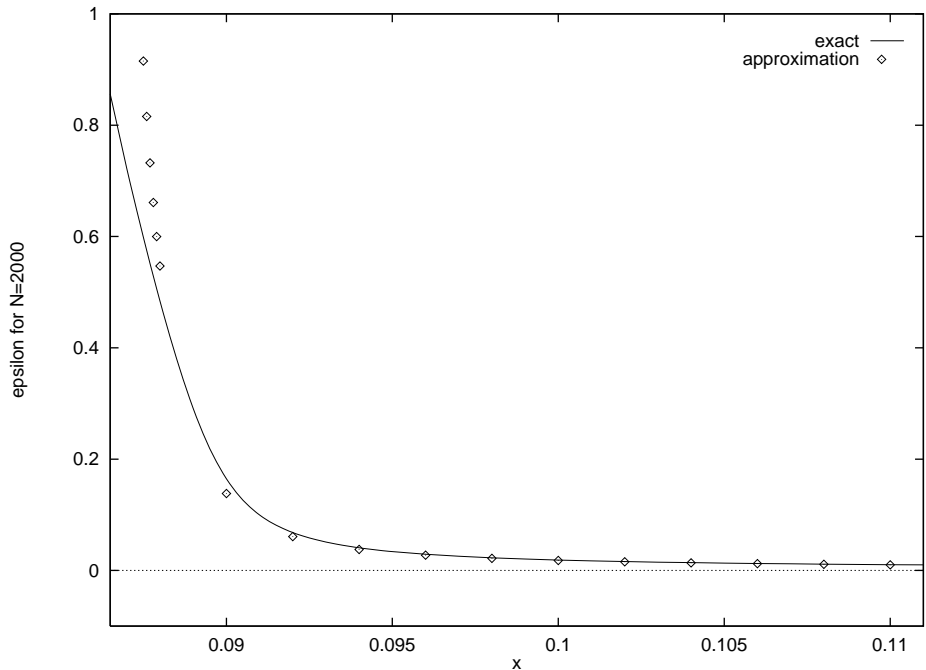}
\end{center}
\caption{\footnotesize This shows $\epsilon$ as a function of $x=\hbar\omega/(kT)$. The solid curve shows the result found from a numerical evaluation of the exact harmonic oscillator sums for the isotropic case for $N=2000$. The diamonds show the result of using our analytic approximation.}\label{fig1}
\end{figure}
An equivalent way to obtain Eq.~\ref{3.7.16} is to begin with
\beq
N=\sum_{n_1=0}^{\infty}\sum_{n_2=0}^{\infty}\sum_{n_3=0}^{\infty}
\left\lbrack e^{\beta(E_{\mathbf n}-\mu)}-1\right\rbrack^{-1}
\;,\label{3.7.19}
\eeq
and using Eq.~\ref{3.7.2} (with all frequencies the same) follow through the same steps as led to Eq.~\ref{3.7.13}. The number of particles in the ground state is
\beq
N_{\rm gr}=(e^{\epsilon x}-1)^{-1}\;.\label{3.7.20}
\eeq
Solving for $\epsilon$ results in 
\beq
\epsilon=\frac{1}{x}\ln\left(1+\frac{1}{N_{\rm gr}}\right)\;.\label{3.7.21}
\eeq
Since $N_{\rm gr}<N$ we have
\beq
\epsilon>\frac{1}{x}\ln\left(1+\frac{1}{N}\right)\;.\label{3.7.22}
\eeq
This shows that $\epsilon$ can only reach the critical value $\epsilon=0$ if $x\rightarrow\infty$ (corresponding to $T\rightarrow0$), or if $N\rightarrow\infty$. For a finite particle number, as is the case in any experiment, $\epsilon$ is always positive meaning that BEC as symmetry breaking will never occur. This is in agreement with the more general treatment in Sec.~\ref{secgeneral}.

Even though $\epsilon$ can never reach the value $\epsilon=0$, if $\epsilon$ gets very small then there can be a dramatic increase in the the number of particles in the ground state. In the actual experiments, $\omega/(2\pi)\sim100$ Hz, and $T\sim\ \mu$K. In this case $x$ defined in Eq.~\ref{3.7.8} is small. The particle numbers of interest to the experiments are $N\sim10^3-10^6$. For a specific example we will pick $N=2000$ as in the rubidium experiment \cite{rub} and $\omega/(2\pi)=60$ Hz. It is possible to solve Eq.~\ref{3.7.16} numerically for $\epsilon$ as a function of $x$ for a given particle number. The result is shown in fig.~1.
\begin{figure}[htb]
\begin{center}
\leavevmode
\epsffile{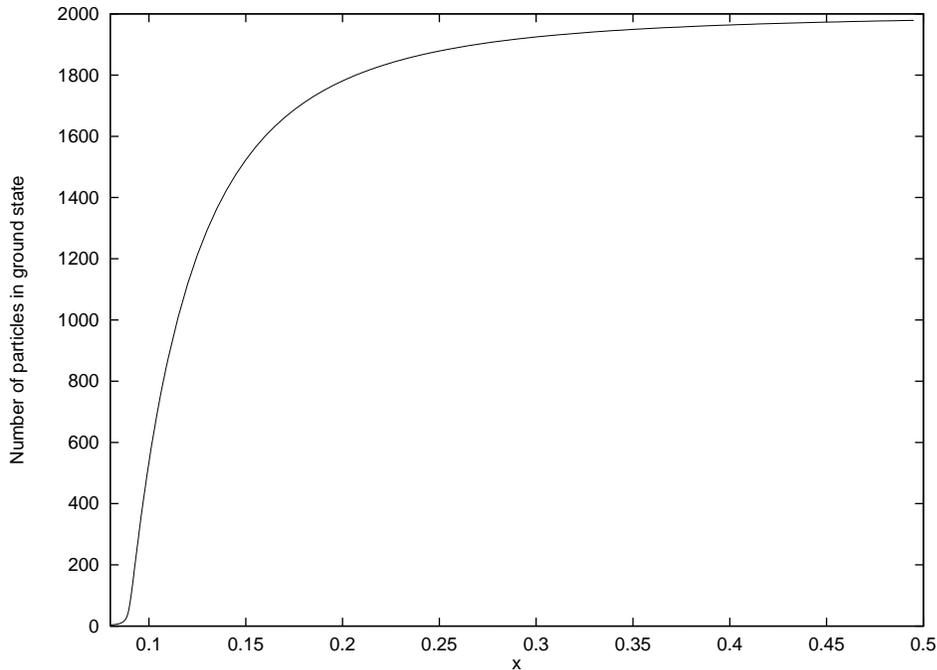}
\end{center}
\caption{\footnotesize This shows the number of particles in the ground state as a function of $x=\hbar\omega/(kT)$.}\label{fig2}
\end{figure}
This calculation shows that over a very small range of $x$, corresponding to a very small range of temperature, $\epsilon$ undergoes a very rapid decrease from large values to very small values. Associated with this sudden change in $\epsilon$ is a sudden rise in the number of particles in the ground state. The ground state occupation number is shown in fig.~2. These numerical results show quite conclusively that BEC  does occur in the sense that a significant number of particles occupy the ground state, and that the ground state occupation number increases very dramatically over a small temperature range. Because the behaviour of both $\epsilon$ and $N_{\rm gr}$ is smooth, there is no unique temperature which signals the onset of BEC. We will return to this point later.

Rather than perform laborious numerical calculations we can try to obtain approximate analytical results. Because the interesting behaviour is found for $x<<1$, the most obvious thing to try is to replace the sums over the discrete energy levels with integrals. However this is like regarding the energy levels as continuous rather than discrete. In the terminology of Sec.~\ref{secgeneral} this effectively replaces $q=0$ with $q=3$, drastically altering the behaviour of the system unless sufficient care is taken in the conversion of sum to integral. We will return to this approach later. Instead of converting the sums to integrals, we will deal with the sums directly. For small $x$ the sums in Eqs.~\ref{3.7.13} and \ref{3.7.16} do not converge particularly rapidly, nor do they display in any transparent way the behaviour for small $x$. However we can use the Mellin-Barnes integral representation in Eq.~\ref{MB} to great effect to obtain an asymptotic expansion valid for small $x$.

Begin with Eq.~\ref{3.7.11}. It is possible to use Eq.~\ref{MB} directly as in Ref.~\cite{KKDJTunpub,KKDJTPRA}; however as the ground state behaviour is of obvious interest it is convenient to separate it off. (It is just the $k=0$ term.) We will write
\beq
q=q_{\rm gr}+q_{\rm ex}\;,\label{3.7.23}
\eeq
where
\beq
q_{\rm gr}=\sum_{n=1}^{\infty}\frac{e^{-n\epsilon x}}{n}=-\ln\left(1-e^{-\epsilon x}\right)\;,\label{3.7.24}
\eeq
is the ground state contribution, and
\beq
q_{\rm ex}=\sum_{k=1}^{\infty}\frac{1}{2}(k+1)(k+2)\sum_{n=1}^{\infty}
\frac{e^{-n(k+\epsilon)x}}{n}\;,\label{3.7.25}
\eeq
represents the contribution from excited states. Using Eq.~\ref{MB} on the RHS of Eq.~\ref{3.7.25} and performing the sums on $n$ and $k$ gives
\beq
q_{\rm ex}&=&\cint\Gamma(\alpha)x^{-\alpha}\zeta_R(\alpha+1)\Big\lbrack
\zeta_H(\alpha-2,1+\epsilon)\nn
&&\quad+(3-2\epsilon)\zeta_H(\alpha-1,1+\epsilon)+(1-\epsilon)
(2-\epsilon)\zeta_H(\alpha,1+\epsilon)\Big\rbrack\;,\label{3.7.27}
\eeq
if we take $c>3$ to avoid the poles of the integrand. If the contour is closed in the left hand plane, there are simple poles at $\alpha=3,2,1,-1,-2,\ldots$ and a double pole at $\alpha=0$. We find
\beq
q_{\rm ex}\simeq\frac{\zeta_R(4)}{x^3}+\Big(\frac{3}{2}-\epsilon\Big)
\zeta_R(3)x^{-2}+\frac{1}{2}(1-\epsilon)(2-\epsilon)\zeta_R(2)x^{-1}
+\cdots\;.\label{3.7.28}
\eeq
(The next term, which is of order $\ln x$, is very complicated and will not be given here.) 

A similar procedure can be used to evaluate the particle number. Defining
\beq
N=N_{\rm gr}+N_{\rm ex}\;,\label{3.7.29}
\eeq
 where $N_{\rm gr}$ was given in Eq.~\ref{3.7.20} and $N_{\rm ex}$ is the contribution from excited states, we find
\beq
N_{\rm ex}&\simeq&\zeta_R(3)x^{-3}+\Big(\frac{3}{2}-\epsilon\Big)\zeta_R(2)
x^{-2}
-\frac{1}{2x}\Big\lbrack\frac{19}{12}+\frac{5}{2}\epsilon-
\frac{3}{2}\epsilon^2\nn
&&+(1-\epsilon)(2-\epsilon)\Big(\ln\,x+
\psi(1+\epsilon)\Big)\Big\rbrack+{\mathcal O}(1)\;.\label{3.7.31}
\eeq
It is possible to extend the calculations to any desired order in the expansion.

We can now test how good the analytic approximation for the particle number is since we have already evaluated the exact expressions numerically. Let $f$ be the fraction of the total number of particles in the ground state~:
\beq
N_{\rm gr}=fN\;.\label{3.7.32}
\eeq
From Eq.~\ref{3.7.21} we have
\beq
\epsilon x=\ln\Big(1+\frac{1}{fN}\Big)\;,\label{3.7.33}
\eeq
which determines $\epsilon$ given $x,f,N$. If we keep only the first two terms in the expansion for $N_{\rm ex}$ we have
\beq
N\simeq N_{\rm gr}+\zeta_R(3)x^{-3}+\Big(\frac{3}{2}-\epsilon\Big)\zeta_R(2)x^{-2}
\;.\label{3.7.34}
\eeq
With Eqs.~\ref{3.7.32} and \ref{3.7.33} this becomes
\beq
(1-f)Nx^3-\frac{3}{2}\zeta_R(2)x+\Big\lbrack\zeta_R(2)
\ln\Big(1+\frac{1}{fN}\Big)-\zeta_R(3)\Big\rbrack=0\;.\label{3.7.35}
\eeq
The result of our approximate evaluation of $\epsilon$ is shown as the diamonds in fig.~1. We have used the exact result for $\epsilon$ to compute $f$, and given $f$ and $N$ we can solve Eq.~\ref{3.7.35} for $x$. The approximate value for $\epsilon$ is found using Eq.~\ref{3.7.33}. As can be seen from this figure the approximation is very good when $\epsilon$ is small, but starts to break down when $\epsilon$ becomes large. This is to be expected because by keeping only the first two terms in Eq.~\ref{3.7.31} we have treated $\epsilon$ as small. (We have really assumed that $\epsilon<<x^{-1/2}$.) We will discuss another method for obtaining results with a wider range of applicability later. Even for $\epsilon\approx0.5$ our approximate result is not too bad.

We mentioned that a plausible approach if we want the behaviour for small $x$ is to replace the sums over the discrete energy levels with integrals. This was done in Ref.~\cite{Bagnato}. The result of this for the particle number is that only the first term in the expansion in Eq.~\ref{3.7.31} is obtained. This can be seen starting from 
\beq
N_{\rm ex}=\frac{1}{2}\sum_{k=1}^{\infty}(k+1)(k+2)\Big\lbrack e^{(k+\epsilon)x}-1\Big\rbrack^{-1}\;,\label{3.7.36}
\eeq
which follows from differentiation of Eq.~\ref{3.7.25}. Assume that there is a well defined transition temperature below which $\epsilon=0$ as for the normal unconfined Bose gas. From our discussion above, $\epsilon$ really only vanishes for $N\rightarrow\infty$ which is often referred to as taking the thermodynamic limit. We prefer, following Ref.~\cite{Kac}, to refer to this as the bulk approximation. The bulk approximation consists of taking $\epsilon=0$ in Eq.~\ref{3.7.36}, along with $(k+1)(k+2)\approx k^2$, and replacing the sum over $k$ with an integral~:
\beq
N_{\rm ex}^{\rm bulk}=\frac{1}{2}\int_{0}^{\infty}dk\,k^2(e^{kx}-1)^{-1}
=\zeta_R(3)x^{-3}\;.\label{3.7.38}
\eeq
(A more systematic way of approximating the sum with an integral will be described later.) The bulk transition temperature is just the temperature at which $N_{\rm gr}=0$. This gives
\beq
x_{\rm bulk}=\left\lbrack\frac{\zeta_R(3)}{N}
\right\rbrack^{1/3}\;.\label{3.7.39}
\eeq
Note that $x_{\rm bulk}=\hbar\omega/(kT_{\rm bulk})$, and that Eq.~\ref{3.7.39} holds for $x\ge x_{\rm bulk}$ corresponding to $T\le T_{\rm bulk}$.
\begin{figure}[htb]
\begin{center}
\leavevmode
\epsffile{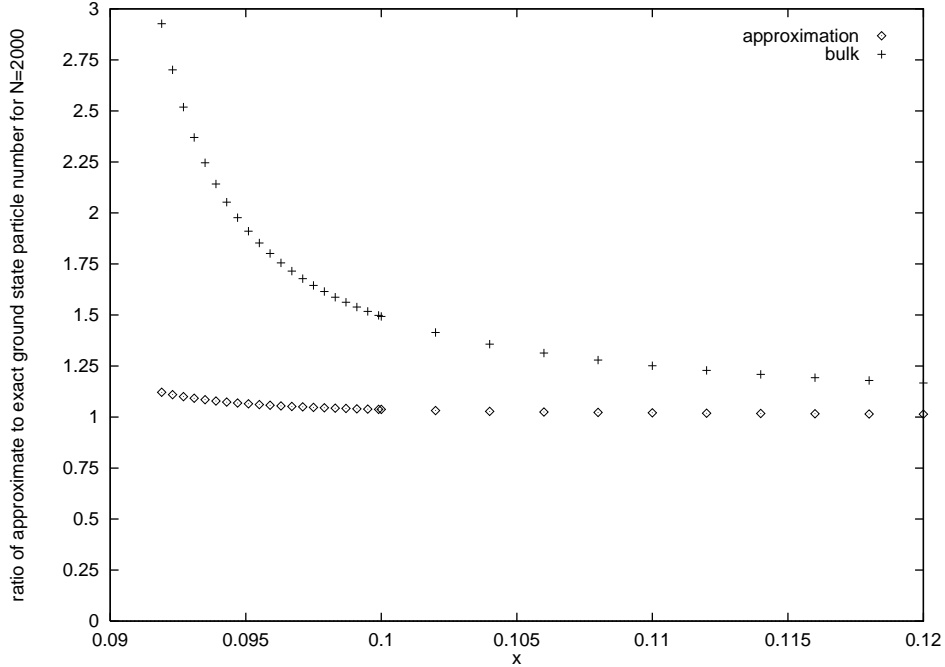}
\end{center}
\caption{\footnotesize The ratio of the approximate to the exact ground state particle number is shown. The diamonds illustrate the result of using our approximation and the crosses denote the results found using the bulk approximation.}\label{fig3}
\end{figure}
The result of using the bulk approximation for the ground state particle number is that for $x\ge x_{\rm bulk}$,
\beq
N_{\rm gr}^{\rm bulk}=N-N_{\rm ex}^{\rm bulk}=N\left\lbrack 1-
\left(\frac{x_{\rm bulk}}{x}\right)^3\right\rbrack\;.\label{3.7.40}
\eeq
Fig.~3 shows a comparison of the bulk result with the approximation described above. The crosses show the ratio of the bulk approximation to the exact result for $N_{\rm gr}$, and the diamonds show the ratio using $N_{\rm gr}$ determined by Eq.~\ref{3.7.34} to the exact result. It is evident that the approximation in Eq.~\ref{3.7.34} is much better than the bulk approximation, especially as $x$ becomes closer to the value where $\epsilon$ undergoes its most rapid change.

Having seen that it is possible to improve the bulk approximation for the ground state particle number significantly, we can see how Eq.~\ref{3.7.39} compares with the temperature quoted in the experiments. From Eq.~\ref{3.7.39} we have
\beq
T_{\rm bulk}=\frac{\hbar\omega}{k}\left\lbrack\frac{N}{\zeta_R(3)}
\right\rbrack^{1/3}\;.\label{3.7.41}
\eeq
For rubidium with $N=2000$ and $\hbar\omega/(2\pi)=60$ Hz (which is the geometric mean of the frequencies) we find $T_{\rm bulk}\simeq$ 34.12nK, which is close to the value quoted in the experiment~\cite{rub}. Given the simplicity of the theoretical model this is perhaps a bit surprising.

It is straightforward to see how the bulk temperature is altered by using the more accurate approximation. We know that $T_{\rm bulk}$ is close to the experimental value, so assume that
\beq
x=x_{\rm bulk}(1+\eta)\;,\label{3.7.42}
\eeq
where $\eta$ is assumed small. Neglecting $\epsilon$ in Eq.~\ref{3.7.34} we find
\beq
\eta\simeq\frac{1}{3}f+\frac{\zeta_R(2)}{2\lbrack\zeta_R(3)\rbrack^{2/3}}
N^{-1/3}\;.\label{3.7.43}
\eeq
Thus
\beq
\frac{T-T_{\rm bulk}}{T}\simeq-\left\lbrack
\frac{1}{3}f+\frac{\zeta_R(2)}{2\lbrack\zeta_R(3)\rbrack^{2/3}}
N^{-1/3}\right\rbrack\;.\label{3.7.44}
\eeq
(If we set $f=0$ in this result we obtain the result in Ref.~\cite{GHPLA,GH2}.) Ignoring $f$, which is typically a few percent, and taking $N=2000$ we find that the more accurate approximation reduces the temperature by around 10\%. The larger the particle number the better the bulk approximation for the temperature becomes. However it should be clear from our study of the ground state particle number that the bulk approximation should be used with caution for other calculations. We will see that this is even more true when we study the specific heat.

Before turning to the calculation of other thermodynamic expressions we will discuss a variation on the method presented above. This makes contact with the approach of ref.~\cite{Haug} which uses the Euler-Maclaurin summation formula to evaluate the exact sums defining the thermodynamic quantities. Previously we removed the ground state contribution from $q$. We will now remove the first excited state contribution as well. We may write
\beq
q_{\rm ex}=q_1+\tilde{q}_{\rm ex}\;,\label{3.7.45}
\eeq
where
\beq
q_1&=&-3\ln\Big\lbrack1-e^{-(1+\epsilon)x}\Big\rbrack\;,\label{3.7.46}\\
\tilde{q}_{\rm ex}&=&-\sum_{k=2}^{\infty}\frac{1}{2}(k+1)(k+2)\ln
\Big\lbrack1-e^{-(k+\epsilon)x}\Big\rbrack\;.\label{3.7.47}
\eeq
(See Eq.~\ref{3.7.10}.) We can expand the logarithm in Eq.~\ref{3.7.47} to obtain
\beq
\tilde{q}_{\rm ex}=
-\sum_{k=2}^{\infty}\frac{1}{2}(k+1)(k+2)\sum_{n=1}^{\infty}
\frac{e^{-n(k+\epsilon)x}}{n}\;.\label{3.7.48}
\eeq
Now use the representation in Eq.~\ref{MB}, but only on part of the exponential. We may write
\begin{displaymath}
e^{-n(k+\epsilon)x}=e^{-n(k-1)x}e^{-n(1+\epsilon)x}\;,
\end{displaymath}
and use Eq.~\ref{MB} on $e^{-n(k-1)x}$ only. This gives the contour integral representation
\beq
\tilde{q}_{\rm ex}&=&\cint\Gamma(\alpha)x^{-\alpha}{\rm Li}_{1+\alpha}
\Big(e^{-(1+\epsilon)x}\Big)\Big\lbrack\frac{1}{2}\zeta_R(\alpha-2)\nn
&&\quad\quad\quad+\frac{3}{2}\zeta_R(\alpha-1)+\zeta_R(\alpha)\Big\rbrack
\;,\label{3.7.49}
\eeq
where ${\rm Li}$ is the polylogarithm function introduced earlier in Eq.~\ref{poly}. We take $c>3$ here to justify the interchange of sum and integral. The contour can be closed and the integral evaluated as usual. ${\rm Li}_{1+\alpha}$ is an analytic function of $\alpha$, so that poles can come only from $\Gamma(\alpha)$ or the Riemann $\zeta$-functions. We find
\beq
\tilde{q}_{\rm ex}&=&x^{-3}{\rm Li}_4\Big(e^{-(1+\epsilon)x}\Big)+
\frac{3}{2}x^{-2}{\rm Li}_3\Big(e^{-(1+\epsilon)x}\Big)\nn
&&+x^{-1}{\rm Li}_2\Big(e^{-(1+\epsilon)x}\Big)+\frac{5}{8}\ln\Big\lbrack
1-e^{-(1+\epsilon)x}\Big\rbrack\nn
&&+\frac{19}{240}x\Big\lbrack e^{(1+\epsilon)x}-1\Big\rbrack^{-1}+\cdots\;,\label{3.7.50}
\eeq
if the first few terms in the expansion are kept, and properties of the polylogarithm are used. A technical point which makes this approach useful for numerical computations is that as $\epsilon$ becomes small the argument of the polylogarithms become approximately $e^{-x}<1$. If we had not removed the first excited state then we would have obtained an expansion similar to that in Eq.~\ref{3.7.50} but where the argument of the polylogarithms was $e^{-\epsilon x}\approx1$. We would then have found that terms in the expansion in Eq.~\ref{3.7.50} would have become large as $\epsilon$ got small, rendering the expansion inaccurate. Combining Eqs.~\ref{3.7.46} and \ref{3.7.50} we find
\beq
q_{\rm ex}&=&x^{-3}{\rm Li}_4\Big(e^{-(1+\epsilon)x}\Big)+
\frac{3}{2}x^{-2}{\rm Li}_3\Big(e^{-(1+\epsilon)x}\Big)\nn
&&+x^{-1}{\rm Li}_2\Big(e^{-(1+\epsilon)x}\Big)-\frac{19}{8}\ln\Big\lbrack
1-e^{-(1+\epsilon)x}\Big\rbrack\nn
&&+\frac{19}{240}x\Big\lbrack e^{(1+\epsilon)x}-1\Big\rbrack^{-1}+\cdots\;.\label{3.7.51}
\eeq
This can be used as the basis for an accurate numerical approximation \cite{Haug}, although some of the simplicity of the earlier purely analytic approximation is lost due to the presence of the polylogarithms which must be evaluated numerically.

\section{Specific Heat for the Isotropic Harmonic Oscillator}
\label{specific}

The specific heat has been used as a criterion for defining a critical temperature in finite systems \cite{Pathriaetal}. A natural criterion is to define the critical temperature as the temperature at which the specific heat has its maximum. The behaviour of the specific heat for the unconfined Bose gas in three spatial dimensions is standard \cite{LLStatPhys,Pathria,Huang,London}. The specific heat is continuous with a maximum coinciding at the BEC transition temperature $T_c$ where the phase transition occurs. The derivative of the specific heat with respect to temperature is discontinuous at $T_c$. The behaviour of the specific heat in dimensions other than 3 is also known \cite{May,Kac}. For $D=4$ the specific heat and its first derivative are continuous at the transition temperature $T_c$, but the second derivative is discontinuous. For $D\ge5$ the specific heat itself is not continuous at the transition temperature. In the cases of $D=1,2$ where no phase transition occurs the specific heat is a smooth function of temperature. We would expect that since no phase transition occurs for the Bose gas confined by the harmonic oscillator potential, the specific heat should be a smooth function of temperature.

The starting point for an evaluation of the specific heat is the internal energy defined by
\beq
U=\sum_{\mathbf n}E_{\mathbf n}\Big\lbrack e
^{\beta(E_{\mathbf n}-\mu)}-1\Big\rbrack^{-1}\;,\label{3.7.52}
\eeq
where the energy levels are given by Eq.~\ref{3.7.2}. For the isotropic harmonic oscillator we find
\beq
\frac{U}{\hbar\omega}=\sum_{k=0}^{\infty}\frac{1}{2}(k+1)(k+2)
(k+\frac{3}{2})\Big\lbrack e^{(k+\epsilon)x}-1\Big\rbrack^{-1}
\;,\label{3.7.66}
\eeq
when written in terms of the dimensionless variables $x$ and $\epsilon$ defined in Eqs.~\ref{3.7.8} and \ref{3.7.9}. The term which involves $3/2$ in Eq.~\ref{3.7.66} may be recognized as the total particle number. We have
\beq
\frac{U}{\hbar\omega}=\frac{3}{2}N+\frac{1}{2}\sum_{k=1}^{\infty}k(k+1)(k+2)
\Big\lbrack e^{(k+\epsilon)x}-1\Big\rbrack^{-1}
\;.\label{3.7.67}
\eeq
If we binomially expand the factor in square brackets in Eq.~\ref{3.7.67} we find
\beq
\frac{U}{\hbar\omega}=\frac{3}{2}N+\frac{1}{2}\sum_{k=1}^{\infty}k(k+1)(k+2)
\sum_{n=1}^{\infty} e^{-n(k+\epsilon)x}
\;.\label{3.7.68}
\eeq
The sum over $k$ is easily performed with the result
\beq
\frac{U}{\hbar\omega}=\frac{3}{2}N+3u_1\;,\label{3.7.69}
\eeq
where
\beq
u_1=\sum_{n=1}^{\infty}e^{-n(1+\epsilon)x}\Big(1-e^{-nx}\Big)^{-4}\;.
\label{3.7.70}
\eeq

The specific heat is defined by
\beq
C=\left(\frac{\partial U}{\partial T}\right)_{N,\omega}\;.\label{3.7.71}
\eeq
Here $\omega$ plays the role of the volume for the free Bose gas. Because $N$ is held fixed when we differentiate $U$, only the term $3u_1$ will contribute to the specific heat. Because the temperature enters $x=\beta\hbar\omega$, we have
\beq
\frac{C}{k}=-3x^2\left(\frac{\partial u_1}{\partial x}\right)_{N,\omega}
\;.\label{3.7.72}
\eeq
Care must be taken because $N$ is fixed rather than $\epsilon$. From Eq.~\ref{3.7.16} we have
\beq
N=\sum_{n=1}^{\infty}e^{-n\epsilon x}\Big(1-e^{-nx}\Big)^{-3}\;,
\label{3.7.73}
\eeq
which defines $\epsilon$ implicitly as a function of $N$ and $x$. Differentiating both sides of Eq.~\ref{3.7.73} keeping $N$ fixed gives us
\beq
\left(\frac{\partial(\epsilon x)}{\partial x}\right)_{N}=-3\frac{S_2}{S_1}
\;,\label{3.7.74}
\eeq
where
\beq
S_1&=&\sum_{n=1}^{\infty}ne^{-n\epsilon x}\Big(1-e^{-nx}\Big)^{-3}\;,
\label{3.7.75}\\
S_2&=&\sum_{n=1}^{\infty}ne^{-n(\epsilon+1)x}\Big(1-e^{-nx}\Big)^{-4}\;.
\label{3.7.76}
\eeq
Performing the differentiation in Eq.~\ref{3.7.72} and using Eq.~\ref{3.7.74} results in
\beq
\frac{C}{k}=3x^2\left\lbrace4S_3+S_2-3\frac{S_2^2}{S_1}\right\rbrace\;,
\label{3.7.77}
\eeq
with
\beq
S_3&=&\sum_{n=1}^{\infty}ne^{-n(\epsilon+2)x}\Big(1-e^{-nx}\Big)^{-5}\;.
\label{3.7.78}
\eeq
\begin{figure}[ht]
\begin{center}
\leavevmode
\epsffile{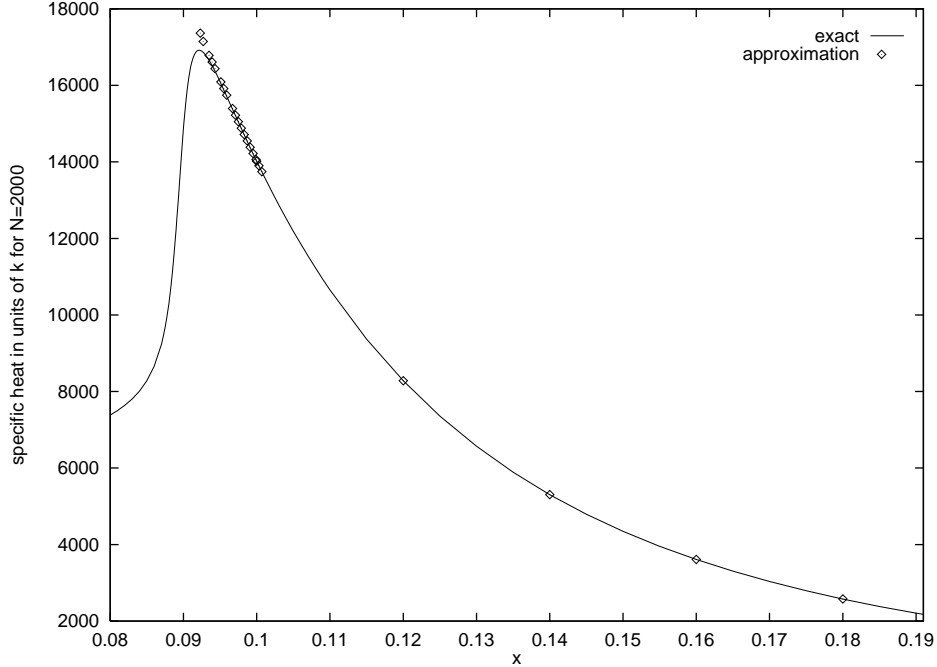}
\end{center}
\caption{\footnotesize The specific heat computed numerically for the isotropic harmonic oscillator is shown as the solid curve. The diamonds show the result using our approximation. The units for the specific heat are in factors of the Boltzmann constant $k$. The particle number is $N=2000$. The maximum occurs for $x\simeq0.0921$. Our approximation breaks down for $x$ below the point where the specific heat maximum occurs.}
\end{figure}

This formula may be used to compute the specific heat numerically. The procedure used is to first solve Eq.~\ref{3.7.73} for $\epsilon$ for a given $N$ and $x$. Knowing $\epsilon$ and $x$ it is possible to evaluate the sums $S_1,S_2$ and $S_3$ and hence determine the specific heat. The 
result of this calculation is shown in fig.~4 for N=$2000$. This figure shows that the specific heat is a perfectly smooth function of temperature in contrast to the result for the free Bose gas which has a discontinuous first derivative. A feature which the confined gas has in common with the free gas is that there is a well defined maximum. The maximum can be seen to occur at a value of $x$ which is very close to the place where $\epsilon$ undergoes its most rapid decrease and the population of the ground state starts to rise. We can therefore identify the specific heat maximum with the onset of BEC.

Rather than perform detailed numerical calculations we can try to obtain an analytical approximation for the specific heat by using the Mellin-Barnes representation to approximate the sums in Eq.~\ref{3.7.78}. If we assume $x<<1$ and $\epsilon<<1$ we find
\beq
\frac{C}{k}&\simeq&12\zeta_R(4)x^{-3}+9\zeta_R(3)x^{-2}+2\zeta_R(2)x^{-1}
-12\epsilon\zeta_R(3)x^{-2}\nn
&&-18\epsilon^2\zeta_R(2)\zeta_R(3)x^{-3}-9\epsilon^2\zeta_R^2(3)x^{-4}+
9\epsilon^4\zeta_R^2(3)x^{-6}\;.\label{3.7.79}
\eeq
The result of using Eq.~\ref{3.7.79} to calculate the specific heat is shown as the diamonds in fig.~4. It can be seen that at temperatures below the point where the specific heat has its maximum the approximation is quite good. For values of $x<x_m\simeq0.0921$ the assumption that $\epsilon$ is small is no longer valid and the simple approximation breaks down. It is possible to obtain a better approximation using polylogarithms.

We can also see what happens if the bulk approximation is used. From Eq.~\ref{3.7.68} if we replace the sum on $k$ with an integral and use $k(k+1)(k+2)\approx k^3$ we obtain
\beq
\frac{U_{\rm bulk}}{\hbar\omega}&=&\frac{3}{2}N+\frac{1}{2}
\sum_{n=1}^{\infty}e^{-n\epsilon x}\int_{0}^{\infty}dk\,k^3e^{-nkx}\nn
&=&\frac{3}{2}N+3x^{-4}{\rm Li}_4(e^{-\epsilon x})\;.\label{3.7.80}
\eeq
For $T<T_{\rm bulk}$ we have $\epsilon=0$ so
\beq
\frac{U_{\rm bulk}^<}{\hbar\omega}=
\frac{3}{2}N+3x^{-4}\zeta_R(4)\;,\label{3.7.81}
\eeq
gives the internal energy. The specific heat for $T<T_{\rm bulk}$ is
\beq
C_{\rm bulk}^<=\left(\frac{\partial U_{\rm bulk}^<}{\partial T}
\right)_{N,\omega}=12k\zeta_R(4)x^{-3}\;.\label{3.7.82}
\eeq
This is observed to correspond to only the first term on the right hand side of Eq.~\ref{3.7.79}.

The bulk approximation for the specific heat when $T>T_{\rm bulk}$ is more involved. We first need the number of particles when $\epsilon\ne0$. This can be obtained exactly as in Eq.~\ref{3.7.36} by approximating the sum over $k$ with an integral. The result is
\beq
N_{\rm bulk}=x^{-3}{\rm Li}_3(e^{-\epsilon x})\;.\label{3.7.83}
\eeq
We define $T_{\rm bulk}$ as before by setting $\epsilon=0$ in Eq.~\ref{3.7.83} and obtaining
\beq
N=x_{\rm bulk}^{-3}\zeta_R(3)\;.\label{3.7.84}
\eeq
Eliminating $N$ between Eq.~\ref{3.7.83} and Eq.~\ref{3.7.84} results in
\beq
{\rm Li}_3(e^{-\epsilon x})=\zeta_R(3)\left(\frac{T_{\rm bulk}}{T}
\right)^3\;.\label{3.7.85}
\eeq
We can now use Eq.~\ref{3.7.80} to give
\beq
C_{\rm bulk}^>=3k\left(\frac{k}{\hbar\omega}\right)^3
\frac{\partial}{\partial T}
\left\lbrace T^4{\rm Li}_4(e^{-\epsilon x})\right\rbrace
\Big|_{N,\omega}\;.\label{3.7.87}
\eeq
By differentiating the polylogarithm it is easy to see that
\beq
\left.\frac{\partial}{\partial T}
{\rm Li}_4(e^{-\epsilon x})\right|_{N,\omega}=-\left(
\frac{\partial(\epsilon x)}{\partial T}\right)_{N,\omega}
{\rm Li}_3(e^{-\epsilon x})\;.\label{3.7.89}
\eeq
Differentiating Eq.~\ref{3.7.85} with respect to $T$ holding $N$ and $\omega$ fixed results in
\beq
\left(
\frac{\partial(\epsilon x)}{\partial T}\right)_{N,\omega}=
\frac{3\zeta_R(3)T_{\rm bulk}^3T^{-4}}{{\rm Li}_2(e^{-\epsilon x})}\;.
\label{3.7.90}
\eeq
Putting these results together yields
\beq
\frac{C_{\rm bulk}^>}{kN}=12\frac{{\rm Li}_4(e^{-\epsilon x})}
{{\rm Li}_3(e^{-\epsilon x})}-9\frac{{\rm Li}_3(e^{-\epsilon x})}
{{\rm Li}_2(e^{-\epsilon x})}\;.\label{3.7.91}
\eeq
For comparison, we can use Eq.~\ref{3.7.84} in Eq.~\ref{3.7.82} to obtain
\beq
\frac{C_{\rm bulk}^<}{kN}=12\frac{\zeta_R(4)}{\zeta_R(3)}
\left(\frac{T}{T_{\rm bulk}}\right)^3\;.\label{3.7.86}
\eeq

If we take the limit $T\rightarrow T_{\rm bulk}$ in Eqs.~\ref{3.7.91} and \ref{3.7.86} we find
\beq
\frac{C_{\rm bulk}^<}{kN}&\rightarrow&
12\frac{\zeta_R(4)}{\zeta_R(3)}\label{3.7.92}\\
\frac{C_{\rm bulk}^>}{kN}&\rightarrow&
12\frac{\zeta_R(4)}{\zeta_R(3)}-9\frac{\zeta_R(3)}{\zeta_R(2)}
\;.\label{3.7.93}
\eeq
The bulk approximation leads to a discontinuous result for the specific heat. This contrasts with both the free Bose gas, which had a continuous but not smooth behaviour, and the exact result which was shown in fig.~4 which is continuous and smooth. Although $T_{\rm bulk}$ is close to the BEC temperature determined by the true specific heat maximum, the bulk approximation for the specific heat shows features not present in the true system. 

\begin{figure}[ht]
\begin{center}
\leavevmode
\epsffile{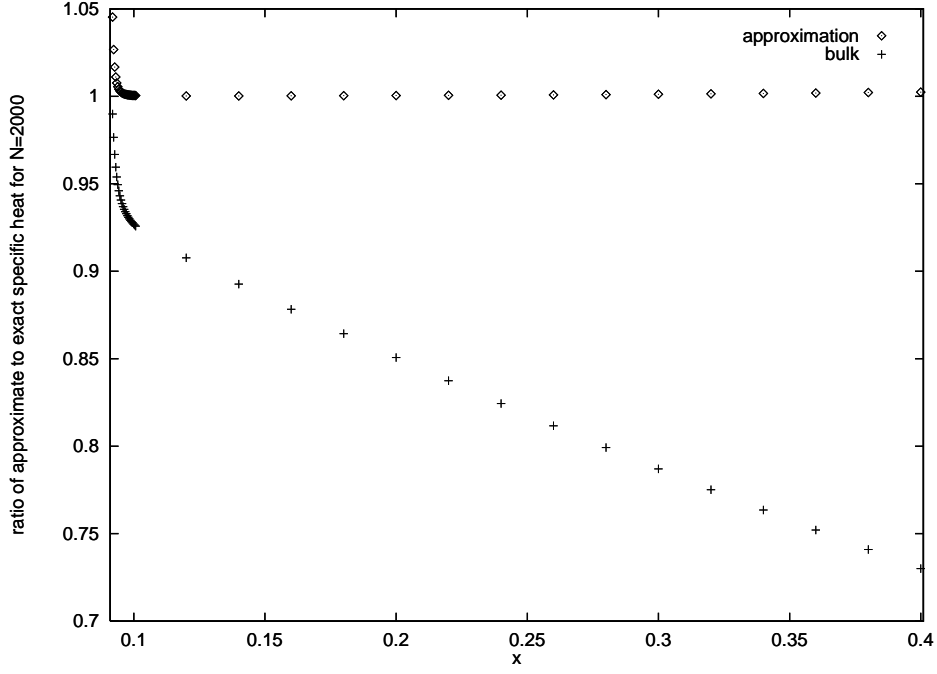}
\end{center}
\caption{\footnotesize The diamonds show the ratio of our approximation for the specific heat to the exact result. The crosses denote the ratio of the bulk specific heat to the exact value. $N=2000$ is taken. Both results become increasingly inaccurate below the specific heat maximum.}
\end{figure}
Leaving aside the fact that the bulk approximation for the specific heat is discontinuous, we can see how good an approximation it is when compared with the true result. This is shown in fig.~5. The crosses show $C_{\rm bulk}/C_{\rm exact}$ and the diamonds show the ratio of the approximation in Eq.~\ref{3.7.79} to the exact result. Although the agreement between the bulk value and the exact value is good near the specific heat maximum, it is off by about 25\% once $x$ has reached 0.4. In contrast Eq.~\ref{3.7.79} is within about 1\% of the true value at this point.

\section{Anisotropic Harmonic Oscillator Potential}

The case where all of the oscillator frequencies are different is more difficult to analyze than the isotropic case. Nevertheless the Mellin-Barnes representation can still be used to great effect \cite{KKDJTPRA}. We have
\beq
E_{\mathbf n}=\hbar(n_1\omega_1+n_2\omega_2+n_3\omega_3)+\mu_c\;,
\label{3.7.94a}
\eeq
if Eqs.~\ref{3.7.2} and \ref{3.7.3} are used. We will define dimensionless variables $x_1,x_2,x_3$ by
\beq
x_i=\beta\hbar\omega_i\;,\label{3.7.94}
\eeq
analogously to $x$ in Eq.~\ref{3.7.8}. We will also define
\beq
\Omega=\frac{1}{3}(\omega_1+\omega_2+\omega_3)\;,\label{3.7.95}
\eeq
to be the average frequency, and
\beq
\alpha=\hbar\beta\Omega=\frac{1}{3}(x_1+x_2+x_3)\;.\label{3.7.96}
\eeq
Finally we will define
\beq
\mu=\hbar\Omega\Big(\frac{3}{2}-\epsilon\Big)\label{3.7.97}
\eeq
analogously to Eq.~\ref{3.7.9}. We then have
\beq
\beta(E_{\mathbf n}-\mu)=
\sum_{i=1}^{3}n_ix_i+\alpha\epsilon\;.\label{3.7.98}
\eeq
The $q$-potential in Eq.~\ref{3.7.5} reads
\beq
q=-\sum_{n_1=0}^{\infty}\sum_{n_2=0}^{\infty}\sum_{n_3=0}^{\infty}
\ln\left\lbrack1-e^{-(n_1x_1+n_2x_2+n_3x_3+\alpha\epsilon)}\right\rbrack
\;,\label{3.7.99}
\eeq
when it is written in terms of dimensionless variables defined above.

As before, we expand the logarithm in its Taylor series
\beq
q&=&\sum_{n=1}^{\infty}\frac{1}{n}
\sum_{n_1=0}^{\infty}\sum_{n_2=0}^{\infty}\sum_{n_3=0}^{\infty}
e^{-n(n_1x_1+n_2x_2+n_3x_3+\alpha\epsilon)}
\label{3.7.100}\\
&=&\sum_{n=1}^{\infty}\frac{1}{n}
e^{-n\alpha\epsilon}(1-e^{-nx_1})^{-1}(1-e^{-nx_2})^{-1}(1-e^{-nx_3})^{-1}
\;,\label{3.7.101}
\eeq
where the second line has followed from performing the sums over $n_1,n_2,n_3$. We can separate off the ground state from the sums as we did in the isotropic case, but this gets slightly messy here so is left as an exercise for the reader. Instead we will apply Eq.~\ref{MB} directly to the exponential in Eq.~\ref{3.7.100} to obtain
\beq
q=
\frac{1}{2\pi i}
\int_{c-i\infty}^{c+i\infty}\!\!\!\!\!\!\!\!\!dz\,\Gamma(z)\zeta_R(z+1)
\zeta_B(z,\alpha\epsilon|{\mathbf x})\;,\label{3.7.101b}
\eeq
where $\zeta_B$ is the three dimensional Barnes $\zeta$-function defined in Eq.~\ref{Barnes1}. This function is analytic for $\Re(z)>3$ with simple poles at $z=3,2,1$. Using the values of the residues at these poles given in Eqs.~\ref{Barnes4}--\ref{Barnes6} we find
\beq
q&\simeq&\frac{\zeta_R(4)}{x_1x_2x_3}+\frac{\zeta_R(3)}{x_1x_2x_3}\Big(
\frac{3}{2}-\epsilon\Big)\alpha\nn
&&+\frac{\zeta_R(2)}{12\,x_1x_2x_3}\lbrack6\alpha^2\epsilon^2-18\alpha^2
\epsilon+9\alpha^2+x_1x_2+x_2x_3+x_3x_1\rbrack+\cdots\;.\label{3.7.102}
\eeq
The particle number is
\beq
N=\beta^{-1}\left(\frac{\partial q}{\partial\mu}\right)_{T,\omega_i}=
-\frac{1}{\alpha}\left(\frac{\partial q}{\partial\epsilon}\right)_{x_i}\;,
\label{3.7.103}
\eeq
if we use the fact that fixed $T$ and $\omega_i$ corresponds to fixed $x_i$, and hence fixed $\Omega$ and $\alpha$; thus, $\delta\mu=-\hbar\Omega\delta\epsilon=-\beta^{-1}\alpha\delta\epsilon$. From Eq.~\ref{3.7.100}
we find easily that
\beq
N&=&
\sum_{n_1=0}^{\infty}\sum_{n_2=0}^{\infty}\sum_{n_3=0}^{\infty}
e^{-n(n_1x_1+n_2x_2+n_3x_3+\alpha\epsilon)}\label{3.7.104}\\
&=&
\frac{1}{2\pi i}
\int_{c-i\infty}^{c+i\infty}\!\!\!\!\!\!\!\!\!dz\,\Gamma(z)\zeta_R(z)\zeta_B(z,\alpha\epsilon|{\mathbf x})\label{3.7.105}\\
&\simeq&
\frac{\zeta_R(3)}{x_1x_2x_3}+\frac{\zeta_R(2)}{x_1x_2x_3}\Big(
\frac{3}{2}-\epsilon\Big)\alpha+\frac{1}{\alpha\epsilon}+\cdots
\;.\label{3.7.106}
\eeq
The first two terms on the right hand side of Eq.~\ref{3.7.106} are easy to obtain, and in fact follow directly from differentiating the expression for $q$ in Eq.~\ref{3.7.102}. The third term $(\alpha\epsilon)^{-1}$ requires some explanation. Its origin is the pole of the integrand of Eq.~\ref{3.7.105} at $z=1$ which is a double pole since $\zeta_R(z)=1/(z-1)+\gamma+\cdots$ near $z=1$ and $\zeta_B$ has a pole at $z=1$. To evaluate the third term completely we need to know the finite part of $\zeta_B$ at $z=1$ in addition to the pole part. This is not known to my knowledge; however, if we are interested simply in the case where $\epsilon\alpha$ is small, which is the case on physical grounds, then we may use the fact that $\zeta_B(z,a|{\mathbf x})\approx a^{-z}$ gives the leading contribution (coming from the ${\mathbf n}={\mathbf 0}$ term in the sum). Keeping only the dominant term for small $\alpha\epsilon$ results in Eq.~\ref{3.7.106}. Another way to check the validity of Eq.~\ref{3.7.106} is to remove the ground and first excited state from the sums in Eq.~\ref{3.7.104} in the same way as we did for the isotropic oscillator, and express $N$ as in Eq.~\ref{3.7.105} but where polylogarithms arise in place of the Riemann $\zeta$-function. This avoids the problem of the double  pole at $z=1$ because the polylogarithm is analytic there. Furthermore, this allows the limit $\epsilon\rightarrow0$ to be taken because the argument of the polylogarithm involves $e^{-x_{\rm min}}$ where $x_{\rm min}$ is the smallest of the $x_i$.

A calculation similar to that outlined for the isotropic harmonic oscillator may be performed here. The details of this may be found in Ref.~\cite{KKDJTPRA}. Again the results of the analytic approximation may be compared with the exact results. The agreement between the two is as good as in the isotropic case. For $N=2000$ and $\omega_1=\omega_2=240\pi/\sqrt{8}\ {\rm s}^{-1} $ and $\omega_3=240\pi\ {\rm s}^{-1}$ the specific heat maximum is found numerically to occur at $\alpha=0.106$. Using Eq.~\ref{3.7.96} this may be shown to correspond to a temperature of $30.9$~nK which can be compared to $31.3$~nK if we approximate the system by an isotropic harmonic oscillator whose frequency is the geometric mean of the three frequencies, and to $34.1$~nK if the bulk approximation is used. The fact that the oscillator is anisotropic only changes the temperature by a few percent from the simpler isotropic calculation.

A final point we wish to discuss is that the $(\alpha\epsilon)^{-1}$ term in Eq.~\ref{3.7.106} is observed to be the first term in the expansion of
\beq
N_{\rm gr}=(e^{\epsilon\alpha}-1)^{-1}\;,\label{3.7.107}
\eeq
in powers of $\alpha\epsilon$. Thus we can interpret the first two terms in Eq.~\ref{3.7.106} as the contributions from the excited states~:
\beq
N_{\rm ex}\simeq
\frac{\zeta_R(3)}{x_1x_2x_3}+\frac{\zeta_R(2)}{x_1x_2x_3}\Big(
\frac{3}{2}-\epsilon\Big)\alpha\;.\label{3.7.108}
\eeq
An analysis similar to that presented for the isotropic harmonic oscillator may now be given.
\beq
N_{\rm ex}\approx
\frac{\zeta_R(3)}{x_1x_2x_3}+\frac{3\zeta_R(2)}{2\,x_1x_2x_3}
\alpha\;,\label{3.7.109}
\eeq
if we assume that $\epsilon<<3/2$. By saying that at the onset of BEC we have $N_{\rm ex}\approx N$ we obtain the equation
\beq
N\approx
\frac{\zeta_R(3)}{x_1x_2x_3}+\frac{3\zeta_R(2)}{2\,x_1x_2x_3}
\alpha\;,\label{3.7.110}
\eeq
which determines $T$. (Setting $\alpha=0$ in Eq.~\ref{3.7.110} is the bulk approximation.) Using Eqs.~\ref{3.7.94} and \ref{3.7.95} results in the cubic equation
\beq
N\approx aT^3+bT^2\;,\label{3.7.111}
\eeq
where
\beq
a&=&\frac{k^3\zeta_R(3)}{\hbar^3\omega_1\omega_2\omega_3}
\;,\label{3.7.112}\\ 
b&=&\frac{3k^2\Omega\zeta_R(2)}{2\hbar^2\omega_1\omega_2\omega_3}
\;.\label{3.7.113}
\eeq
Since $N$ is large and $T$ is small we may assume
\beq
T=\left(\frac{N}{a}\right)^{1/3}(1+\eta)\;,\label{3.7.114}
\eeq
with $\eta$ small. It is straightforward to obtain
\beq
kT\approx\hbar\omega\left(\frac{N}{\zeta_R(3)}\right)^{1/3}\left\lbrack
1-\frac{\Omega\zeta_R(2)}{2\omega\zeta_R^{2/3}(3)}\,N^{-1/3}\right\rbrack\;
\label{3.7.115}
\eeq
where $\omega=(\omega_1\omega_2\omega_3)^{1/3}$. As before, the second term in Eq.~\ref{3.7.115} represents the correction to the bulk result. Because we have assumed $N_{\rm ex}=N$, there will be a small correction to this if we compare the result with the specific heat maximum due to the fact that at the specific heat maximum there is a non-zero occupation for the ground state particle number. Taking $N=2000$ and the frequencies quoted before gives $T\simeq31.9$~nK which is about a 6\% reduction from the bulk value. More details on the anisotropic oscillator may be found in Ref.~\cite{KKDJTPRA}.

\section{Density of States Method}

We have already discussed the bulk approximation in which sums are replaced with integrals. As a rough approximation for calculating the BEC transition temperature this might be sufficient; however we saw that there was a more accurate way of obtaining approximate results. In Refs.~\cite{GHPLA,GH2} it was discussed how an improved approximation in which sums were converted to integrals could be achieved by including a modified density of states factor for the harmonic oscillator. For the isotropic oscillator we had the energy levels $E_k=(k+3/2)\hbar\omega$ with degeneracy $\frac{1}{2}(k+1)(k+2)$. (See Sec.~\ref{harm}.) Grossmann and Holthaus suggested that the sum over $k$ could be converted to an integral over $k$ but with a density of states factor $\frac{1}{2}k^2+\frac{3}{2}k$ in place of simply $\frac{1}{2}k^2$ as in the bulk approximation. By a change of variable from $k$ to $E$, the density of states factor becomes
\beq
\rho(E)=\frac{1}{2}\,\frac{E^2}{(\hbar\omega)^3}+\frac{3}{2}\,
\frac{E}{(\hbar\omega)^2}\;.\label{3.8.1}
\eeq
For the anisotropic oscillator the situation is more complicated, and the density of states was parametrized by
\beq
\rho(E)=\frac{1}{2}\,\frac{E^2}{(\hbar\omega)^3}+\gamma\,
\frac{E}{(\hbar\omega)^2}\;,\label{3.8.2}
\eeq
in Refs.~\cite{GHPLA,GH2} where $\gamma$ is a constant which depends on the frequencies and which had to be evaluated numerically. (In Eq.~\ref{3.8.1} $\omega$ is the geometric mean of the three frequencies. It was later shown \cite{KKDJTPLA,KKDJTcav} how $\gamma$ could be arrived at analytically, and it is this approach we will discuss here. Furthermore it is possible to set the problem of conversion of the sums to integrals in a much wider context leading to other possible applications.

Let $\hat{H}$ be the Hamiltonian operator for some system with $E_n$ the energy levels. Let ${\mathcal N}(E)$ be the number of energy levels for which $E_n\le E$. We can write
\beq
{\mathcal N}(E)=\sum_n\theta(E-E_n)\;,\label{3.8.3}
\eeq
where $\theta(x)$ is the Heaviside distribution (or step function) defined by
\beq
\theta(x)=\left\lbrace\begin{array}{c}1,\ x>0\\
0,\ x<0\\
\frac{1}{2},\ x=0\end{array}\right.\;.\label{3.8.4}
\eeq
The aim is to treat the energy levels, or at least some part of the energy spectrum, as a continuous set rather than a discrete set by introducing the density of states \beq
\rho(E)dE&=&{\mathcal N}(E+dE)-{\mathcal N}(E)\nn
&=&\frac{d{\mathcal N}(E)}{dE}\,dE\;.\label{3.8.5}
\eeq
Since $\theta'(x)=\delta(x)$ we can write
\beq
\rho(E)=\sum_n\delta(E-E_n)\;.\label{3.8.6}
\eeq
Now define
\beq
K(t)=\int_{0}^{\infty}dEe^{-tE}\rho(E)\;,\label{3.8.7}
\eeq
which is just the Laplace transform of the density of states. If we use Eq.~\ref{3.8.6}, then $K(t)$ becomes
\beq
K(t)=\sum_ne^{-tE_n}\;,\label{3.8.8}
\eeq
The inversion formula for Laplace transforms gives
\beq
\rho(E)=\frac{1}{2\pi i}
\int_{c-i\infty}^{c+i\infty}\!\!\!\!\!\!\!\!\!dt\,e^{tE}K(t)
\;,\label{3.8.9}
\eeq
for $c\in\reals$ with $c>0$.

The importance of Eq.~\ref{3.8.7} or Eq.~\ref{3.8.9} is that if we can evaluate $K(t)$ given in Eq.~\ref{3.8.8} then the density of states may be found. The problem of evaluating Eq.~\ref{3.8.8} for a given $E_n$ is sometimes not too difficult, as in the case of the harmonic oscillator. We will do this in Sec.~\ref{secdenharm} below.

Even if it is not possible to perform the sum in Eq.~\ref{3.8.8} exactly, it is still possible to obtain a knowledge of the density of states. This is because the density of states is determined by the asymptotic behaviour of $K(t)$ as $t\rightarrow0$ which is known for a wide class of operators. We will assume that as $t\rightarrow0$
\beq
K(t)\simeq\sum_{i=1}^{k}c_it^{-r_i}+{\mathcal O}(t^{-r_k+1})
\;,\label{3.8.10}
\eeq
for some coefficients $c_i$ and powers $r_i$ with $r_1>r_2>\cdots r_k>0$. Noting that
\beq
 t^{-r_i}=\frac{1}{\Gamma(r_i)}\int_{0}^{\infty}dE\,E^{r_i-1}e^{-tE}
\;,\label{3.8.11}
\eeq
for $r_i>0$ it is easy to see from Eq.~\ref{3.8.7} that
\beq
\rho(E)\simeq\sum_{i=1}^{k}\frac{c_i}{\Gamma(r_i)}\,E^{r_i-1}
\;.\label{3.8.12}
\eeq
Thus a knowledge of the coefficients $c_i$ in Eq.~\ref{3.8.10} gives us the density of states. (The result may be extended to $r_i<0$ but it requires a more powerful method than Laplace transforms \cite{Brownell,Baltes}.)

\subsection{\it Application to the Simple Harmonic Oscillator}
\label{secdenharm}

We will show how the density of states method may be used to obtain results which agree with those found in Sec.~\ref{harm}. The $q$-potential is
\beq
q=q_{\rm gr}+q_{\rm ex}\;,\label{3.8.13}
\eeq
where
\beq
q_{\rm gr}=-\ln\left(1-ze^{-\beta E_0}\right)\;,\label{3.8.14}
\eeq
is the lowest mode or ground state contribution, and 
\beq
q_{\rm ex}=-\sum_{n>0}\ln\left(1-ze^{-\beta E_n}\right)\;,\label{3.8.15}
\eeq
represents the contribution from the excited states. We can write
\beq
q_{\rm ex}=\sum_{k=1}^{\infty}\frac{e^{k\beta(\mu-\mu_c)}}{k}
\sum_{n>0}e^{-k\beta(E_n-E_0)}\;,\label{3.8.16}
\eeq
since $\mu_c=E_0$. We now convert the sum over $n$ in Eq.~\ref{3.8.16} into an integral with the use of the density of states~:
\beq
q_{\rm ex}\simeq\sum_{k=1}^{\infty}\frac{e^{k\beta(\mu-\mu_c)}}{k}
\int_{E_1-E_0}^{\infty}dE\,\rho(E)e^{-k\beta E}\;.\label{3.8.17}
\eeq
The lower limit on the integral corresponds to the first term in the sum over $n$ in Eq.~\ref{3.8.16}.

We now need $\rho(E)$. To obtain this we first calculate $K(t)$ defined by
\beq
K(t)=\sum_{n>0}e^{-t(E_n-E_0)}\;.\label{3.8.18'}
\eeq
Because we only want the behaviour of $K(t)$ for negative powers of $t$ we can add back the $n=0$ term to obtain
\beq
K(t)&=&\sum_{n_1=0}^{\infty}\sum_{n_2=0}^{\infty}\sum_{n_3=0}^{\infty}
e^{-t\hbar(n_1\omega_1+n_2\omega_2+n_3\omega_3)}-1\nn
&=&\prod_{j=1}^{3}\left(1-e^{-t\hbar\omega_j}\right)^{-1}-1
\;.\label{3.8.12'}
\eeq
It is now straightforward to expand $K(t)$ in powers of $t$ using
\beq
(1-e^{-x})^{-1}=\frac{1}{x}+\frac{1}{2}+\frac{x}{12}+{\mathcal O}(x^3)\;.
\label{3.8.13'}
\eeq
(This expansion just involves the Bernoulli numbers \cite{WW}.) We find
\beq
K(t)\simeq c_1t^{-3}+c_2t^{-2}+c_3t^{-1}+{\mathcal O}(1)\;,\label{3.8.14'}
\eeq
where
\beq
c_1&=&(\hbar^3\omega_1\omega_2\omega_3)^{-1}\;,\label{3.8.15'}\\
c_2&=&\frac{1}{2\hbar^2}\left(\frac{1}{\omega_1\omega_2}+
\frac{1}{\omega_2\omega_3}+\frac{1}{\omega_3\omega_1}\right)\;,
\label{3.8.16'}\\
c_3&=&\frac{1}{12\hbar}\left(\frac{\omega_3}{\omega_1\omega_2}+
\frac{\omega_2}{\omega_1\omega_3}+\frac{\omega_1}{\omega_2\omega_3}+
\frac{3}{\omega_1}+\frac{3}{\omega_2}+\frac{3}{\omega_3}\right)
\;.\label{3.8.17'}
\eeq
We can now use Eq.~\ref{3.8.12} to find
\beq
\rho(E)\simeq\frac{1}{2}c_1E^2+c_2E+c_3\;.\label{3.8.18}
\eeq
The first two terms agree with the ansatz of Ref.~\cite{GHPLA,GH2} and actually represent a direct evaluation of their coefficient $\gamma$. It is easy to see by comparing Eq.~\ref{3.8.2} with Eq.~\ref{3.8.18} that
\beq
\gamma=\frac{1}{2}(\omega_1+\omega_2+\omega_3)
(\omega_1\omega_2\omega_3)^{-1/3}\;.\label{3.8.19}
\eeq
Finally we note that for the isotropic oscillator we have $c_1=(\hbar\omega)^{-3}, c_2=\frac{3}{2}(\hbar\omega)^{-2}$ and $c_3=(\hbar\omega)^{-1}$ which agrees with a direct conversion of the degeneracy factor $\frac{1}{2}(k+1)(k+2)$.

We now return to the evaluation of $q_{\rm ex}$ in Eq.~\ref{3.8.17} using Eq.~\ref{3.8.18}. Before doing the integrals we will make the simplifying assumption that $\beta(E_1-E_0)<<1$ so that we can replace the lower limit on the integral with zero. We then have
\beq
q_{\rm ex}&=&\sum_{k=1}^{\infty}\frac{e^{k\beta(\mu-\mu_c)}}{k}
\int_{0}^{\infty}dE\Big(\frac{1}{2}c_1E^2+c_2E+c_3\Big)e^{-k\beta E}\nn
&=&\sum_{k=1}^{\infty}\frac{e^{k\beta\hbar\Omega\epsilon}}{k}
\left(\frac{c_1}{(k\beta)^3}+\frac{c_2}{(k\beta)^2}+\frac{c_3}{(k\beta)}
\right)\;,\label{3.8.20}
\eeq
if we use Eq.~\ref{3.7.97} with $\mu_c=E_0=\frac{1}{2}\hbar(\omega_1+\omega_2+\omega_3)=
\frac{3}{2}\hbar\Omega$. The sums in Eq.~\ref{3.8.20} are just polylogarithms, and we obtain
\beq
q_{\rm ex}&=&c_1\beta^{-3}{\rm Li}_4(e^{-\beta\hbar\Omega\epsilon})+
c_2\beta^{-2}{\rm Li}_3(e^{-\beta\hbar\Omega\epsilon})\nn
&&+c_3\beta^{-1}{\rm Li}_2(e^{-\beta\hbar\Omega\epsilon})\;.\label{3.8.21}
\eeq
The expansion of the RHS in powers of $\alpha=\beta\hbar\Omega$ generates Eq.~\ref{3.7.102}. The resulting discussion is the same as before.

\subsection{\it Application to BEC in Cavities}

Another problem which can be studied is the situation where a system of spin-0 bosons is confined to a cavity of arbitrary shape. The results for the $q$-potential in Eqs.~\ref{3.8.13}--\ref{3.8.17} still apply, but of course the density of states factor will be different than for the harmonic oscillator potential. We can concentrate on the kernel function defined in Eq.~\ref{3.8.18} as before and evaluate the coefficients $c_i$ appearing in Eq.~\ref{3.8.10} as $t\rightarrow0$. General spectral analysis \cite{Minaketal} gives
\beq
\tilde{K}(\tau)\simeq(4\pi\tau)^{-D/2}\Big\lbrace a_0+a_1\tau^{1/2}+
{\mathcal O}(\tau)\Big\rbrace\;,\label{3.8.22}
\eeq
as $\tau\rightarrow0$, where we have defined 
\beq
\tilde{K}(\tau)={\rm tr}\left(e^{-\tau(-\nabla^2+V(\x))}\right)
\;.\label{3.8.25}
\eeq
The coefficients $a_0$ and $a_1$ appearing in the asymptotic expansion 
Eq.~\ref{3.8.22} are
\beq
a_0&=&V_\Sigma\;,\label{3.8.23a}\\
a_1&=&\frac{b}{2}\sqrt{\pi}\,\partial V_\Sigma\;,\label{3.8.23b}
\eeq
where $\partial V_\Sigma$ denotes the volume of the boundary of the cavity, and $b=-1$ for Dirichlet ({\em ie} fields vanish on $\partial V_\Sigma$) and $b=+1$ for Neumann ({\em ie} normal derivative of fields vanish on $\partial V_\Sigma$). The boundary of the cavity has been assumed to be smooth and we take the spatial dimension $D\ge2$.

The case of interest to us corresponds to $\tau=\frac{t\hbar^2}{2m}$ so that the operator appearing in Eq.~\ref{3.8.25} is the Hamiltonian. The coefficients appearing in Eq.~\ref{3.8.10} are
\beq
 c_1&=&V_\Sigma\left(\frac{m}{2\pi\hbar^2}\right)^{D/2}\ \ \ \  
({\rm with}\ r_1=D/2)\label{3.8.26a}\\
c_2&=&\frac{b}{4}\partial V_\Sigma\left(\frac{m}{2\pi\hbar^2}
\right)^{(D-1)/2}\ \ \ \ ({\rm with}\ r_1=(D-1)/2)\;.\label{3.8.26b}
\eeq
We therefore obtain from Eq.~\ref{3.8.12} the density of states
\beq
\rho(E)\simeq\frac{c_1}{\Gamma(D/2)}E^{D/2-1}+
\frac{c_2}{\Gamma(\frac{D-1}{2})}E^{(D-3)/2}\;,\label{3.8.27}
\eeq
to the order we are working. The first term on the right hand side is sometimes called Weyl's theorem. (See Ref.~\cite{Baltes,Courant} for example.) From Eq.~\ref{3.8.17}, again approximating the lower limit on the integral with 0, gives
\beq
q_{\rm ex}&\simeq&\sum_{k=1}^{\infty}\frac{e^{k\beta(\mu-\mu_c)}}{k}
\left\lbrace c_1(k\beta)^{-D/2}+c_2(k\beta)^{-(D-1)/2}\right\rbrace\nn
&=&c_1\beta^{-D/2}{\rm Li}_{D/2+1}(e^{\beta(\mu-\mu_c)})+
c_2\beta^{-(D-1)/2}{\rm Li}_{(D+1)/2}(e^{\beta(\mu-\mu_c)})\nn
&=&V_\Sigma\lambda_T^{-D}{\rm Li}_{D/2+1}(e^{\beta(\mu-\mu_c)})+
\frac{b}{4}\partial V_\Sigma\lambda_T^{-(D-1)}
{\rm Li}_{(D+1)/2}(e^{\beta(\mu-\mu_c)})\;,\label{3.8.28}
\eeq
where
\beq
\lambda_T=\sqrt{\frac{2\pi\hbar^2\beta}{m}}\;,\label{3.8.29}
\eeq
defines the thermal wavelength. From this it can be seen that the expansion parameter is $\lambda_T/L$ where $L=V_{\Sigma}^{1/D}$ is a typical length scale associated with the cavity. In the infinite volume limit we can ignore the term due to the boundary and the standard result for the free Bose gas found in Sec.~\ref{chargenorel} is recovered.

We are now in a position to see how finite volume effects can modify the temperature associated with BEC in a finite cavity. For a system confined to a finite volume the energy levels will be discrete in general, so that the criterion for BEC as symmetry breaking discussed in Sec.~\ref{secgeneral} is not met and symmetry breaking and a first order phase transition will not occur. However in the bulk limit ({\em ie} the infinite volume limit) if $D\ge3$ we do have BEC as symmetry breaking. Therefore on physical grounds we expect that for a very large, but finite, volume the specific heat should be smooth, but should approach the non-smooth behaviour as the infinite volume limit is taken. For large volumes, although the chemical potential will not be able to reach its critical value, it can become arbitrarily close much as in the harmonic oscillator potential discussed in Sec.~\ref{harm}. This should correspond to a sudden rise in the number of particles in the ground state.

Using $N_{\rm ex}=\beta^{-1}\frac{\partial}{\partial\mu}q_{\rm ex}$ with $q_{\rm ex}$ given in Eq.~\ref{3.8.28} results in 
\beq
N_{\rm ex}\simeq V_\Sigma\,\lambda_T^{-D}
{\rm Li}_{D/2}(e^{\beta(\mu-\mu_c)})+
\frac{b}{4}\partial V_\Sigma\lambda_T^{-(D-1)}
{\rm Li}_{(D-1)/2}(e^{\beta(\mu-\mu_c)})\;.\label{3.8.30}
\eeq
The ground state particle number is
\beq
N_{\rm gr}=\left\lbrack e^{\beta(\mu_c-\mu)}-1\right\rbrack^{-1}\;,
\label{3.8.31}
\eeq
as usual. Our expansion parameter is $\xi=L/\lambda_T$ with $\xi>>1$. Let
\beq
\partial V_\Sigma=\kappa L^{D-1}\;,\label{3.8.32}
\eeq
where $\kappa$ is a dimensionless constant which depends only on the shape of the boundary. If we let $f$ be the fraction of particles in the ground state at the specific heat maximum, then
\beq
\beta(\mu_c-\mu)=\ln\Big(1+\frac{1}{fN}\Big)\;.\label{3.8.33}
\eeq
Assuming that $fN>>1$ we have
\beq
\beta(\mu_c-\mu)\simeq\frac{1}{fN}<<1\;.\label{3.8.34}
\eeq
We will approximate
\beq
N_{\rm ex}\simeq\xi^D{\rm Li}_{D/2}(e^{-\frac{1}{fN}})+\frac{b\kappa}{4}
\xi^{D-1}{\rm Li}_{(D-1)/2}(e^{-\frac{1}{fN}})\;.\label{3.8.35}
\eeq
For $D>3$ the order of the polylogarithm is larger than 1 and we may further simplify this expression to
\beq
N_{\rm ex}\simeq\xi^D\zeta_R(\frac{D}{2})+\frac{b\kappa}{4}
\xi^{D-1}\zeta_R(\frac{D-1}{2})\;.\label{3.8.36}
\eeq
In the case of $D=3$ we use ${\rm Li}_1(z)=-\ln(1-z)$ and approximate
\beq
{\rm Li}_1(e^{-\frac{1}{fN}})\simeq\ln(fN)\;.\label{3.8.37}
\eeq
Thus for $D=3$ we have
\beq
N_{\rm ex}\simeq\xi^3\zeta_R(\frac{3}{2})+\frac{b\kappa}{4}\xi^2\ln(fN)\;.
\label{3.8.38}
\eeq
In either case we have
\beq
N_{\rm ex}=N-N_{\rm gr}=(1-f)N\;.\label{3.8.39}
\eeq

For $D=3$ Eq.~\ref{3.8.38} and Eq.~\ref{3.8.39} give us
\beq
(1-f)N\simeq
\zeta_R(\frac{3}{2})\xi^3+\frac{b\kappa}{4}\xi^2\ln(fN)\;,
\label{3.8.40}
\eeq
resulting in a cubic equation for $\xi$. The bulk value of $\xi$, call it $\xi_0$, is given by
\beq
N=\zeta_R(\frac{3}{2})\xi_{0}^{3}\;.\label{3.8.41}
\eeq
We are assuming here that the boundary corrections to the bulk result are small, so that $\xi$ in Eq.~\ref{3.8.40} should be close to the bulk value $\xi_0$. Write
\beq
\xi=\xi_0(1+\eta)\;,\label{3.8.42}
\eeq
where $\eta$ is small. Working to first order in small quantities we obtain
\beq
(1-f)N\simeq
\zeta_R(\frac{3}{2})\xi_{0}^{3}(1+3\eta)+\frac{b\kappa}{4}
\xi_{0}^{2}\ln(fN)\;.\label{3.8.43}
\eeq
It is easily seen that
\beq
\eta=-\frac{1}{3}-
\frac{\kappa b}{12\lbrack\zeta_R(\frac{3}{2})\rbrack^{2/3}}
\,\frac{\ln(fN)}{N^{1/3}}\;.\label{3.8.44}
\eeq
If $T_0$ denotes the bulk temperature, then
\beq
T\simeq T_0(1+2\eta)\;.\label{3.8.45}
\eeq
(This follows trivially from Eq.~\ref{3.8.42} since $\xi\propto\lambda_T^{-1}\propto T^{1/2}$.) In the case of Neumann boundary conditions where $b=+1$ we can conclude that the temperature is lowered from the bulk value. For Dirichlet boundary conditions the boundary correction to the bulk result tends to increase the temperature, but the overall effect depends upon the relative size of the two terms in Eq.~\ref{3.8.44}.

For spatial dimensions $D>3$ we have in place of Eq.~\ref{3.8.43}
\beq
(1-f)N\simeq\zeta_R(\frac{D}{2})\xi^D+\frac{b\kappa}{4}
\zeta_R(\frac{D-1}{2})\xi^{D-1}\;.\label{3.8.46}
\eeq
The bulk value of $\xi$ is given by
\beq
N=\zeta_R(\frac{D}{2})\xi_{0}^{D}\;.\label{3.8.47}
\eeq
Writing $\xi$ as in Eq.~\ref{3.8.42} and working to first order in $\eta$ gives
\beq
\eta=-\frac{f}{D}-\frac{\kappa b}{4D}\,\frac{\zeta_R(\frac{D-1}{2})}{\lbrack\zeta_R(\frac{D}{2})
\rbrack^{1-1/D}}N^{-1/D}\;.\label{3.8.48}
\eeq
Eq.~\ref{3.8.45} still relates $T$ to the bulk temperature.

BEC in a rectangular cavity has received numerous treatments. (See Refs.~\cite{Pathriaetal} for a review of the earlier work, and Ref.~\cite{GHcav} for a recent analysis.) More details on the approach presented here can be found in Ref.~\cite{KKDJTcav}.


\section*{Acknowledgements}

I would like to thank my collaborator Klaus Kirsten for his great help during the course of our joint work. He has improved my understanding of the material presented in these lectures enormously. I am also grateful to John Smith for many useful discussions we have had during the time he was my Ph.~D. student. In particular I would like to thank him for emphasizing the usefulness of Weyl's theorem. Much help on numerical methods and other things was provided by L~~D.~L.~Brown. Finally I would like to thank Mahler (1988--1996) for her devoted friendship over this period. She helped me in more ways than she could possibly know, and I dedicate these lectures to her memory.

\end{document}